\documentclass[12pt,notitlepage,nofootinbib,superscriptaddress,showpacs,showkeys]{revtex4-2}

\usepackage[T1]{fontenc}
\usepackage{lmodern}
\usepackage{amsmath,amssymb,amsthm}
\usepackage{array}
\usepackage{textcomp}
\usepackage{setspace}
\usepackage{graphicx}
\usepackage{float}
\usepackage[bottom]{footmisc}
\usepackage{needspace}

\graphicspath{{./figures/}}
\usepackage{hyperref}          

\usepackage{mathrsfs}

\begin{document}
	
		\title{Cosmological Black hole Candidates: A Detailed Analysis of McVittie, Culetu, Sultana-Dyer, and Glass-Mashhoon Spacetimes}
	
	\author{Mahdi Esfandiar}
	\thanks{Electronic address: \href{mahdi.esfandiar@ut.ac.ir}{mahdi.esfandiar@ut.ac.ir}}
	\affiliation{Department of Physics, University of Tehran,
		Tehran , Iran}
	
	\author{Fatimah Shojai}
	\thanks{Electronic address: \href{fshojai@ut.ac.ir}{fshojai@ut.ac.ir (Corresponding author)}}
	\affiliation{Department of Physics, University of Tehran,
		Tehran , Iran}

	\author{Omid Zamani Jamshidi}
	\thanks{Electronic address: \href{omidjamshidi@ut.ac.ir}{omidjamshidi@ut.ac.ir}}
	\affiliation{Department of Physics, University of Tehran,
		Tehran , Iran}

	\author{Soran Zoorasna}
\thanks{Electronic address: \href{zoorasna@ut.ac.ir}{zoorasna@ut.ac.ir}}
\affiliation{Department of Physics, University of Tehran,
	Tehran , Iran}
	
	\setstretch{1.6}
	\begin{abstract}
	This paper investigates the existence of cosmological black holes by analyzing the properties of trapping horizons in detail, based on Hayward's formalism of future outer and past inner trapping horizons, in several dynamical spacetimes embedded in an expanding universe. Through a detailed examination of the McVittie, Culetu, and Sultana--Dyer metrics, as well as the generalized Glass--Mashhoon solution, we evaluate the existence and characteristics of trapping horizons and energy conditions. The Glass--Mashhoon solution provides an analytical model for spherical stellar collapse. However, it is shown that, as long as certain conditions are satisfied, it lacks suitable future outer trapping horizons, meaning it does not represent a cosmological black hole. As a result, the McVittie class of solutions also fails to describe a cosmological black hole. Conversely, the Culetu and Sultana--Dyer spacetimes can describe a cosmological black hole in the matter-dominated early universe, provided that the relevant energy conditions are satisfied.
	\end{abstract}

	\maketitle
	
	\setstretch{1}
	\newpage

	\clearpage
	
		\tableofcontents
	\newpage 
	\section{Introduction}
Cosmological black holes (BHs) \cite{Novikov1967},\cite{hawking1971} are crucial for understanding the interplay between BH physics and cosmology. While isolated BHs are traditionally described by stationary and asymptotically flat metrics  \cite{stationary BH1},\cite{stationary BH2},\cite{stationary BH3},\cite{hawking1973},  embedded BHs in expanding universes require a more dynamic framework \cite{Faraoni}. These solutions demonstrate the effects of the cosmic expansion on both the BH horizon and the energy conditions (ECs) of the BH source. 

This paper investigates candidates for cosmological BHs within Hayward’s quasi-local framework.\cite{Hayward} In our approach, a cosmological BH is understood as a dynamical BH \cite{Hayward},\cite{Ashtekar Krishnan 2004} embedded in an expanding universe \footnote{typically matter dominated}. Quasi-locally, the dynamical BH is characterized by the presence of a future outer trapping horizon (FOTH), while the cosmological expansion is associated with a cosmological trapping horizon which, in expanding backgrounds, is usually of past type (typically a past inner trapping horizon, PITH).
 The motivation for adopting Hayward's terminology is the compatibility of the concept of the FOTH with the general laws governing dynamical BHs. This type of horizon is a three-dimensional spacelike or null hypersurface foliated by outer marginally trapped surfaces (MTSs) and provides a suitable candidate for defining an evolving horizon in the context of dynamical BHs.
FOTHs possess cross-sections with non-decreasing area. This behavior corresponds to a generalized version of the "second law of thermodynamics" for dynamical BHs. Furthermore, variations in the area along the FOTH are determined by the "trapping gravity" which is a generalization of surface gravity to spacetimes lacking a timelike Killing vector field, as well as by the energy flux. These concepts provide a local formulation of the "first law of thermodynamics" of evolving horizons.
Finally, the total trapping gravity associated with a compact outer MTS admits an upper bound, which serves as an analogue of the "zeroth law of thermodynamics" for dynamical BHs.\cite{stationary BH1},\cite{stationary BH3},\cite{Faraoni},\cite{Hayward 1998},\cite{Hayward 2004}
Another key motivation is the effort to define a notion of a dynamical BH in situations where an event horizon may not exist. An event horizon is a non-local concept that requires the tracing of future-directed null geodesics to future null infinity, and thus, determining the global causal structure of spacetime. This is generally not feasible in dynamical spacetimes.\cite{stationary BH1},\cite{stationary BH3},\cite{Hayward 2006}
Accordingly, the concept of a dynamical BH and its associated laws, independent of the event horizon, necessitates  introducing  the local notion of a trapping horizon (TH), which is accessible to physical observers. 

On the other hand, to define a quasi-local analogue of cosmological horizons in an expanding universe within Hayward’s formalism, we introduce past type trapping horizons (PTHs). A PTH is a three-dimensional hypersurface foliated by past marginal two-surfaces. It separates an untrapped region from a past-trapped (anti-trapped) region and plays the role of a quasi-local cosmological-horizon analogue.\cite{Faraoni,Hayward}
	More concretely, since we consider spherically symmetric dynamical solutions that are asymptotically expanding FLRW, we expect that the spacetime asymptotically contains the FLRW cosmological TH. Therefore, in addition to an FOTH, we also expect the existence of a PTH interpreted as a cosmological TH. In general, in an expanding universe the cosmological TH is represented by a past inner trapping horizon (PITH) \cite{Hayward}, \cite{Thakurta BH}\footnote{p.2}, particularly in a matter-dominated background. Along increasing areal radius (with the ``outward'' direction defined by the global asymptotic structure), the spacetime transitions from an untrapped region to an anti-trapped region, where the future-directed ingoing and outgoing radial null rays both diverge.
	Moreover, if the null energy condition (NEC) holds on the PITH over a given time interval, then, using the definition of a PITH together with the Raychaudhuri equation \cite{Faraoni}, one can show that its area (more precisely, its area form) increases along the intrinsic evolution direction of the horizon \footnote{The direction tangent to the horizon and normal to the marginal two-spheres}. The area is constant only if the null energy flux and the internal shear along the ingoing null direction vanish. In that case the PITH is null. In the generic case with NEC satisfied, the PITH is typically timelike and has increasing area \cite{Hayward}.
	Therefore, a PITH can serve as an appropriate analogue of a cosmological TH, especially for spherically symmetric dynamical spacetimes in an early matter-dominated universe. Imposing the NEC on the PITH further supports interpreting this horizon as a physically acceptable cosmological TH.
	
	We note, however, that there are examples in which a past outer trapping horizon (POTH) can also represent a cosmological TH, depending on the matter distribution of the spacetime \cite{new3}. In general, to determine whether a given POTH corresponds to a white-hole (WH) TH or a cosmological TH, one may use additional global and/or asymptotic diagnostics. If the TH surrounds an observer’s Hubble patch in an asymptotically FLRW region and asymptotes to the FLRW cosmological TH, it should be classified as a cosmological TH. In contrast, if there is a spacetime singularity to its past together with a Kruskal-like causal structure, it should be classified as a WH TH. For the spacetimes studied in this work in an early matter-dominated universe, we adopt the PITH as the cosmological TH.

Given that dynamical BH laws can be formulated using the FOTH concept, and that cosmological THs in expanding backgrounds can be formulated as P(I)THs, studying cosmological BHs within the Hayward formalism appears to be a particularly promising approach. In this framework, one assumes the existence of at least an FOTH (associated with the inner curvature) together with a P(I)TH, which provides a concrete setting to analyze the physical properties of cosmological BH candidates in an expanding universe.

Prominent examples of cosmological BH solutions include the McVittie metric \cite{McVittie1933} and its extensions \cite{glass1976}, which describe spherically symmetric configurations embedded in Friedmann-Lemaître-Robertson-Walker (FLRW) spacetime. 
The McVittie spacetime, introduced in 1933 \cite{McVittie1933}, provides a well-studied solution that interpolates between the Schwarzschild and FLRW metrics depending on the scale factor evolution. However, its global structure and interpretation as a BH in an expanding universe have been extensively debated \cite{kaloper2010},\cite{McVittie de sitter},\textbf{\cite{nolan2017}. }

Our analyses here, in isotropic and Painlev\'e-Gullstrand (PG) coordinates, have provided insights into the THs and causal structures of the spatially flat McVittie spacetime in an expanding universe. Our studies suggest that this spacetime does not represent a cosmological BH within our Hayward-inspired quasi-local framework, primarily because it does not admit an FOTH.

In addition to McVittie's solution, there are other candidates for cosmological BHs. Recently, it is shown that, the Thakurta spacetime \cite{Thakurta}, which is a conformal Schwarzschild spacetime constructed with the Schwarzschild coordinates, does not describe a cosmological BH \cite{Thakurta BH}. Similarly, the McClure-Dyer solution \cite{McClure-Dyer solution} which is obtained by rewriting the Thakurta spacetime in isotropic coordinates, also does not describe a cosmological BH \cite{sato}.

Other spacetimes such as the Sultana-Dyer \cite{sultana} and Culetu \cite{culetu} metrics  have been proposed as cosmological BH models. The Culetu spacetime, constructed as a conformal Schwarzschild solution in PG coordinates, 
	does describe a cosmological BH. However, its source---comprising a mixture of homogeneous and inhomogeneous fluids---
	only partially satisfies the ECs at sufficiently early times in a matter-dominated universe \cite{culetu,Culetu2018}. 
	The Sultana--Dyer spacetime, constructed as a conformal Schwarzschild solution in Kerr--Schild coordinates, 
	incorporates a combination of perfect and null fluids as its source \cite{sultana}. 
	Similar to the Culetu case, although it satisfies the condition for describing a BH in an expanding universe, 
	its source only partially satisfies the ECs at sufficiently early times in the matter-dominated universe. \cite{sato}.

The Glass-Mashhoon solution \cite{glass1976} provides  an extended class of spherically symmetric metrics derived from Einstein's equations under McVittie's assumption. Under certain conditions, this solution asymptotes to FLRW and Schwarzschild spacetimes. In this paper, we examine the properties of the Glass-Mashhoon solution,  such as calculating density, pressure, and the Misner-Sharp-Hernandez (MSH) mass. We also examine the ECs and show that their validity in the general case is restricted. 
In a matter--dominated universe, fixing the sign of certain free parameters in the Glass-Mashhoon solution, yields regions where these the ECs are explicitly satisfied. 
However, as we calculated in detail for a special case, namely the matter--dominated 
spatially curved McVittie spacetime with positive curvature, 
there are additional regions beyond these explicit ones where the ECs are still satisfied.
Due to the presence of free parameters in the Glass--Mashhoon solution, 
it appears that certain conditions could be satisfied for it to describe 
a cosmological BH in Hayward's terminology. 
However, we have shown that it possesses only past-type THs 
in an expanding universe. 
Therefore, it fails to represent a cosmological BH.
By fixing some of the parameters and studying specific cases, one can analyze the properties of new subclasses of the solution by  investigating various types of MTSs.
This paper examines the spatially flat and curved McVittie solutions \cite{McVittie1933}. The analysis includes identifying and classifying THs and evaluating the associated ECs.


In this paper, we analyze several cosmological BH candidates, focusing on their TH structures, ECs, and physical interpretations. The analysis employs different coordinate systems tailored to each spacetime: the isotropic coordinates are used for the 
 McVittie and Glass-Mashhoon spacetimes, PG-Schild coordinates for the Culetu spacetime, and Kerr-Schild coordinates for the Sultana-Dyer spacetime. 

It is well known that the location and types of THs do not change under different spherical foliations of spacetime. This is due to the fact that the expansion parameters and their Lie derivatives are scalars on the two-dimensional spacetime submanifold described by time and radial coordinates \cite{faraoni TH}. Moreover, the signs of the expansion parameters and their Lie derivatives are independent of the choice of the ingoing and outgoing null vector fields. It is only necessary that these vectors are well defined on the THs \cite{expansion}. For the sake of clarity, and where applicable, the interpretation of MTSs and THs is also provided in PG coordinates, in addition to the coordinate systems mentioned above, for certain spacetimes considered in this work.

The organization of the paper is as follows: Section \ref{II} studies the THs of the spatially flat McVittie spacetime in isotropic and PG coordinates. Section \ref{III} explores the Culetu spacetime, and Section \ref{IV} focuses on the Sultana--Dyer metric. Finally, in Section \ref{V}, analyzes the Glass--Mashhoon solution and its physical implications. We summarize our results in Section \ref{VI}.
	\section{Spatially flat McVittie spacetime} \label{II}
		\subsection{Trapping horizons in isotropic coordinates} \label{II.A}
		McVittie spacetime is in fact a generalization of Schwarzschild-deSitter spacetime \cite{McVittie de sitter}. It describes a central body immersed in a Friedmann universe that is not necessarily deSitter spacetime. The McVittie geometry is time-dependent and its source is a perfect fluid that is assumed to have no radial flow, i.e. $T^1_0=0$ \cite{McVittie1933}.  Spatially flat McVittie spacetime is described by the following line element in the isotropic coordinates:
			\begin{equation}\label{new2}
				ds^2 = - A^2(\overline{r}, t)	dt^2 + B^2(\overline{r}, t) d\overline{r}^2 + R^2 d\Omega^2
			\end{equation}
where:
\begin{align}
A (\bar{r}, t) 
&= \frac{1 - \frac{m_0}{2 \bar{r} a(t)}}{1+ \frac{m_0}{2 \bar{r} a(t)}}
\label{new3a} \\[6pt]
B(\bar{r}, t) 
&= a(t) \left( 1 + \frac{m_0}{2 \bar{r} a(t)}\right)^2
\label{new3b} \\[6pt]
R(\bar{r}, t) 
&= \bar{r} B(\bar{r}, t)
\label{new3c}
\end{align}
and $m_{0}$ is the mass constant of the metric which is assumed to be positive here. This spacetime  contains a spacelike, weak gravitational curvature singularity\footnote{At this singularity, the determinant of the spatial part of \ref{new2} and thus the differential volume is nonzero.} at $\bar{r}={m_0}/{(2 a(t))}$.  Another singularity is located at $\bar{r} = 0$. McVittie originally interpreted the metric \eqref{new2} as representing a point mass situated at $\bar{r} = 0$. However, in general, this point mass is enclosed by the singular surface at $\bar{r} = {m_0}/{(2 a(t))}$ which has an unclear physical interpretation \cite{McVittie1933}, \cite{McVittie singularity}, \cite{sussman1985},\cite{ferraris1996}. Metric \eqref{new2} is asymptotic to the spatially flat FLRW spacetime when $m_{0}=0$ and approaches the Schwarzschild metric when $a(t)=1$.

In the following, we address the ECs of the matter distribution in the spatially flat McVittie spacetime outside the curvature singularity.  
The Einstein equations corresponding to the line element \eqref{new2}, with a perfect fluid source, yield the following expressions for the energy density and pressure:\cite{Faraoni,McVittie energy condition}
	\begin{align}
	\rho &= \frac{3H^2}{8\pi} , \label{NM1} \\
	P &= -\frac{1}{8\pi}\left(3H^2 + 2\dot{H}\,\frac{1+\tfrac{m_0}{2a(t)\bar{r}}}{1-\tfrac{m_0}{2a(t)\bar{r}}}\right) . \label{NM2}
\end{align}
	\noindent
\textbf{NEC:}
The NEC for the energy--momentum tensor of a perfect fluid \cite{Faraoni}, corresponding to Eqs.~\eqref{NM1} and \eqref{NM2}, simplifies 
to the following form
	\begin{equation}
	\rho + P = -\frac{\dot{H}}{4\pi}\,\frac{1+\tfrac{m_0}{2a(t)\bar{r}}}{1-\tfrac{m_0}{2a(t)\bar{r}}} \geq 0 . 
	\label{NM3}
\end{equation}
Outside the curvature singularity 
\(\left(\bar{r} > {m_0}/{2a(t)}\right)\), 
for a decelerating universe \((\dot{H} \leq 0)\), 
the NEC for the spatially flat McVittie spacetime is always satisfied. 
An example is the matter-dominated flat McVittie case,
$a(t) \sim t^{2/3}, 
\, H = {2}/{3t}, 
\, \dot{H} = -{3H^2}/{2}$,
where the NEC is always satisfied.

	\noindent
\textbf{WEC:} The WEC for the energy--momentum tensor of a perfect fluid \cite{Faraoni}, 
is equivalent to the non-negativity of Eqs.~\eqref{NM1} and \eqref{NM3}.

It is clear that based on Eq.~\eqref{NM1} since $\rho \geq 0$, the WEC is equivalent to the NEC. 
As we have shown, outside the curvature singularity of McVittie, 
for a decelerating universe $(\dot{H} \leq 0)$, including the matter-dominated spatially flat McVittie case, 
this condition will be satisfied.  

	\noindent
\textbf{DEC:} The DEC for the energy-–momentum tensor of a perfect fluid is equivalent to the validity of the following relations \cite{Faraoni}:
\begin{equation}
	\rho + P \geq 0 \, , \quad \rho - P \geq 0
	\label{NM4}
\end{equation}  
which, in the regions outside the curvature singularity, simplify according to Eqs.~\eqref{NM1}, \eqref{NM2}, and \eqref{NM3} as follows:
	\begin{equation}
	-3H^2 \,\frac{1-\tfrac{m_0}{2a(t)\bar{r}}}{1+\tfrac{m_0}{2a(t)\bar{r}}} \leq \dot{H} \leq 0 \, .
	\label{NM5}
\end{equation}
As an example, in the matter-dominated spatially flat McVittie background, based on Eq.~\eqref{NM5}, for 
\(\bar{r} \geq \tfrac{3m_0}{2a(t)}\) (or equivalently, in terms of the areal radius \eqref{new3c}, 
\(R \geq {8}/{3} m_0\)), the DEC holds at all times. 
Moreover, in the interval \({m_0}/{2a(t)} < \bar{r} < {3m_0}/{2a(t)}\), 
or equivalently \(2m_0 < R < {8}/{3} m_0\), this condition is violated at all times.

\noindent
\textbf{SEC:} The SEC for the energy–momentum tensor of a perfect fluid \cite{Faraoni}, 
	under the validity of Eq.~\eqref{NM3}, 
	simplifies as follows:
	\begin{equation}
	\rho + 3P = -\frac{3}{4\pi} \left(H^2 + \dot{H}\,\frac{1+\tfrac{m_0}{2a(t)\bar{r}}}{1-\tfrac{m_0}{2a(t)\bar{r}}}\right) \geq0 
	\label{NM6} 
	\end{equation}
According to Eqs.~\eqref{NM3} and \eqref{NM6}, the SEC outside the curvature singularity will hold provided the following relation is satisfied:

	\begin{equation}
	\dot{H} \leq -H^2 \,\frac{1-\tfrac{m_0}{2a(t)\bar{r}}}{1+\tfrac{m_0}{2a(t)\bar{r}}} \leq 0 \, .
	\label{NM7}
\end{equation}
In the matter-dominated spatially flat McVittie spacetime, it can be shown that, according to Eq.~\eqref{NM7}, this condition is always satisfied.


To investigate the existence and nature of the THs of metric \eqref{new2}, it is first necessary to compute the corresponding outgoing and ingoing radial null vector fields. Setting ${ds}^2={d\Omega}^2=0$ in \eqref{new2}, we obtain:
						\begin{equation}
				\frac{d\bar{r}}{dt} = \pm \frac{A}{B}
\label{new4}
			\end{equation}
			 Noting Eqs.~\eqref{new3a} and \eqref{new3b}, ${A}/{B}>0$ outside the curvature singularity region, i.e., $\bar{r}>{m_0}/(2 a(t))$. Accordingly, the outgoing and ingoing null vector fields are, respectively:
						\begin{equation}
				\begin{cases}
					l^p = (\frac{1}{A}, \frac{1}{B}) \\
					n^p = (\frac{1}{A}, -\frac{1}{B})
				\end{cases}
\label{new5}
			\end{equation}
			with $p=0,1$ and the normalization is set to $ l_p n^p = -2$\footnote{Moreover, the ingoing and outgoing null vector fields satisfy  $g_{ab} n^a n^b = g_{ab} l^a l^b = 0$, (where $a,b=1,2$). Therefore, only one free function ultimately remains out of the four degrees of freedom associated with these two null vector fields. This has no effect on the final results.
}.
		
It is useful to remember that the expansions along the null vector fields are given by:\footnote{According to Eqs.~\eqref{new3a}, \eqref{new3b}, and \eqref{new4},  $ {d\bar{r}}/{dt} = 0 $ in the limit of $a(t)\rightarrow \infty$. Consequently, the light cone closes and it is impossible to distinguish between the ingoing and outgoing null vector fields. As a result, the definition of the null expansion parameter in Eq.~\eqref{new6} becomes meaningless in this coordinate system. Therefore, one should analyze the behavior of the null vector fields in an appropriate coordinate system, where the definition of the outgoing and ingoing vector fields is not indeterminate \cite{McVittie de sitter}.}.
						\begin{equation}
				\begin{cases}
					\theta_+ = \frac{1}{S}\mathcal{L}_+ S = \frac{2}{R} l^p \nabla_p R \\
					\theta_- = \frac{1}{S} \mathcal{L}_- S = \frac{2}{R} n^p \nabla_p R
				\end{cases}
				\label{new6}
			\end{equation}
where $\pm$ denotes the outgoing and ingoing expansion parameters, respectively.
			Here,	$\mathcal{L}_+$ and $\mathcal{L}_-$ represent the Lie derivatives along the respective null vectors. $S=4\pi R^2$ is the area of the 2-spheres of symmetry of the metric \eqref{new2} with the areal radius $R$. By definition, a MTS is defined by $\theta_+\theta_-=0$. It is given by $\nabla^pR\nabla_pR=0$ or $R=2M_{MSH}$ where the MSH mass \cite{hernandez1966} is given by  
	\begin{equation}
M_{MSH}=\frac{R}{2} (1- h^{\alpha \beta} \nabla_{\alpha} R \nabla_{\beta} R)
\label{new32}
\end{equation}
	$h_{\alpha \beta}$ is the  metric of the 2-space normal to the 2-spheres of symmetry:	
	$ds^2=h_{\alpha \beta} dx^{\alpha} dx^{\beta} +R^2 d\Omega^2$
 ($\alpha$, $\beta =0,1$). Thus, the MSH mass can be evaluated for the spatially flat McVittie spacetime \eqref{new2} as \cite{Faraoni}
			\begin{equation}\label{m}
		  M_{MSH}=m_0+\frac{1}{2}H^2R^3
			\end{equation}
			which approaches the MSH mass of the spatially flat FLRW spacetime when $m_0=0$ \cite{FLRW}.
			
									A TH is the closure of a hypersurface foliated by MTSs, classified by the signs of the Lie derivatives along the null vectors \cite{Hayward}. Taking into account the symmetry between $\theta_{+}$ and $\theta_{-}$, consider  that the two-surface $\Sigma$ is described by $\theta_{-} = 0$. A two-surface is called \textit{future marginally trapped} if $\theta_{+}|_{\Sigma}$ is negative, \textit{bifurcating} if $\theta_{+}|_{\Sigma} = 0$, and \textit{past marginally trapped} if $\theta_{+}|_{\Sigma}$ is positive. Additionaly,  the marginally trapped two-surface is classified as \textit{outer} if $\mathcal{L}_{+} \theta_{-}|_{\Sigma}$ is negative, \textit{degenerate} if $\mathcal{L}_{+} \theta_{-}|_{\Sigma}$ is zero, or \textit{inner} if $\mathcal{L}_{+} \theta_{-}|_{\Sigma}$ is positive, based on the sign of $\mathcal{L}_{+} \theta_{-}|_{\Sigma}$.
			
			Using Eqs.~\eqref{new5} and \eqref{new6},  the corresponding expansion parameters of the spatially flat McVittie metric \eqref{new2}, can be derived as:
						\begin{equation}
				\theta_\pm = 2 \left( H(t) \pm \frac{A}{\overline{r} B} \right) 
				\label{new7}
			\end{equation}		
				\needspace{8\baselineskip}
				For positive $H$, according to Eq.~\eqref{new7}, $\theta_+>\theta_-$.
			Furthermore, $\theta_-=0$ and $\theta_+>0$ for the THs obtained by solving the equation $H={A}/{(\bar{r}B)}$. \footnote{For the spatially flat McVittie metric, there is an explicit relationship
				between the metric components and the areal radius, obtained by combining
				Eqs.~\eqref{new3b} and \eqref{new3c}:
				\begin{equation*}
					\left(\frac{1-\dfrac{m_0}{2\bar r a(t)}}{1+\dfrac{m_0}{2\bar r a(t)}}\right)^2
					= 1-\frac{2m_0}{R}.
				\end{equation*}
				As a result of this useful relation, it is easy to verify that the position of the TH(s)
				corresponding to Eq.~\eqref{new7} is consistent with $R=2M_{MSH}$.}
			
						Since $\theta_+|_{\theta_-=0}>0$, the corresponding THs are PTHs.

			In the following, we solve $\theta_{-}=0$ to determine the THs
				of this spacetime outside the curvature singularity, i.e., for 
				$\bar{r} \geq {m_{0}}/{2a(t)}$.
					For simplicity, we rewrite Eq.~\eqref{new7} as:
			\begin{equation}
				\tilde{\theta}_{\pm} \;=\; 2\left(\tilde{H}(\tilde{t}) \,\pm\, \frac{A}{\tilde{\bar{r}}\,B}\right),
				\label{NN1}
			\end{equation}
	where all quantities with a tilde are normalized. For example, in the case of the spatially flat McVittie solution 
			with a matter-dominated background (i.e., $\tilde{H}(\tilde{t})={2}/{3\tilde{t}}$), we have
			$\tilde{\bar{r}} = \tfrac{\bar{r}}{m_{0}}, 
			\,
			\tilde{t} = \tfrac{t}{m_{0}}, 
			\, 
			\tilde{\theta}_{\pm} = m_{0}\,\theta_{\pm}$.
			
			To determine the roots of $\theta_{-}=0$, we first analyze the sign of 
			$\theta_{-}$ in the interval 
			$\tilde{\bar{r}} \in \bigl({1}/{2a(\tilde{t})}, +\infty\bigr)$. 
			At both ends of this interval, according to Eq.~\eqref{NN1}, and for finite times 
			with a positive Hubble parameter, one finds
			\begin{equation}
				\tilde{\theta}_{-}\big|_{\tilde{\bar{r}}\to(1/2a)^{+}} = 2\tilde{H} > 0, 
				\qquad 
				\tilde{\theta}_{-}\big|_{\tilde{\bar{r}}\to +\infty^{-}} = 2\tilde{H} > 0 .
				\label{NN2}
			\end{equation}
		Therefore, for $\tilde{t}>0$, the quantity $\theta_{-}$ is always positive at 
			the boundaries $(1/2a)^{+}$ and $+\infty^{-}$. 
			From Eq.~\eqref{NN1}, the radial derivative of $\tilde{\theta}_{-}$ can be written as:
			\begin{equation}
				\partial_{\tilde{\bar{r}}}\tilde{\theta}_{-}
				= \frac{ 8 a(\tilde{t}) \bigl(4 \tilde{\bar{r}}^{2} a(\tilde{t})^{2}-8a(\tilde{t})\tilde{\bar{r}}+1\bigr)}
				{\bigl(1+2a(\tilde{t})\tilde{\bar{r}}\bigr)^{4}} .
				\label{NN3}
			\end{equation}
						A straightforward calculation shows that in the interval 
			$\tilde{\bar{r}}\in\bigl({1}/{2a(\tilde{t})}, \tilde{\bar{r}}_{(-,D)}(\tilde{t})\bigr)$, 
			one always has $\partial_{\tilde{\bar{r}}}\theta_{-}<0$. 
			At $\tilde{\bar{r}}=\tilde{\bar{r}}_{(-,D)}(\tilde{t})={(2+\sqrt{3})}/{2a(\tilde{t})}$, 
			, $\partial_{\tilde{\bar{r}}}\theta_{-}=0$. 
			For $\tilde{\bar{r}}\in\bigl(\tilde{\bar{r}}_{(-,D)}(\tilde{t}),+\infty\bigr)$, 
			one finds $\partial_{\tilde{\bar{r}}}\theta_{-}>0$.
			
		Therefore, as a function of $\tilde{\bar{r}}$, $\theta_{-}$ first decreases and then increases, 
			possessing a local minimum at $\tilde{\bar{r}}_{(-,D)}(\tilde{t})$. 
			Since, according to Eq.~\eqref{NN2}, $\theta_{-}$ is positive at both ends of the interval 
			$(1/2a(\tilde{t}), +\infty)$, we examine the sign of $\theta_{-}$ at its local minimum. 
			Substituting $\partial_{\tilde{\bar{r}}}\theta_{-}=0$ into Eq.~\eqref{NN1} and simplifying, we obtain
			\begin{equation}
				\min_{\tilde{\bar{r}}\in(1/2a(\tilde{t}),\,+\infty)} \tilde{\theta}_{-} 
				= \tilde{\theta}_{-}\big|_{\tilde{\bar{r}}=\tilde{\bar{r}}_{(-,D)}}
				= \tilde{H}(\tilde{t}) - \frac{1}{3\sqrt{3}} .
				\label{NN4}
			\end{equation}
						Hence, for times when $\tilde{H}(\tilde{t})>{1}/{3\sqrt{3}}$, 
			we have $\theta_{-}>0$ throughout $\tilde{\bar{r}}\in(1/2a(\tilde{t}),+\infty)$, and no root exists. 
			At $\tilde{H}(\tilde{t})={1}/{3\sqrt{3}}$, the minimum value $\theta_{-}=0$ occurs at 
			$\tilde{\bar{r}}_{(-,D)}$. Thus, $\tilde{H}={1}/{3\sqrt{3}}$ marks the time of formation of the first root 
			of $\theta_{-}=0$. This root is degenerate, since both $\theta_{-}=0$ and 
			$\partial_{\tilde{\bar{r}}}\theta_{-}=0$ occur simultaneously.
					For $\tilde{H}(\tilde{t})<{1}/{3\sqrt{3}}$, the minimum of $\theta_{-}$ in the interval 
			$(1/2a(\tilde{t}),+\infty)$ becomes negative, and two distinct roots of $\theta_{-}=0$ appear, 
			which separate from each other as time evolves.
			
		Therefore, the spatially flat McVittie spacetime with positive Hubble parameter possesses two THs, 
			which begin to form at a specific time when the two horizons coincide.
			
		In the special case of the matter-dominated spatially flat McVittie spacetime, where 
			$\tilde{H}={2}/{3\tilde{t}}$, we find that for $0<\tilde{t}<\tilde{t}_{*}=2\sqrt{3}$, 
			no root of $\theta_{-}=0$ exists, and hence the spacetime admits no THs. 
		At critical time, $\tilde{t}=\tilde{t}_{*}=2\sqrt{3}$, a degenerate root of $\theta_{-}=0$ forms, corresponding 
			to a Past Outer TH (POTH). \footnote{Since, according to Eqs.~\eqref{NN1} and \eqref{NN4}, we have
				$\min_{\tilde{\bar{r}}\in(1/2a(\tilde{t}),+\infty)}\theta_{-}>0 \quad 
				\text{for } 0<\tilde{t}<\tilde{t}_{*}$,
				while 
				$\min_{\tilde{\bar{r}}\in(1/2a(\tilde{t}),+\infty)}\theta_{-}<0 
				$ for $\tilde{t}>\tilde{t}_{*}$,
				it follows that
				$\partial_{\tilde{t}}\theta_{-}\big|_{\tilde{\bar{r}}=\tilde{\bar{r}}_{(-,D)}(\tilde{t}_{*})}<0$.
				
				Consequently, at the moment of formation of the degenerate root $\theta_{-}=0$, we obtain
				$\mathcal{L}_{+}\theta_{-}\Big|_{\theta_{-}(\tilde{\bar{r}}_{(-,D)}(\tilde{t}_{*}))=0}
				=\left(\frac{1}{A}\,\partial_{\tilde{t}}\theta_{-}\right)\Big|_{\tilde{\bar{r}}=\tilde{\bar{r}}_{(-,D)}(\tilde{t}_{*})}<0 $,
				which corresponds to a POTH.}
			For $\tilde{t}>\tilde{t}_{*}$, there exist two distinct roots corresponding 
			to two PTHs, which are formed at $\tilde{t}=\tilde{t}_{*}$ at 
			$\tilde{\bar{r}}_{(-,D)}(\tilde{t}_{*})$.
			According to Eq.~\eqref{NN1}, it can be seen that the smaller root 
			(i.e., $\tilde{\bar{r}}_{-}^{l}$) tends to $(1/2a(\tilde{t}))^{+}$, 
			while the larger root (i.e., $\tilde{\bar{r}}_{-}^{u}$) grows 
			proportionally to $\tilde{t}^{1/3}$.
			
		Following Fig.~\ref{F0}, the evolution of the THs of the matter-dominated spatially flat McVittie spacetime is shown as a function of $\tilde{t}$.
						\begin{figure}[H]
				\centering
				
				\includegraphics[width=0.6\textwidth]{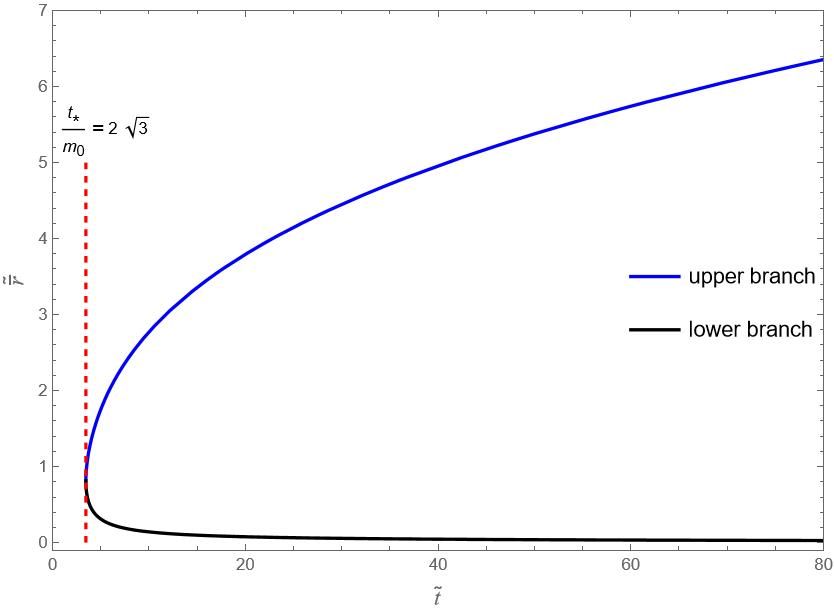}
				
				\caption{The time evolution of the THs of the matter-dominated spatially flat McVittie spacetime in isotropic coordinates $\tilde{\bar{r}}$. 
					The upper branch corresponds to $\tilde{\bar{r}}_{-}^u$, while the lower branch corresponds to $\tilde{\bar{r}}_{-}^l$. 
					The critical time 
					$\tilde{t}_{*} = t_*/m_0$
					represents the formation of the degenerate root of this spacetime, at which the two horizons coincide. At times before the critical time, there are no THs and
					after this time, the THs begin to separate from each other.
				}
				
				\label{F0}
			\end{figure}
					 To determine whether they are Past Inner THs (PITH) or POTH, we need to know  the sign of $\mathcal{L}_+{\theta_-|}_{\theta_-=0}$ outside the curvature singularity, i.e. $\bar{r}>{m_0}/{(2 a(t))}$. A simple calculation leads to:
						\begin{equation}
		\left .		\frac{1}{2} \mathcal{L}_+ \theta_-|_{\theta_- = 0} = \left(\frac{\dot{H}}{A} + \frac{2 H^2}{1 - \left( \frac{m_0}{2a \overline{r}}\right)} - \frac{m_0 H^3}{A^3} \left[3 - \left(\frac{m_0}{2 a \overline{r}}\right)\right]\right)\right |_{\theta_- = 0}
				\label{new8}
			\end{equation}
			As seen above, there is no straightforward way to determine whether the sign of Eq.~\eqref{new8} is positive or negative unless we know the time dependence of the Hubble parameter. In the matter-dominanted universe, $H = {2}/{(3t)}$ and the situation is somewhat easier to handle:
				\begin{equation}
												\frac{1}{2} \mathcal{L}_+ \theta_-|_{\theta_- = 0} = \left.\left(H^2 \frac{1 - 3 \frac{m_0}{2 a \overline{r}}}{2 - \frac{m_0}{a \overline{r}}} - \frac{m_0 H^3}{A^3} \left(3 - \frac{m_0}{2 a \overline{r}}\right)\right)\right |_{\theta_- = 0}
						\label{new9}
			\end{equation}
			Then, we use the identity $\frac{1-{m_0}/{(2 \bar{r} a(t))}}{1+{m_0}/{(2 \bar{r} a(t))}} = \sqrt{1-{2 m_{0}}/{R}}$ to simplify Eq. \eqref{new9} as:
\begin{equation}
		\frac{1}{2} \mathcal{L}_+ \theta_-|_{\theta_- = 0}=\left.\left(\frac{H}{2R} (2H R -1) - 2m_0 \frac{2H R +1}{R ^3 (1+ HR)}\right)\right |_{\theta_- = 0}\label{new10}
\end{equation}
For the THs (as shown in Fig \ref{F0}), $R=2M_{MSH}$ or equivalently $HR=\sqrt{1-{2m_0}/{R}}$. This gives the  areal radius of the THs of the spatially flat McVittie spacetime as: 
		\begin{equation}
		\begin{cases}
		R_1=\frac{2}{\sqrt{3} H} \sin (\frac{1}{3} \sin^{-1} ( 3\sqrt{3} m_0 H))\\
		R_2=\frac{1}{H} \cos (\frac{1}{3} \sin^{-1} ( 3\sqrt{3} m_0 H))-\frac{1}{\sqrt{3}H} \sin (\frac{1}{3} \sin^{-1} ( 3\sqrt{3} m_0 H)).
		\end{cases}
		\label{new11}
		\end{equation}
From Eq.~\eqref{new11}, it can be seen that the THs of the spatially flat McVittie spacetime are located outside the curvature singularity, i.e. $R=2m_0$ or $\bar{r}={m_0}/{(2 a(t))}$. The evolution of $\mathcal{L}_{+} \theta_{-}|_{\theta_{-}=0}$ from Eq.~\eqref{new10} for each of the THs of Eqs. \eqref{new11} is shown in the Fig. \ref{fig I}.
\begin{figure}[H] 
   \centering
      \includegraphics[width=1\textwidth]{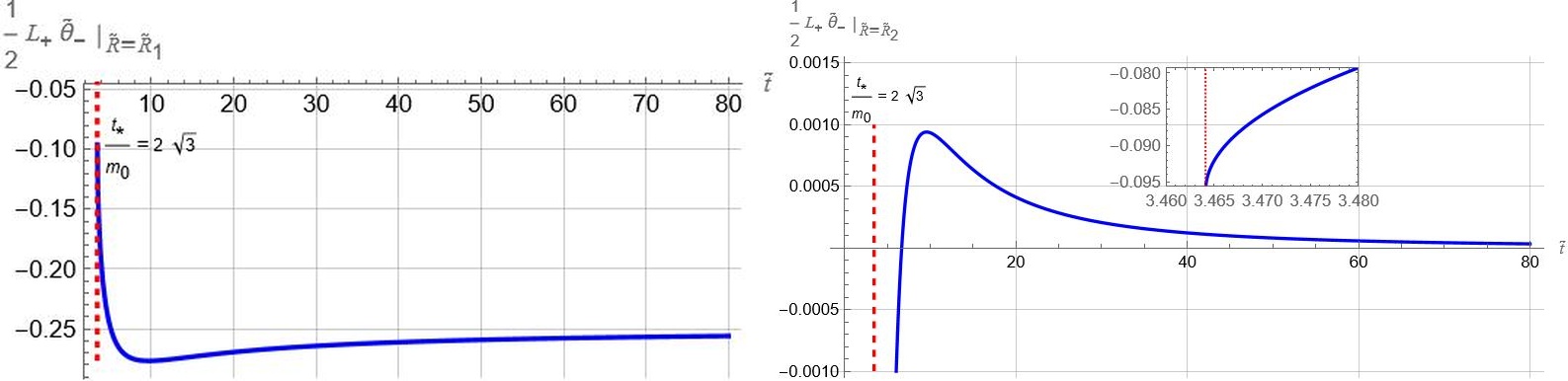}
        \caption{ The time evolution of $\mathcal{L}_{+} \tilde{\theta}_{-}|_{\tilde{\theta}_{-}=0}$ for the matter-dominated spatially flat McVittie spacetime on the THs, $\tilde{R}_1 \equiv R_{1}/m_0$ and  $\tilde{R}_2 \equiv R_2/m_0$. $\tilde{t}_{*}=t_*/m_0=2\sqrt 3$, is the critical time when these horizons coincide. As it can be seen from Eqs.~\eqref{new10} and \eqref{new11}, for $t_1 \sim 2t_*$, the corresponding graph for $R_2$ becomes zero and for $t=t_*$ both graphs approach to ${-1}/{(6\sqrt{3})}$.}
        \label{fig I}
\end{figure}
Thus, $R_1$ corresponds to a WH TH, i.e. a POTH, while $R_2$ is also a POTH only for times closer to $t_* $ ($ t_*<t<t_1$, where $t_1 \sim 2t_*$). For $t>t_1$, $R_2$ becomes a PITH and therefore corresponds to a cosmological TH. Based on the sign of both $\theta_-$ and $\theta_+$, the regions $2m_0<R<R_1$ and $R>R_2$ are past trapped where both $\theta_-$ and $\theta_+$ are positive. The region $R_1<R<R_2$ is untrapped since $\theta_+>0$ and $\theta_-<0$. This behavior of the expansion parameters \eqref{new7} in the specified regions is consistent with our expectation based on the definition of trapped and untrapped regions.\footnote{According to Eq.~\eqref{new7}, the positivity of $\theta_{+}$ is obvious. Furthermore, by using Eq.~\eqref{new7} and the identity $A(\bar{r},t)=\sqrt{1-{2 m_0}/{R}}$, the sign of $\theta_{-}$ is proportional to $H^2 R^3-R+2m_0=2M_{MSH}-R$. Based on Eq.~\eqref{new11}, it can be seen that $\theta_{-} >0$ for $R<R_1$ and $R>R_2$, while $\theta_- <0$ for $R_1<R<R_2$.}. 
  The results are summarized in Table I.
\begin{table}[h!]
	\centering
	\begin{tabular}{|c|c|c|c|}
		\hline
		& $t_* \le t <t_1$ & $ t = t_1 $ & $t>t_1$ \\
		\hline
		$R_1$ & POTH & POTH & POTH \\
		\hline
		$R_2$ & POTH & PDTH & PITH \\
		\hline
	\end{tabular}
	\caption{Types of the THs in the matter-dominated spatially flat McVittie spacetime, $t_*/m_0=2\sqrt{3}$, $t_1 \sim 2 t_*$, for different ranges of $t$. At $t=t_1$,  $R_2$ is a past degenerate TH (PDTH) for which  $\mathcal{L}_{+} \theta_{-}|_{\theta_{-}=0}=0$.}
\end{table}
Here, we examine the signature of the THs. To do this, it is convenient to write the McVittie line element \eqref{new2} in PG-like coordinates\footnote{We call these coordinates  "PG-like" because the coefficient of $dR^2$ is not unity. Metric \eqref{new12} can be obtained by rewriting McVittie metric \eqref{new2} in terms of the areal radius \eqref{new3c}.  }  \cite{Faraoni}
\begin{equation}
	ds^2 = - \left(1 - \frac{2 m_0}{R} - H^2 R^2\right) dt^2 + \frac{dR^2}{1-\frac{2 m_0}{R}} - \frac{2HR}{\sqrt{1 - \frac{2m_0}{R}}} dR dt + R^2 d\Omega^2
\label{new12}
\end{equation}
The TH described by $\phi\left(t,R\right)=R-R_{TH}(t)=0$ has the normal vector $N_\mu=\nabla_\mu \phi$ with the squared norm:
\begin{equation}
	N^2|_{R=R_{TH}} = \left( \nabla_\mu \phi \right)^2 |_{R=R_{TH}} = 2 \dot{R}_{TH}(t) - \frac{\dot{R}_{TH}^2(t)}{H^2 R^2_{TH}(t)}
\label{new13}
\end{equation}
Using Eq.~\eqref{new11}, the time evolution of the squared norm \eqref{new13} is illustrated in Fig. \ref{fig2} and then summarized in Table II.
\begin{figure}[H]
    \centering
        \includegraphics[width=\textwidth]{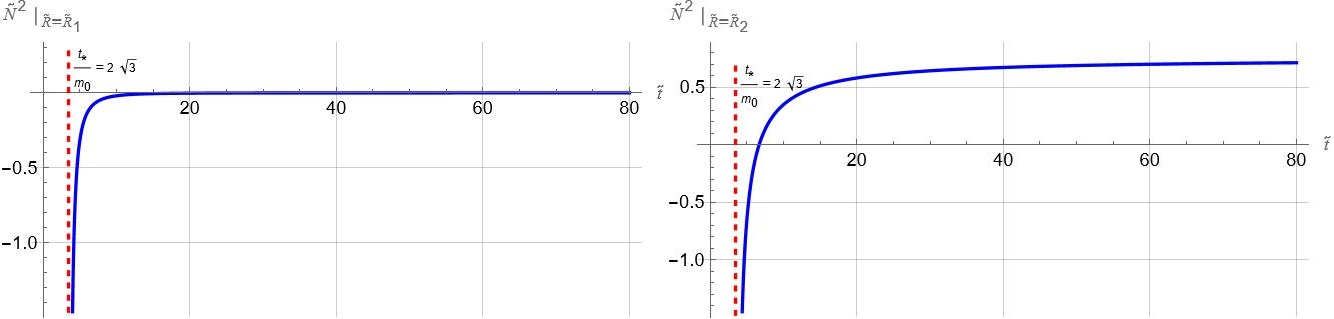}
    \caption{The time evolution of the normalized squared norm of the normal vector of the THs $\tilde{R}_1 \equiv R_{1}/m_0$ and $\tilde{R}_2 \equiv R_2/m_0$ for the matter-dominated spatially flat McVittie spacetime.}
    \label{fig2}
\end{figure}
\begin{table}[h!]
	\centering
	\begin{tabular}{|c|c|c|c|}
		\hline
		& $t_* \le t <t_1$ & $ t = t_1$ & $t>t_1$ \\
		\hline
		$R_1$ & SL & SL & SL \\
		\hline
		$R_2$ & SL & Null & TL \\
		\hline
	\end{tabular}
	\caption{Signatures of the THs in the matter-dominated spatially flat McVittie spacetime for different ranges of $t$.  $SL$ and $TL$ denote to spacelike and timelike, respectively. $t_*/m_0=2\sqrt{3}$ and $t_1 \sim 2 t_*$.}
\end{table}
As expected, both THs are spacelike and POTH near $t_*$, this is consistent with the fact that these horizons coincide at $t_*$. As can be seen from Tabel 1 and Table 2, the smaller TH, i.e. $R_1$, asymptotes to the curvature singularity $R=2m_0$ at late times, when $H \to 0$. Thus, $R_1$ is actually a spacelike WH-type TH, or a POTH. The larger TH, $R_2$, which asymptotes to $1/H$ at late times, $t>t_1$, is a timelike cosmological TH. This is consistent with our expectation that $R_2$ is a timelike PITH in a matter-dominated universe. 

In the next section, we will study the McVittie's  THs using PG coordinates.
\subsection{Trapping horizons in Painlev\'e-Gullstrand coordinates}\label{II.B}
To obtain the line element of the McVittie spacetime in PG coordinates, we first note that the general spherically symmetric line element in PG coordinates can be written as follows \cite{nielsen2006}:
\begin{equation}
	ds^2 = - \left[c^2 (R, \tau) - v^2(R, \tau)\right] d\tau^2 + dR^2 + 2v(R, \tau) dR d\tau + R^2 d \Omega^2
	\label{new14}
\end{equation}
where:
\begin{equation}
	\begin{cases}
		c(R,\tau) > 0 \\
		v(R, \tau) = \pm c(R,\tau) \sqrt{\frac{2 M_{MSH}}{R}}
	\end{cases}
	\label{new15}
\end{equation}
For the radial null vector fields ${ds}^2=d\Omega^2=0$, so:
\begin{equation}
	\begin{cases}
		\text{outgoing: } l^p = \left({1}/{c}, {(c-v)}/{c}\right) \\
		\text{ingoing: } n^p = \left({1}/{c}, {(-c-v)}/{c}\right)
	\end{cases}
\label{new16}
\end{equation}
with the normalization $l^p n_p = -2$.
The corresponding expansions along the above null vector fields can be evaluated as:
\begin{equation}
	\begin{cases}
		\theta_+ = \frac{2}{R} \left(1 \mp \sqrt{\frac{2 M_{MSH}}{R}}\right) \\
		\theta_- = -\frac{2}{R} \left(1 \pm \sqrt{\frac{2 M_{MSH}}{R}}\right)
	\end{cases}
	\label{new17}
\end{equation}
The choice of sign for each of the expansion parameters, and thus the type of TH, is actually the choice of sign for $v(\tau,R)$ \footnote{The authors in  \cite{nielsen2006} chose the upper (lower) sign for the $\theta_+$ ($\theta_-$) in Eq.~\eqref{new17}.}. This can be found by evaluating $v(\tau,R)$ at the TH, $R=2M_{MSH}$. The upper (lower) sign is acceptable when $v(\tau,R)$ is positive (negative).

To obtain the form of metric \eqref{new14} for the spatially flat McVittie spacetime,  we use the PG-like line element introduced in Eq.~\eqref{new12}. Then, we perform a coordinate transformation from $(t,R)$ to $(\tau,R)$ to ensure that the resulting metric satisfies the condition $g_{RR} = 1$ with $dt = \mu d\tau - \beta dR$.  This fixes function $\beta$ while leaving the integration factor $\mu$ partially specified. Only ${\partial \mu}/{\partial R} = - {\partial \beta}/{\partial \tau}$ is required.
Replacing $dt$ in Eq.~\eqref{new12} yields:
\begin{equation}
			ds^2 = -f \mu^2 d\tau^2 + dR^2 + 2\mu\left(\beta f - \frac{H R}{\sqrt{1 - \frac{2 m_0}{R}}}\right) d\tau dR + R^2 d\Omega^2 
			\label{new18}
\end{equation}
	where $f \equiv 1 - {2 m_0}/{R} - H^2 R^2$ and $\beta$ is given by 
	\begin{equation}
	- f \beta^2 + \frac{1}{1 -\frac{ 2m_0}{R}} +\frac {2 H R\beta}{\sqrt{1 - \frac{2 m_0}{R}}}  = 1
	\label{new19}
	\end{equation}
	Comparing Eq.~\eqref{new14} with Eq.~\eqref{new18}, gives:
\begin{equation}
	\begin{cases}
		v(R, \tau) \equiv \mu( \beta f - \frac{HR}{\sqrt{1 - \frac{2 m_0}{R}}}) \\
		c(R, \tau) \equiv \sqrt{f \mu^2 + v^2} 
		\label{new20}
	\end{cases}
\end{equation}
where we have chosen the positive sign for the function $c(R, \tau)$ without loss of generality. Then, using Eq.~\eqref{new19} to simplify $v\left(\tau,R\right)$, we get:
\begin{equation}
	\frac{v}{\mu} =  \pm \sqrt{1-f}
	\label{new21}
\end{equation}
We are now ready to consider the sign of $v\left(\tau,R\right)$. On the THs, $R=2M_{MSH}$ or equivalently $f=0$, if one assumes that $\beta\vert_{f=0}$ is finite, then $(\beta f)\vert_{f=0}=0$ and
 $v\left(\tau,R\right)$ is negative according to the RHS of Eq.~\eqref{new20}. This is because as we mentioned before, we are interested in solutions with $H>0$ and $\mu \equiv {\partial t}/{\partial \tau}$ can also be assumed to be positive. Thus, the negative sign in Eq.~\eqref{new21} is acceptable: 
 \footnote{As mentioned, Eq.~\eqref{new22} is only valid for the case $H>0$. In the Schwarzschild limit, where $a(t)\to 1$ (and hence $H\to 0$), the assumption that $\beta\vert_{f=0}$ is finite breaks down due to Eqs.~\eqref{new20} and~\eqref{new21}. Therefore, Eq.~\eqref{new22} is not valid in the $H=0$ limit. In this limit, to determine which sign in Eq.~\eqref{new21} is admissible, one must first specify which Schwarzschild PG patch (outgoing or ingoing) is being used. For example, in the ingoing PG patch, according to the standard definition in \cite{stationary BH1,Faraoni}, the positive sign in Eq.~\eqref{new21} is admissible. Consequently, the upper signs for the null expansions in Eq.~\eqref{new17} apply. These give the FTH at $r=2m_0$ as the null boundary separating the untrapped region (region~I, $r>2m_0$) from the trapped region (region~II, $0<r<2m_0$), as expected for the Schwarzschild spacetime.
 }
\begin{equation}
	\frac{v}{\mu}|_{f=0} =  -1 <0
\label{new22}
\end{equation}
We can also easily check that the corresponding definitions of $v$ and $c$ in Eq.~\eqref{new20} are consistent with Eq.~\eqref{new15}:
\begin{equation}
	\begin{cases}
		v(R, \tau) = -\mu( \sqrt{\frac{2 M_{MSH}}{R}}) \\
		c(R, \tau) = \mu 
	\end{cases}
\label{new23}
\end{equation}
Finally, since $v\left(\tau,R\right)<0$, the lower sign in Eq.~\eqref{new17} is acceptable and the null expansions along $l$ and $n$ are obtained as:
\begin{equation}
	\begin{cases}
		\theta_+ = \frac{2}{R} \left(1 + \sqrt{\frac{2 M_{MSH}}{R}}\right) \\
		\theta_- = - \frac{2}{R} \left(1 - \sqrt{\frac{2 M_{MSH}}{R}}\right)
\label{new24}
	\end{cases}
\end{equation}
Therefore, our results in PG coordinates again show that the only possible THs for spatially flat McVittie spacetime are PTHs, which can be either a WH or a cosmological TH. As deduced in the previous section, using $R=2M_{MSH}$, it can be easily shown that the smaller TH corresponds to the WH-type TH, while the larger one corresponds to a cosmological TH. Our results are in agreement with the expansion parameters of the spatially flat McVittie spacetime derived in PG-like coordinates \cite{McVittie de sitter}. 

Based on Eqs.~\eqref{new3b} and \eqref{new3c}, in the late-time limit $(t \to \infty)$
 and the comoving radius remains finite, the physical radius \( R \) diverges to infinity. Furthermore, according to Eqs.~\eqref{new17} and \eqref{new23}, and  the MSH mass of the spatially flat McVittie spacetime \eqref{m}, if at asymptotic times, $H \to H_0=\text{cons.} > 0 $, then in the PG coordinate system, the outgoing and ingoing radial null vector fields become  indistinguishable. This renders the definition of the expansion parameter meaningless.
As discussed in \cite{kaloper2010,McVittie de sitter}, in PG-like coordinates \eqref{new12}, the maximally extended spatially flat McVittie spacetime features both a BH and a WH event horizon, if the scale factor satisfies \( a(t) \approx \exp(H_0 t) \) for  sufficiently large \( t \), where \( H_0 \) is a positive constant. This form of the scale factor asymptotically matches that of the \(\Lambda\)CDM cosmology. When the spatially flat McVittie solution asymptotically approaches the Schwarzschild-de Sitter solution, such as the Schwarzschild-de Sitter spacetime itself, the geometry possesses a timelike Killing vector that vanishes on a bifurcation two-sphere as \( t \to +\infty \). Therefore, this solution includes the inner bifurcation two-sphere characteristic of the Schwarzschild-de Sitter spacetime. This implies that the BH and WH-type event horizons together constitute a partial boundary of this McVittie solution. Accordingly, when the spatially flat McVittie solution asymptotically approaches a constant positive Hubble parameter \( H \to H_0 > 0 \), it reduces to the Schwarzschild-de Sitter spacetime  \cite{kaloper2010,McVittie de sitter}.
For $H>0$ with $\lim_{t\to\infty}H(t)=0$ and $\Lambda=0$, the flat McVittie spacetime \footnote{which includes, in particular, our matter-dominated flat McVittie model.},as noted in \cite{nolan2017}, there are no trapped regions in the flat McVittie spacetime and hence no FTH exists, which is in agreement with our results. In addition, as shown in \cite{McVittie de sitter} and \cite{nolan2017}, for $H>0$ with $\lim_{t\to\infty}H(t)=0$ and $\Lambda=0$ there exists an analogue of a BH event horizon whose generators approach $(t\to\infty,\ R\to 2m_0)$. The limiting surface $(t=\infty,\ R=2m_0)$ is not a weak null curvature singularity, contrary to earlier claims \cite{nolan2017,Nolan2014}. Moreover, it is stated that there is no bifurcation two-sphere in this McVittie spacetime, and that there is a BH horizon to the future of the singularity and a WH horizon in the past, which together form a wedge \cite{McVittie de sitter}. Hence, from the global point of view, flat McVittie with $\lim_{t\to\infty}H(t)=0$ and $\Lambda=0$ contains BH and WH event horizons, but it does not contain a finite-radius cosmological event horizon at late times. Therefore, if by ``cosmological BH'' one means a spacetime that possesses both a BH event horizon and a finite-radius cosmological event horizon at late times, then the matter-dominated flat McVittie spacetime does not describe a cosmological BH in this sense either.

 Similarly, within our Hayward-inspired quasi-local framework \cite{Hayward}, neither solution qualifies as a cosmological BH because they do not possess both an FOTH and a PITH simultaneously.

	

In summary, according to our results done in both isotropic and PG coordinates, the spatially flat McVittie metric does not describe a cosmological BH  in a matter-dominated background.
\section{Culetu spacetime}\label{III}
		\subsection{Trapping horizons in Conformal Painlev\'e-Gullstrand-Schild coordinates}\label{III.A}
The Culetu spacetime is defined by the following line element:\cite{culetu},
\begin{equation}
	ds^2=a(\tau)^2\left(-(1-\frac{2m}{r})d\tau^2+2\sqrt{\frac{2m}{r}} d\tau dr+ dr^2+ r^2 d\Omega^2\right)
	\label{new25}
\end{equation}
which is a conformally PG-Schwarzschild spacetime and asymptotically approaches the spatially flat FLRW in $r>>2m$. In the Culetu spacetime, $\tau>0$  is a timelike coordinate, and the THs are analyzed by assuming $a(\tau)\sim \tau^2$ \footnote{In a matter-dominated universe, $a(t) = t^{{2}/{3}}$, defining the conformal time as $dt = a \, d\tau$, yields $a(\tau) \sim \tau^2$. }.

The Culetu spacetime is regular everywhere except at $r=0$\footnote{According to the metric \eqref{new25}, $r=0$ corresponds to a strong curvature singularity since the determinant of the three-dimensional spatial metric vanishes.
} and $\tau=0$ which corresponds to big-bang singularity for $a(\tau)\sim \tau^\alpha$, with $\alpha>0$. As Culetu proved, the surface $r=2m$ is not a curvature singularity. This surface is actually an event horizon \cite{sato}. It represents a null surface, and by solving the outgoing radial null rays equation ${dr}/{d\tau}=1-\sqrt{{2m}/{r}}$, it can be shown that as $\tau \to -\infty$, we have $r \to 2m$ and $(\tau_{E},r_{E})=(-\infty,2m)$ forms an extendable boundary.

The energy-momentum tensor associated with the Culetu spacetime, as shown in \cite{sato}, 
belongs to the Hawking--Ellis Type I classification. It can be interpreted as a superposition of a homogeneous perfect fluid and an inhomogeneous anisotropic fluid.
Furthermore, in the vicinity of the strong curvature singularity at $r=0$, the mentioned energy--momentum tensor violates all ECs assuming $m>0$ and ${da}/{d\tau}>0$.

To investigate the nature and signature of the THs, we first rewrite the line element \eqref{new25} using $dt=a d\tau$:
\begin{equation}
	ds^2=-(1-\frac{2m}{r})dt^2+2\sqrt{\frac{2m}{r}} adt dr+a^2 dr^2+a^2 r^2 d\Omega^2
	\label{new26}
\end{equation} 
and derive the equations for the outgoing and ingoing null rays by setting ${ds}^2={d\Omega}^2=0$ in the line element \eqref{new26}:
\begin{equation}
	\frac{dr}{dt}=\frac{1}{a} (-\sqrt{\frac{2m}{r}} \pm 1)
	\label{new27}
\end{equation}
Using Eq.~\eqref{new27}, the outgoing and ingoing null vector fields can be determined as follows:
\begin{equation}
	\begin{cases}
		l^p=(1, \frac{1}{a}(1-\sqrt{\frac{2m}{r}})) \\
		n^p=(1,-\frac{1}{a} (1+\sqrt{\frac{2m}{r}}))
	\end{cases}
	\label{new28}
\end{equation}
where $p=0,1$ and the normalization is set to $l_{p} n^{p}=-2$. From the above and Eq.~\eqref{new6}, the expansion parameters along the null vector fields can be obtained as 
\begin{equation}
\theta_{\pm}=\frac{2}{R} \left (HR\pm (1\mp \sqrt{\frac{2ma}{R}} ) \right )
\label{new29}
\end{equation}
where the areal radius of the Culetu spacetime is $R=ar$ according to Eq.~\eqref{new26}
and $H>0$ is assumed. For the location of the marginally trapped surface, we set $\theta_{+} \theta_{-}=0$. According to Eq.~\eqref{new29}, it appears that both the outgoing and ingoing null expansion parameters can vanish:
\begin{equation}
	\begin{cases}
\theta_{+}=0 \Rightarrow HR_{+}=-1+\sqrt{\frac{2ma}{R_{+}}}\\
\theta_{-}=0 \Rightarrow HR_{-}=1+\sqrt{\frac{2ma}{R_{-}}}
	\end{cases}
	\label{new30}
\end{equation}
Clearly $\theta_{+}$ vanishes only when $0<R_{+}<2ma$  or $0<r<2m$ and for $\theta_{-}=0$, it is obvious that $HR_{-} \geq 1$ is always satisfied. As expected, $\theta_{+} \theta_{-}=0$ must be occur at the roots of the equation $R=2M_{MSH}$. By Eq.~\eqref{new32}, $M_{MSH}$
can be easily calculated for the Culetu spacetime as \footnote{In \cite{sato}, the MSH mass of the Culetu spacetime is obtained as: $M_{MSH} = {a_0 \tau^{\alpha-2} r} \left( \tau \sqrt{{2m}/{r}} - \alpha r \right)^2/2$. This is in full agreement with Eq.~\eqref{new33} by considering that $aH =  {da}/{(ad\tau)} = {\alpha}/{\tau}$ and the definition of the areal radius, i.e., $R = ar $.}:
\begin{equation}
M_{MSH}=\frac{R}{2} \left(HR-\sqrt{\frac{2ma}{R}}\right)^2
\label{new33}
\end{equation}
which is always non-negative and, for $R>>2ma$ or $r>>2m$, asymptotically approaches ${H^2 R^3}/{2}$, the MSH mass of the flat FLRW spacetime \cite{FLRW}.
Then, we can use Eq.~\eqref{new33} to find the solutions of $R=2M_{MSH}$ :
\begin{equation}
(HR-\sqrt{\frac{2ma}{R}})^2=1 \Rightarrow HR \pm1 = \sqrt{\frac{2ma}{R}}
\label{new34}
\end{equation}
which is fully consistent with Eq.~\eqref{new30}. The above equation can be expressed as a cubic equation:
\begin{equation}
P_{\pm}=H^2 R^3\pm 2HR^2+R-2ma=0
\label{new35}
\end{equation}
where the positive sign corresponds to $\theta_{+}=0$, and the negative sign corresponds to $\theta_{-}=0$.
To determine the number and sign of the roots of Eq.~\eqref{new35}, we compute the discriminant of this cubic algebraic equation in terms of \(R\):
\begin{equation}
	\Delta_{\pm} = \mp 4maH^3 (2 \pm 27maH)
	\label{35}
\end{equation}
As a result, based on Eq.~\eqref{35}, since \(maH > 0\), we always have \(\Delta_{+} < 0\). Therefore, the equation \(P_{+} = 0\) necessarily admits a real root, which according to Eq.~\eqref{new30}, is positive.
On the other hand, according to Eq.~\eqref{35}, \(\Delta_{-} < 0\) does not always hold. Thus, at first glance, the equation \(P_{-} = 0\) may appear to admit more than one real root. However, since the roots of \(P_{-}\) must be positive and also satisfy the condition \(HR_{-} \geq 1\), it can be seen that \({\partial P_{-}}/{\partial R} \geq 0\) is satisfied. Additionally, we have \(P_{-}\rvert_{HR \to 1} < 0\), and therefore, the equation \(P_{-} = 0\) can only admit a single positive root that satisfies the condition \(HR_{-} \geq 1\).
Thus, \(\theta_{+} = 0\) and \(\theta_{-} = 0\) correspond to the two roots \(R_{+}\) and \(R_{-}\), respectively, which represent the THs of the Culetu spacetime.

According to Eqs.~\eqref{new29} and~\eqref{new30}, one can see that the conditions \(\theta_{+}\rvert_{\theta_{-} = 0} > 0\) and \(\theta_{-}\rvert_{\theta_{+} = 0} < 0\) hold, which correspond to a PTH and an FTH, respectively. Furthermore, based on Eq.~\eqref{new30}, the inequality \(0 < R_{+} < R_{-}\) always holds.
We now examine the ECs near and on the THs obtained earlier, in order to determine the time intervals during which the Culetu spacetime can be interpreted as a solution containing a cosmological BH.
In \cite{sato}, the following characteristic times were introduced:
\begin{equation}
\tau_{N}(r) = 3 \sqrt{\frac{r^3}{2m}}, \quad
\tau_{W}(r) = 2 \sqrt{\frac{r^3}{2m}}, \quad
\tau_{D}(r) = \frac{6}{7} \sqrt{\frac{r^3}{2m}}, \quad (m > 0)
\label{36}
\end{equation}
It was shown that, for the matter-dominated Culetu spacetime with \( a(\tau) \sim \tau^2 \), the energy--momentum tensor satisfies the NEC and SEC in the interval \( 0 < \tau \leq \tau_{N}(r) \), the WEC in the interval \( 0 < \tau \leq \tau_{W}(r) \), and the DEC in the interval \( 0 < \tau \leq \tau_{D}(r) \).

According to Eq.~\eqref{new30}, and using the relation \( aH = da/a d\tau=2/\tau \), we obtain the following expressions for the THs:
\begin{equation}
	\begin{cases}
		\tau_{+}(r) = \dfrac{2r}{\sqrt{2m/r} - 1} \\[4pt]
		\tau_{-}(r) = \dfrac{2r}{\sqrt{2m/r} + 1}
	\end{cases}
	\label{37}
\end{equation}
 Using Eqs.~\eqref{36} and \eqref{37}, it is evident that the NEC, WEC and SEC hold on the PTH, i.e., \(R_{-}\), for all \(r > 0\). However, it can be seen that the DEC holds only for \(r \geq {32m}/{9}\). Therefore, all ECs are satisfied on the PTH of the matter-dominated Culetu spacetime for \(r \geq {32m}/{9}\).
The analysis of ECs on the FTH, i.e., \(R_{+}\), is somewhat more involved. It can be seen from Eq.~\eqref{37}, along with \(\tau_{D}(r)\) and \(\tau_{W}(r)\) in Eq.~\eqref{36}, that the WEC and DEC are violated throughout the domain of definition of this FTH, i.e., for \(0 < r < 2m\). On the other hand, the NEC and SEC are satisfied in the range \(0 < r \leq {2m}/{9}\), but are again violated in the interval \({2m}/{9} < r < 2m\). Consequently, all ECs are violated on the FTH of the matter-dominated Clifton spacetime for \({2m}/{9} < r < 2m\).

As discussed in~\cite{sato}, for the Culetu metric to appropriately describe a physical spacetime, the NEC must be satisfied on and outside the event horizon. This implies that \(0 < \tau \leq \tau_{N}(2m)\), or equivalently \(0 < \tau \leq 6m\). Therefore, at early times, the Culetu solution may be expected to represent a physical spacetime.

\begin{figure}[H]
 	\centering
 	
 	\includegraphics[width=0.8\textwidth]{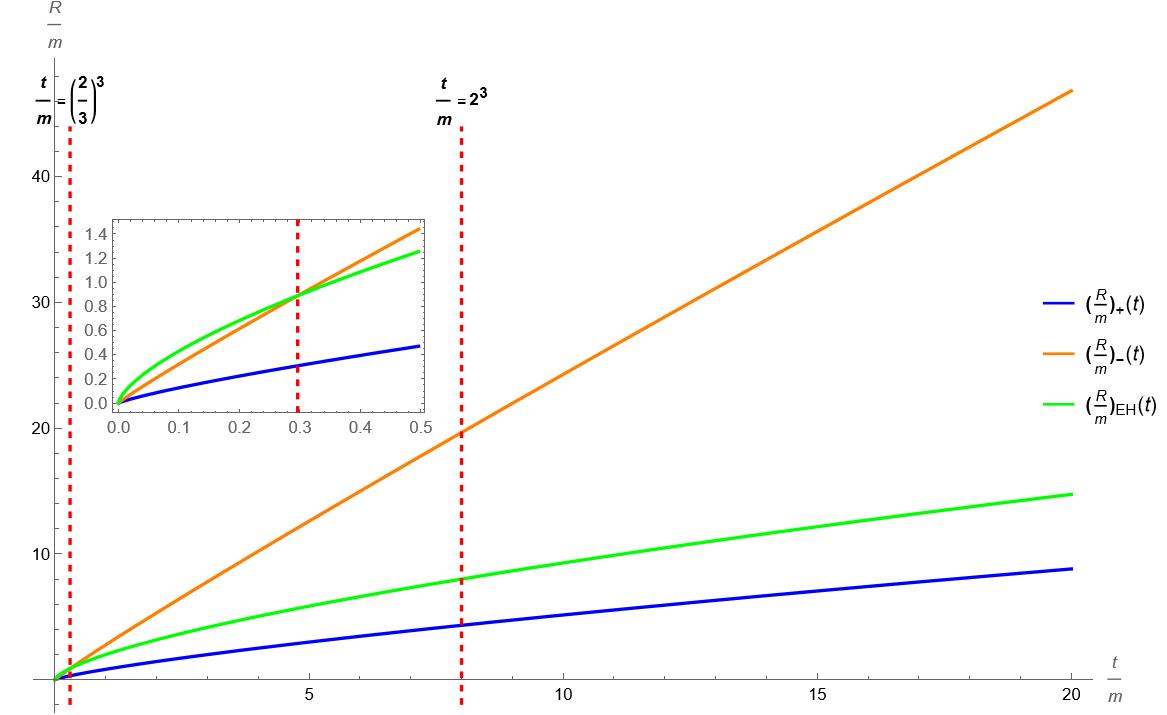}
 	
 	\caption{The time evolution of the horizons in the matter-dominated Culetu spacetime with \( a(t) = \tilde{t}^{2/3} \). The time and radius are normalized as \( \tilde{t} = t/m \) and \( \tilde{R} = R/m \), with \( t_0 = m \), or equivalently \( \tilde{t}_0 = 1 \).
 			In the interval \( 0 < \tilde{t} < (2/3)^3 \), \( 0 < R_{+} < R_{-} < R_{\mathrm{EH}} \) holds. For \( \tilde{t} > (2/3)^3 \), the ordering becomes \( 0 < R_{+} < R_{\mathrm{EH}} < R_{-} \). The time \( \tilde{t} = 2^3 \) is the upper limit of the interval over which the NEC holds both on and outside the event horizon.
 		Since $R_-$ lies outside the event horizon after a certain period of time, it follows from \cite{nielsen2009} that the NEC must be violated in some region of the Culetu spacetime.} 
 \label{figIII}
 \end{figure}

Fig.~\ref{figIII} illustrates the evolution of the THs and event horizon of the Culetu spacetime according to \eqref{new35}.
 To determine the nature of THs in the Culetu spacetime, we examine the signs of \(\mathcal{L}_{-} \theta_{+}\rvert_{\theta_+ = 0}\) and \(\mathcal{L}_{+} \theta_{-}\rvert_{\theta_- = 0}\):
\begin{equation}
	\begin{cases}
		\frac{1}{2} \mathcal{L}_{+} \theta_{-}\rvert_{\theta_{-}=0} =  \frac{1}{R_{-}^2} \left( - 2H^2 R_{-}^2 + \frac{7}{2} H R_{-} - 1 \right) \\
		\frac{1}{2} \mathcal{L}_{-} \theta_{+}\rvert_{\theta_{+}=0} = - \frac{1}{R_{+}^2} \left( 2H^2 R_{+}^2 + \frac{7}{2} H R_{+} + 1 \right)
	\end{cases}
	\label{38}
\end{equation}
It is evident that \(\mathcal{L}_{-} \theta_{+}\rvert_{\theta_+ = 0} < 0\), and since \(\theta_{-}\rvert_{\theta_+ = 0} < 0\), \(R_+\) corresponds to an FOTH, which is equivalent to a BH-type TH.

Regarding \(R_{-}\), for which \(\theta_{+}\rvert_{\theta_- = 0} > 0\), the term \(\mathcal{L}_{+} \theta_{-}\rvert_{\theta_- = 0}\) can be negative, positive, or zero. Based on Eq.~\eqref{38}, we find that \(\mathcal{L}_{+} \theta_{-}\rvert_{\theta_- = 0} < 0\) when \(H R_{-} > C\), it vanishes when \(H R_{-} = C\), and it becomes positive when \(H R_{-} < C\). Here, \(C = ({7 + \sqrt{17}})/{8}\).

Finally, using Eq.~\eqref{new30} and analyzing the sign of \(\mathcal{L}_{+} \theta_{-}\) in terms of \(H R_{-}\), we find that in the interval \(0 < \tilde{t} < \tilde{t}_c\), the surface \(R_{-}\) corresponds to a POTH; at \(\tilde{t} = \tilde{t}_c\), it corresponds to a PDTH; and for \(\tilde{t} > \tilde{t}_c\), it becomes a PITH. The value of $\tilde{t}_{c}$ is given by $\tilde{t}_c = \left( {4}/({3C(C - 1)^2} )\right)^3 \approx 249.13$.
By restricting the time interval to \(0 < \tilde{t} < 2^3 < \tilde{t}_c\), the \(R_{-}\) remains a POTH. However, since Eq.~\eqref{new30} implies that $H R_{-}\to 1$ as $r\to +\infty$, which corresponds to the FLRW cosmological TH, Culetu's PTH can be interpreted as a cosmological TH.

To determine the signature of the THs, one can evaluate the 2-dimensional line element of the Culetu spacetime Eq.~\eqref{new25} on the hypersurfaces \(\tau = \tau_{\pm}(r)\), in a matter-dominated universe:\cite{sato}.
\begin{equation}
	\begin{cases}
		ds^2\big|_{\tau = \tau_{-}(r)} = \dfrac{a(\tau_{-})^2\, dr^2}{(1+X)^3} (4X+3)(4X^2 + X - 1) \\
		ds^2\big|_{\tau = \tau_{+}(r)} = \dfrac{a(\tau_{+})^2\, dr^2}{(1+X)^3} (4X-3)(4X^2 - X - 1)
	\end{cases}
	\label{39}
\end{equation}
Here, $X = \sqrt{2m/r}$ is evaluated at the location of the THs. For \(R_{+}\), since \(X > 1\), Eq.~\eqref{39} implies that \(ds^2|_{\tau = \tau_{+}(r)} > 0\), indicating that \(R_{+}\) is always spacelike. 
For \(R_{-}\), Eq.~\eqref{39} shows that the line element is negative for \(0 < X < X_c\), vanishes at \(X = X_c\), and becomes positive for \(X > X_c\), where \(X_c = C - 1 = ({-1 + \sqrt{17}})/{8}\). Therefore, using Eq.~\eqref{new30}, we conclude that \(R_{-}\) is spacelike for \(0 < \tilde{t} < \tilde{t}_c\), null at \(\tilde{t} = \tilde{t}_c\), and timelike for \(\tilde{t} > \tilde{t}_c\).

The following table summarizes the signature and nature of the THs in the Culetu spacetime:
\begin{table}[h!] 
	\centering
	\begin{tabular}{|c|c|c|c|}
		\hline
		& $0 \le \tilde{t} <\tilde{t}_{c}$ & $ \tilde{t} = \tilde{t}_{c} $ & $\tilde{t}>\tilde{t}_c$ \\
		\hline
		$R_{-}$ & POTH -SL& PDTH-Null & PITH-TL \\
		\hline
		$R_{+}$ & FOTH-SL & FOTH-SL & FOTH-SL \\
		\hline
	\end{tabular}
	\caption{The sign and nature of the THs of the matter-dominated Culetu spacetime for different time intervals $\tilde{t}$. 
		SL and TL refer to spacelike and timelike, respectively. }
		\label{TIII}
\end{table}
In the Culetu spacetime, the null expansion parameters in the conformal PG-Schild coordinates can both vanish, as demonstrated in \cite{sato}, which is fully consistent with our findings in Eq.~\eqref{new30}. Moreover, \cite{sato} showed that on \(\tau_{+}(r)\), where the outgoing null expansion parameter \(\theta_+\) vanishes, it corresponds to a spacelike FOTH. In contrast, \(\tau_{-}(r)\), where the ingoing null expansion parameter \(\theta_-\) vanishes, describes a spacelike POTH for \(0 < r < r_{d(C)}\), a null PDTH at \(r = r_{d(C)}\), or a timelike PITH for \(r > r_{d(C)}\).
These results are in full agreement with those obtained in this section. \footnote{According to Eq.~\eqref{new30}) and the definition of \(R\), at the PTH of the Culetu spacetime in a matter-dominated universe,  we obtain: $HR=1+\sqrt{2m/r}$.  Based on Eqs.~\eqref{38}, \eqref{39} and table \ref{TIII}, \(R_{-}\) represents a spacelike POTH for \(0<r< r_{d(C)}\). At \(r= r_{d(C)}\) it corresponds to a null PDTH, and for \(r> r_{d(C)}\) it describes a timelike PITH. The authors of \cite{sato} state that \(r_{d(C)}=2m ({1}/({C-1}))^2=2m ({8}/({-1+\sqrt{17}}))^2\) in a matter-dominated universe which is consistent with our result.
}.

 \subsection{Trapping horizons in Painlev\'e-Gullstrand coordinates}\label{III.B}
For the Culetu spacetime, the transition to the PG line element is straightforward. We begin with the transformation of the variables $(t, r)$ to $(t, R)$ in line element \eqref{new26}, where $R$ is the areal radius, we therefore substitute $dR = a \, dr + H a r\, dt $ into line element \eqref{new26}:  
\begin{equation}
ds^2=-\left(1-(HR-\frac{2ma}{R})^2\right)dt^2 + 2\left(\sqrt{\frac{2ma}{R}}- HR\right) dt dR+dR^2+R^2 d\Omega^2
\label{new42}
\end{equation}
As seen in metric \eqref{new42}, the condition $g_{RR}=1$  is satisfied, and no additional coordinate transformation is necessary.  
Comparing line elements \eqref{new14} and \eqref{new42} gives:  
\begin{equation}
\begin{cases}
 c(R,t)=1\\
 v(R,t)=\sqrt{\frac{2ma}{R}}- HR=\pm \sqrt{\frac{2M_{MSH}}{R}}
\end{cases}
\label{new43}
\end{equation}  
The above equations are consistent with Eq.~\eqref{new33}\footnote{According to Eq.~\eqref{new33}, $\sqrt{{2M_{MSH}}/{R}} = \left| HR - \sqrt{{2ma}/{R}} \right|$.  The sign of the expression within the absolute value determines the acceptable sign for $v$.}, where the explicit form of the MSH mass was derived. 

According to line elements \eqref{new14} and \eqref{new42}, the temporal component of metric is given by $g_{tt} = \left( 1 - {2M_{MSH}}/{R} \right)$  in the PG coordinate system. Consequently, the location of the THs in this coordinate system is determined by $ R = 2M_{MSH} $, which is consistent with the results obtained in Eq.~\eqref{new34}.

Now, it is necessary to determine which sign in Eq.~\eqref{new43} is acceptable.
Using \(aH = 2/\tau\) together with \(R = a r\), the quantity \(\nu(R,t)\) in Eq.~\eqref{new43} can be rewritten as
\begin{equation}
	\nu \;=\; \sqrt{\frac{2m}{r}}\;-\;\frac{2r}{\tau}.
	\label{42}
\end{equation}
Consequently, the condition \(\nu \le 0\) holds for \(0 < \tau \le 2r\sqrt{r/(2m)}\), while \(\nu > 0\) is satisfied for \(\tau > 2r\sqrt{r/(2m)}\).
This behavior coincides exactly with the validity or violation of the WEC in the matter-dominated Culetu spacetime, as shown in \cite{sato}.
From Eqs.~\eqref{36} and \eqref{37} we see that at the PTH
\(R_{-}\) (i.e.\ \(\tau = \tau_{-}\)) the inequality
\(0 < \tau_{-} < \tau_{W}(r)\) holds. Hence the WEC is always satisfied on
\(R_{-}\) and \(\nu|_{R_{-}} < 0\).
Conversely, at the FTH \(R_{+}\)
(i.e.\ \(\tau = \tau_{+}(r)\)) we have \(\tau_{+} > \tau_{W}(r)\),
so the WEC is invariably violated on \(R_{+}\) and \(\nu|_{R_{+}} > 0\).

As a result, in the interval \(0 < \tau < \tau_{W}(r)\)
(which contains the PTH \(R_{-}\)) Eq.~\eqref{new43} yields
\(\nu = -\sqrt{2M_{\mathrm{MSH}}/R}\); therefore in
Eq.~\eqref{new17} the positive sign must be chosen for
\(\theta_{+}\) and the negative sign for \(\theta_{-}\).
In the complementary regime \(\tau > \tau_{W}(r)\)
(which contains the FTH \(R_{+}\)) one has
\(\nu = +\sqrt{2M_{\mathrm{MSH}}/R}\), so the negative sign applies to
\(\theta_{+}\) and the positive sign to \(\theta_{-}\) in Eq.~\eqref{new17}.

In summary, for the Culetu spacetime expressed in PG
coordinates one can write:
\begin{equation}
	\left\{
	\begin{aligned}
		c(R,t) &= 1 \\
		\nu(R,t) &= - \sqrt{\frac{2M_{\mathrm{MSH}}}{R}}
	\end{aligned}
	\right.
	\quad \text{for } 0<\tau<\tau_{W}(r)
	\quad , \quad
	\left\{
	\begin{aligned}
		c(R,t) &= 1 \\
		\nu(R,t) &= + \sqrt{\frac{2M_{\mathrm{MSH}}}{R}}
	\end{aligned}
	\right.
	\quad \text{for } \tau > \tau_{W}(r)
	\label{43}
\end{equation}
Thus, based on Eq.~\eqref{new17}, the null expansion parameters are given by:
\begin{equation}
	\left\{
	\begin{alignedat}{2}
		\theta_{+} & = \frac{2}{R}(1+\sqrt{\frac{2 M_{MSH}}{R}}) \\
		\theta_{-} & = -\frac{2}{R}(1-\sqrt{\frac{2 M_{MSH}}{R}})
	\end{alignedat}
	\right.
	\quad \text{for } 0<\tau<\tau_{W}(r)
	\quad \text{,}
	\quad
	\left\{
	\begin{alignedat}{2}
		\theta_{+} & = \frac{2}{R}(1-\sqrt{\frac{2 M_{MSH}}{R}}) \\
		\theta_{-} & = -\frac{2}{R}(1+\sqrt{\frac{2 M_{MSH}}{R}})
	\end{alignedat}
	\right.
	\quad \text{for } \tau>\tau_{W}(r)
	\label{44}
\end{equation}
By comparing Eqs.~\eqref{new29} and \eqref{44}, we can see that the results of the ingoing and outgoing null expansions of the Culetu spacetime are consistent in both the conformal PG-Schild and PG coordinates.
In the calculations related to the outgoing and ingoing null expansions, Eqs.~\eqref{new29}, \eqref{new30}, and \eqref{44}, we have shown that both $\theta_{\pm}$ can vanish simultaneously in both coordinate systems. This represents an FOTH at $R_+$ and a POTH, PDTH or PITH at $R_-$ based on Eq.~\eqref{38}. 

Therefore, within our Hayward-Inspired quasi-local framework, the Culetu solution can describe a cosmological BH. It should be noted that this geometry admits only two types of THs: a BH-type TH and a cosmological TH. The cosmological TH behaves as a POTH at early times and as a PITH at later times.

	\section{Sultana-Dyer spacetime}\label{IV}
		\subsection{Trapping horizons in Conformal Kerr-Schild coordinates} \label{IV.A}
The Sultana-Dyer spacetime is described by the following line element:\cite{sultana}
\begin{equation}
ds^2=a^2(\eta)\left(-(1-\frac{2m}{r})d\eta^2+\frac{4m}{r} d\eta dr+(1+\frac{2m}{r})dr^2+r^2 d\Omega^2\right)
\label{new48}
\end{equation}
  In Kerr-Schild coordinates, this metric is conformally Schwarzschild and asymptotes to the flat FLRW metric in $r>>2m$ ($m>0$).  $\eta>0$  is a timelike coordinate, and in the following, we analyze the determination of the signature and nature of the THs in a matter-dominated universe ($a(\eta) \sim \eta^2$)\footnote{For a matter-dominated universe,  
$a(t) \propto  t^{{2}/{3}}$. By defining the conformal time as $dt = a \, d\eta$ it follows that $a(\eta) \propto  \eta^2$.}.

 The Sultana-Dyer spacetime is regular except at $r=0$ \footnote{The determinant of the three-dimensional spatial metric vanishes at $r = 0$. So it corresponds to a strong curvature.} and $\eta=0$ where the latter corresponds to the big-bang singularity. Also, $r = 2m$  is not a curvature singularity but describes the event horizon of the Sultana-Dyer spacetime.  It represents a null surface, and by solving the outgoing radial null rays equation ${dr}/{d\eta}=({1-{2m}/{r}})/({1+{2m}/{r}})$, it can be shown that $\eta \to -\infty$ gives $r \to 2m$ and $(\eta_{E},r_{E})=(-\infty,2m)$ forms an extendable boundary \cite{sato}. 

Assuming $a(\eta) \sim \eta^{\alpha}$, Sultana and Dyer~\cite{sultana} demonstrated that, 
	for the specific case $\alpha = 2$ (i.e., a matter-dominated universe), 
	the source of metric \eqref{new48} can be expressed as a mixture of a null dust and a dust fluid. 
	However, this interpretation is not globally valid. 
	As shown in~\cite{sato}, this decomposition holds only within the domain 
	$\eta \leq \eta_{\text{max}}$. 
	They further proved that the global decomposition of the energy--momentum tensor 
	associated with the Sultana--Dyer class can instead be interpreted as a combination 
	of a homogeneous perfect fluid and an inhomogeneous type-II null fluid.
It can also be seen that, if the condition \(\rho_{F} + p_{F} \geq 0\) holds in the Sultana--Dyer class of spacetimes, 
 then the corresponding energy-momentum tensor belongs to the Hawking--Ellis type I classification, provided that \(m \geq 0\) and \({da}/{d\eta} \geq 0\). Furthermore, if \(m > 0\) and \({da}/{d\eta} > 0\), then the energy-momentum tensor violates all ECs in the vicinity of the  curvature singularity at \(r = 0\).

For the matter-dominated universe, in terms of the cosmic time, we write the line element \eqref{new48} as:  
\begin{equation}
ds^2=-(1-\frac{2m}{r})dt^2+\frac{4ma}{r} dt dr+a^2(1+\frac{2m}{r})dr^2+a^2 r^2 d\Omega^2
\label{new49}
\end{equation}  
For the radial null rays:  
\begin{equation}
\frac{dr}{dt}=\frac{\pm 1-\frac{2m}{r}}{a(1+\frac{2m}{r})}
\label{new50}
\end{equation}  
Thus, the outgoing and ingoing null vector fields are given by: 
\begin{equation}
\begin{cases}
l^p=(1+\frac{2m}{r},\frac{1}{a}(1-\frac{2m}{r})) \\
n^p=(1,-\frac{1}{a})
\end{cases}
\label{new51}
\end{equation}  
where $ p = 0, 1 $, and the normalization is set to $l^p n_{p}=-2$.  
Finally, using Eq.~\eqref{new6}, we can compute the expansion along the null vector fields,  $l^p$ and $n^p$, as:  
\begin{equation}
\begin{cases}
\theta_{+}=\frac{2}{R} [(1+\frac{2ma}{R})HR+(1-\frac{2ma}{R})] \\
\theta_{-}=\frac{2}{R}(HR-1)
\end{cases}
\label{new52}
\end{equation}  
where $H={\dot{a}}/{a}$, $\dot{a} = {da}/{dt}$, $R = ar$ and $H>0$ is assumed.  
We see that, as in the Culetu spacetime, both outgoing and ingoing null expansions can vanish.  
This gives the THs as:  
\begin{equation}
\begin{cases}
\theta_{+}=0 \Rightarrow R_{+}=\frac{2ma(1-HR_{+})}{1+HR_{+}} \\
\theta_{-}=0 \Rightarrow R_{-}=\frac{1}{H}
\end{cases}
\label{new53}
\end{equation}  
Clearly, $\theta_{+}=0$ occurs only if $0<R<2ma$ or $0<r<2m$. 

\begin{figure}[H]
   \centering

      \includegraphics[width=0.8\textwidth]{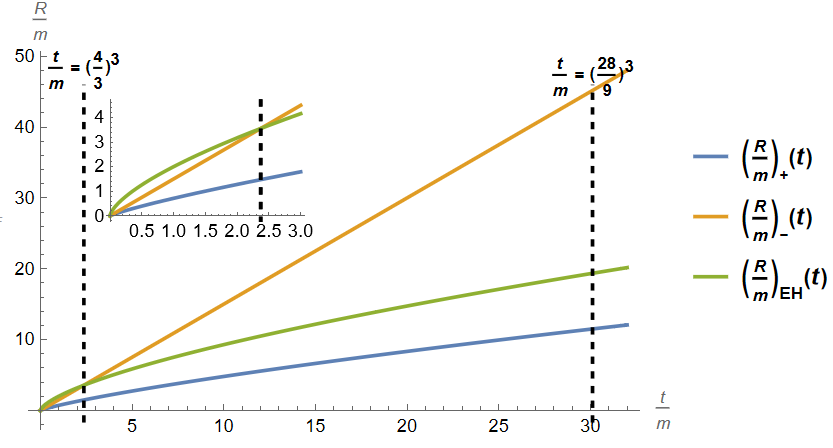}
  
          \caption{The time evolution of the horizons in the matter-dominated Sultana-Dyer spacetime, $a(t)=\tilde{t}^{{2}/{3}}$. The time and radius of the horizons are normalized as $ \tilde{t}={t}/{m}$, $  \tilde{R}={R}/{m}$ and $ t_0 = m $, with $ \tilde{t}_0$ set to unity.
        In the interval $0< \tilde{t}<({4}/{3})^3$, we have $0<R_{-}<R_{+}<R_{EH}$  and $0<R_{-}<R_{EH}<R_{+}$ in the interval $\tilde{t}>({4}/{3})^3$.  $\tilde{t}=({28}/{9})^3$ is the upper limit of $\eta$. 
         Since $R_-$ lies outside the event horizon after a certain period of time, it follows from \cite{nielsen2009} that the NEC must be violated in some region of the Sultana-Dyer spacetime.} 
        \label{fig4aa}
\end{figure}

It can also be shown that the solutions of Eqs.~$R=2M_{MSH}$ and $\theta_{+} \theta_{-}=0$ are the same.  
According to Eq.~\eqref{new32}, the MSH mass of the Sultana-Dyer spacetime is obtained as\footnote{The MSH mass of the Sultana-Dyer spacetime is given by  $M_{MSH} =ma(1-{2 \alpha r}/{\eta}+{\alpha^2 r^2}/{\eta^2}+{\alpha^2 r^3}/{(2m \eta^2)})$ \cite{sato}. Considering that $aH = {{da}/{(a d\eta)}} = {\alpha}/{\eta}$ and using the definition of the areal radius, i.e., $R = ar $, this relation agrees with ~\eqref{new55}.}:
\begin{equation}
M_{MSH}=ma(1-HR)^2+\frac{H^2 R^3}{2}
\label{new55}
\end{equation}  
Obviously, assuming $m>0$, the MSH mass is always positive and in the limit $r>>2m$, it  asymptotically approaches ${H^2 R^3}/{2}$ which is the MSH mass of the flat FLRW spacetime \cite{FLRW}.  
We will now solve the THs equation $R=2M_{MSH}$: 
\begin{equation}
(R(1+HR)-2ma(1-HR))(HR-1)=0
\label{new56}
\end{equation}  
which is completely the same as Eq.~\eqref{new53}. The solutions of Eq.~\eqref{new56} are:  
\begin{equation}
\begin{cases}
R_{+}=\frac{-1-2maH+\sqrt{(1+2maH)^2+8maH}}{2H}\\
R_{-}=\frac{1}{H}
\end{cases}
\label{new57}
\end{equation}  
From Eq.~\eqref{new57}, we immediately see that $0<R_{+}<2ma$ always holds. Therefore, the Sultana-Dyer spacetime has two THs: Future TH ($\theta_{+}=0,\theta_{-}|_{\theta_{+}=0}<0$) and Past TH ($\theta_{-}=0,\theta_{+}|_{\theta_{-}=0}>0$) according to Eq.~\eqref{new52}.  Furthermore, it is obvious that $0<R_{+}<R_{-}$ is always satisfied.  
We now examine the ECs near and on the obtained THs to determine the time intervals in which the Sultana-Dyer spacetime has a cosmological BH interpretation. 
In \cite{sato}, by defining $ \eta_{N}(r)={2r(2r+3m)}/{3m}$ and $\eta_{W}(r)={r(r+2m)}/{2m}$ and assuming $m>0$, it is shown that for a matter-dominanted universe, ($a(\eta) \sim \eta^2$), the energy-momentum tensor of the Sultana-Dyer spacetime,  satisfies the NEC and SEC in the interval $0<\eta<\eta_{N}(r)$, while the WEC and DEC are satisfied in the interval $0<\eta<\eta_{W}(r)$.
Considering Eq.~\eqref{new53} and using $Ha={2}/{\eta}$, we find that:  
\begin{equation}
\begin{cases}
\eta_{+}(r)=2r \frac{2m+r}{2m-r}\\
\eta_{-}(r)=2r
\end{cases}
\label{new58}
\end{equation}
on the THs. By comparing Eq.~\eqref{new58} with the definitions of $\eta_{N}(r)$ and $\eta_{W}(r)$, we conclude that all ECs are violated on $R_{+}$, while the NEC and SEC are satisfied in the interval $0 < r < 2m$ on $R_{-}$, but the  WEC and DEC are violated. However,  all ECs are satisfied on $R_{-}$ for $r > 2m$.  
As discussed in \cite{sato}, for the Sultana-Dyer metric to describe a suitable a physical spacetime, NEC must hold on the event horizon and outside it. This implies that $0<\eta<\eta_{N}(2m)$ or equivalently $0<\eta<{28m}/{3}$. Therefore, at least in early stages, the Sultana-Dyer solution describes a physical spacetime.  
Fig. \ref{fig4aa} illustrates the evolution of the Sultana-Dyer THs according to \eqref{new57}.
To determine the nature of the THs of the Sultana-Dyer spacetime, we analyze the signs of $\mathcal{L}_{+} \theta_{-}|_{\theta_{-}=0}$ and $ \mathcal{L}_{-} \theta_{+}|_{\theta_{+}=0}$.
A simple calculation using Eqs.~\eqref{new51}, \eqref{new53}, and \eqref{new57} shows that  
\begin{equation}
\begin{cases}
\frac{1}{2} \mathcal{L}_{+} \theta_{-}|_{\theta_{-}=0}=\frac{1}{2} H^2(1-6maH)\\
\frac{1}{2} \mathcal{L}_{-} \theta_{+}|_{\theta_{+}=0}=2H^2 \frac{-1-14maH+\sqrt{(1+2maH)^2+8maH}}{(-1-2maH+\sqrt{(1+2maH)^2+8maH})^3}
\end{cases}
\label{new59}
\end{equation}  
We see that $ \mathcal{L}_{-} \theta_{+}|_{\theta_{+}=0}<0$, and since $\theta_{-}|_{\theta_{+}=0}<0$, then $R_{+}$ is an FOTH representing a BH-type TH. However, for the PTH $R_{-}$, ($\theta_{+}|_{\theta_{-}=0}>0$), $\mathcal{L}_{+} \theta_{-}|_{\theta_{-}=0}$ can be positive, negative, or zero. With respect to the normalized time $\tilde{t}={t}/{t_0}$ and according to Eq.~\eqref{new59}, $\mathcal{L}_{+} \theta_{-}|_{\theta_{-}=0}<0$ in the interval $0< \tilde{t}<4^3$, and so  $R_{-}$ is a POTH. At $\tilde{t}=4^3$, $R_{-}$ is a PDTH since $\mathcal{L}_{+} \theta_{-}|_{\theta_{-}=0}=0$. For $\tilde{t}>4^3$, $\mathcal{L}_{+} \theta_{-}|_{\theta_{-}=0}>0$,  so $R_{-}$ is a PITH. Restricting the time interval to $0<\tilde{t}<(\frac{28}{9})^3<4^3$, 
 $R_{-}$ is a POTH. However, based on Eq.~\eqref{new53}, the PTH of the Sultana--Dyer spacetime coincides with the FLRW cosmological TH. Since this spacetime is asymptotically FLRW, its PTH can be interpreted as a cosmological TH.

To compute the signature of the THs, we take the sign of the normal vector field of the hypersurface, $\phi(t,r)=ar-R_{TH}(t)=0$. Using the line element Eq.~\eqref{new49} for $R_{-}$
\begin{equation}
\begin{cases}
 \phi_{-}=ar-R_{-}(t)=0, \quad N_\mu=\nabla_\mu \phi_{-}\\
N^2|_{R=R_{-}}=\frac{9}{2}(\frac{1}{6}-maH)
\end{cases}
\label{new60}
\end{equation}  
Therefore, 
 $R_{-}$  is spacelike for $0<\tilde{t}<4^3$. At $\tilde{t}=4^3$, it becomes null, and for $\tilde{t}>4^3$, it is timelike. If the time interval is restricted to $0<\tilde{t}<(\frac{28}{9})^3<4^3$, $R_{-}$ is spacelike.  A similar calculation for $R_{+}$ yields:
\begin{equation}
	\begin{cases}
	 \phi_{+}=ar-R_{+}(t)=0, \quad M_\mu=\nabla_\mu \phi_{+}\\
	M^2|_{R=R_{+}}=-(1+\frac{2ma}{R_{+}})(HR_{+}-\dot{R}_{+})^2-\frac{4ma}{R_{+}}(-HR_{+}+\dot{R}_{+})+(1-\frac{2ma}{R_{+}})
	\end{cases}
\label{new61}
\end{equation}  
Now, we will show that $M^2|_{R=R_{+}}<0$ at all times. The first  term in $M^2|_{R=R_{+}}$ is clearly negative. The same is true for the third term due to $0<R_+<2ma$, as seen in Eq.~\eqref{new53}. 
The coefficient of the second term is $-HR_{+}+\dot{R}_{+}$  which is proportional to $\dot{r}_{+}$, and $ R = ar$.
In a matter-dominated universe, using Eq.~\eqref{new57}, $\tilde{r}_{+}(\tilde{t})=-1+(-3+\sqrt{9+{8(2+9 \tilde{t}^{1/3})}{\tilde{t}^{{-2/3}}}})\tilde{t}^{{1/3}}/4$, where $\tilde{r}={r}/{t_0}$ and $\tilde{t}={t}/{t_0}$. Since the function $\tilde{r}_{+}(\tilde{t})$ increases with time, $\dot{r}_{+}>0$, and therefore, $-HR_{+}+\dot{R}_{+}>0$. This indicates that $R_{+}$ is always spacelike.  

The table below summarizes the results for the signature and nature of the Sultana-Dyer spacetime's THs:
\begin{table}[h!] 
	\centering
	\begin{tabular}{|c|c|c|c|}
		\hline
		& $0 \le \tilde{t} <\tilde{t}_0$ & $ \tilde{t} = \tilde{t}_0 $ & $\tilde{t}>\tilde{t}_0$ \\
		\hline
		$R_{-}$ & POTH -SL& PDTH-Null & PITH-TL \\
		\hline
		$R_{+}$ & FOTH-SL & FOTH-SL & FOTH-SL \\
		\hline
	\end{tabular}
	\caption{The sign and nature of the THs of the matter-dominated Sultana-Dyer spacetime for different time intervals $\tilde{t}$. 
	SL and TL refer to spacelike and timelike, respectively. }
	\label{TIV}
\end{table}
The null expansion parameters of the Sultana-Dyer spacetime in the conformal Kerr-Schild coordinates can both vanish \cite{sato}, which is in full agreement with our results in \eqref{new53}. 
 Furthermore, \cite{sato} demonstrated that, \(\eta_{+}(r)\), on which the outgoing null expansion parameter \(\theta_+\) vanishes, corresponds to a spacelike FOTH.
 In contrast,  \(\eta_{-}(r)\), on which the ingoing null expansion parameter \(\theta_-\) vanishes, can describe a spacelike POTH for \(0<r<r_{d(SD)}\), a null PDTH for \(r= r_{d(SD)}\), or a timelike PITH for \(r>r_{d(SD)}\).
 These results are in full agreement with those obtained in this section\footnote{According to Eq.~\eqref{new57}) and the definition of \(R\), at the PTH of the Sultana-Dyer spacetime in a matter-dominated universe,  we obtain: $maH = m/r$.  Based on Eqs.~\eqref{new59}, \eqref{new60} and table \ref{TIV}, \(R_{-}\) represents a spacelike POTH for \(0<r<6m\). At \(r=6m\) it corresponds to a null PDTH, and for \(r>6m\) it describes a timelike PITH. The authors of \cite{sato} state that \(r_{d(SD)}=6m\) in a matter-dominated universe which is consistent with our result.
}.
\subsection{Trapping horizons in Painlev\'e-Gullstrand coordinates}\label{IV.B}
To derive the line element of the Sultana-Dyer spacetime in PG coordinates \eqref{new14}, we start with Eq.~\eqref{new49} and perform a coordinate transformation from $(t,r)$ to $(t,R)$, where $R=ar$ is the areal radius. The result is:  
\begin{equation}
\begin{cases}
ds^2=-A' dt^2+(1+\frac{2ma}{R})dR^2+2B' dRdt+R^2 d\Omega^2\\
A'=(1-\frac{2ma}{R})-H^2R^2(1+\frac{2ma}{R})+4maH \\
B'=\frac{2ma}{R}-HR(1+\frac{2ma}{R})
\end{cases}
\label{new62}
\end{equation} 
Next, we perform another transformation, from $(t,R)$ to $(\tau,R)$, to ensure that the final form of the metric satisfies the condition $g_{RR}=1$. This transformation is given by $dt=Fd\tau-\gamma dR$, where the right-hand side is a total differential due to the integration factor $F$.
Substituting $dt$ into line element \eqref{new62} yields:  
\begin{equation}
ds^2=-A' F^2 d\tau^2+(-A' \gamma^2+2B' \gamma+1+\frac{2ma}{R})dR^2+2F(\gamma A'+B')d\tau dR+R^2 d\Omega^2
\label{new63}
\end{equation}  
Using Eqs.~\eqref{new55} and \eqref{new62}, we find that $A'=1-{2M_{MSH}}/{R}$ and $B'^2+{2maA'}/{R} =  {2M_{MSH}}/{R}$. The condition $g_{RR}=1$ also yields the following equation for $\gamma$:  
\begin{equation}
A' \gamma^2+2B' \gamma-\frac{2ma}{R}=0
\label{new64}
\end{equation}  
On the other hand, comparing Eqs.~\eqref{new14} and \eqref{new63} yields:
\begin{equation}
\begin{cases}
 c(R,\tau)=\sqrt{A'F^2+v^2}\\
 v(R,\tau)=F(\gamma A'+B')
\end{cases}
\label{new65}
\end{equation}  
The positive sign for $c(R,\tau)$ has been considered without compromising the generality of the problem. Using Eqs.~\eqref{new64} and \eqref{new65} , we can simplify $v(R,\tau)$ as:  
\begin{equation}
\frac{v}{F}=\pm \sqrt{1-A'}
\label{new66}
\end{equation}  
Similar to the flat-McVittie spacetime analysis in section \ref{II.B}, we compute $v(R,\tau)$ on the THs, which are given by $A'=0$. We also assume that $\gamma$  is finite there to determine the sign of $v(R,\tau)$. Using Eqs. \eqref{new65} and \eqref{new66}, we obtain:
\begin{equation}
\frac{v}{F}|_{A'=0}=\pm 1=B'|_{A'=0}
\label{new67}
\end{equation}  
The solutions for $A'=0$ are obtained in Eq.~\eqref{new57}. Using these solutions and the definition of $B'$ in Eq.~\eqref{new62}, we have:  
\begin{equation}
\begin{cases}
 B'|_{R=R_+}=+1\\
  B'|_{R=R_-}=-1
\end{cases}
\label{new68}
\end{equation}  
As expected, for $R_{+}$, which was shown to be a FTH in section \ref{IV.A}, $v|_{R_{+}}>0$ holds. This means that the negative sign should be chosen for $\theta_{+}$  and the positive sign for $\theta_{-}$ in Eq.~\eqref{new17}. However, for $R_{-}$, which was shown to be a PTH in  section \ref{IV.A}, $v|_{R_{-}}<0$ holds. Therefore,  the positive sign should be chosen for $\theta_{+}$ and the negative sign for $\theta_{-}$ in \eqref{new17}. Ultimately, the final forms of $c(R,\tau)$ and $v(R,\tau)$  in the Sultana-Dyer spacetime can be summarized as follows:  
\begin{equation}
\begin{cases}
 c(R,\tau)=F\\
 v(R,\tau)=\pm F \sqrt{ \frac{2M_{MSH}}{R}}
\end{cases}
\label{new69}
\end{equation}
Since $R_{+}<R_{-}$ , we expect the positive sign for $v(R,\tau)$  to be acceptable in any interval $(R_{0},R_{1})$, where $0<R_{0}<R_{+}<R_{1}<R_{-}$ is assumed.  Similarly, in any interval $(R_{1},R_{2})$, where $R_+<R_{1}<R_{-}<R_{2}$, the negative sign for  $v(R,\tau)$  is acceptable. Thus, the null expansion parameters are given by:  
\begin{equation}
\left\{
\begin{alignedat}{2}
\theta_{+} & = \frac{2}{R}(1-\sqrt{\frac{2 M_{MSH}}{R}}) \\
\theta_{-} & = -\frac{2}{R}(1+\sqrt{\frac{2 M_{MSH}}{R}})
\end{alignedat}
\right.
\quad \text{for } R \in (R_0, R_1)
\quad \text{,}
\quad
\left\{
\begin{alignedat}{2}
\theta_{+} & = \frac{2}{R}(1+\sqrt{\frac{2 M_{MSH}}{R}}) \\
\theta_{-} & = -\frac{2}{R}(1-\sqrt{\frac{2 M_{MSH}}{R}})
\end{alignedat}
\right.
\quad \text{for } R \in (R_1, R_2)
\label{new70}
\end{equation}
The signs of $\theta_{+}$ and $\theta_{-}$ at different intervals, are summarized in table~\ref{table4}.  
 \begin{table}[h!]
\centering 
\begin{tabular}{|c|c|c|}
\hline
$R$ & $\theta_{+}$ & $\theta_{-}$ \\ \hline
$(R_{0},R_{+})$ & N & N \\ \hline
$R_{+}$ & Zero & N \\ \hline
$(R_{+},R_{1})$ & P & N \\ \hline
$(R_{1},R_{-})$ & P & N \\ \hline
$R_{-}$ & P & Zero \\ \hline
$(R_{-},R_{2})$ & P & P \\ \hline
\end{tabular}
\caption{The behavior of $\theta_{+}$ and $\theta_{-}$ in different regions of $R$. 
$R < 2M_{{MSH}}$ corresponds to the region outside of the two THs and $R > 2M_{{MSH}}$ corresponds to the region between the two THs. The regions $R_{0}<R<R_{+}$ and $R_{-}<R<R_{2}$ are outside the two THs, where $\theta_{+} \theta_{-} > 0$ and correspond to the trapped regions. The region $R_{+}<R<R_{-}$, where $\theta_{+} \theta_{-} <0$, is an untrapped region. When $\theta_{+} \theta_{-} = 0$, the two surfaces $R_{+}$ and $R_{-}$ represent the marginally trapped surfaces. These results obtained are consistent with those obtained in Kerr-Schild coordinates.}
\label{table4}
\end{table}
The results of the ingoing and outgoing null expansion calculations in the conformal Kerr-Schild 
 and PG coordinate systems for the Sultana-Dyer spacetime are consistent. According to Eqs.~\eqref{new52}, \eqref{new53},   \eqref{new57} and  \eqref{new70},  we have demonstrated that $\theta_{+} = 0$ and $\theta_{-} = 0$  can occur simultaneously in both coordinate systems. This represents 
an FOTH at $R_{+}$ and a POTH, PDTH or PITH at $R_{-}$ \eqref{new59}, respectively.

In summary, the Sultana-Dyer solution can describe a cosmological BH based on our Hayward-Inspired quasi-local framework. It should be noted that this geometry includes two types of THs: a BH-type TH and a cosmological TH. At early times in an expanding universe, the cosmological TH behaves as a POTH, while at later times it behaves as a PITH.

\section{Glass-Mashhoon spacetime}\label{V}
\subsection{A class of dynamical spherically symmetric spacetimes}\label{V.A}
The Glass-Mashhoon solution of Einstein's equations describes a spherically symmetric stellar system with a collapsed core surrounded by a matter distribution in gravitational collapse. In a special case, it reduces to the relativistic Plummer model \cite{McVittie energy condition,fackerell1971,Plummer}, resulting in smooth density and pressure profiles that vanish at the boundary of star. During gravitational collapse, a trapped surface dynamically forms and eventually the Schwarzschild horizon appears. This construction is one of the earliest analytic models for BH formation from an astrophysically realistic matter distribution \cite{glass1976}. The Glass-Mashhoon spacetime is described by 
the following line element in the isotropic coordinates:
	\begin{equation} 
		ds^2 = - e^{\Sigma(r,t)}dt^2 + e^{\nu(r,t)}(dr^2+r^2d\Omega^2) \label{new71}
	\end{equation}
Suppose that the above geometry is generated by a perfect fluid describing the distribution of matter around a central object.
Substituting line element \eqref{new71} into Einstein's field equations together with McVittie's assumption \cite{McVittie1933} that the background matter has no radial flux in the isotropic coordinate, i.e. $G^{t}_{r}=0$,  yields the equations necessary to determine the metric components, as well as the energy density and pressure of the corresponding perfect fluid.
According to \cite{McVittie1933}, combining Einstein's equations yields the following two equations, which include the metric components:
		\begin{equation} 
		G^{t}_{r}= e^{-\Sigma} (\dot{\nu}^\prime-\frac{1}{2}\dot{\nu}\Sigma^\prime)=8\pi T^{t}_{r}= 0 \label{new72}
	\end{equation}
		\begin{equation} 
		G^{\phi }_{\phi }(=G^{\theta}_{\theta})-G^{r }_{r }= e^{-\nu} (\Sigma^{\prime\prime}+\nu^{\prime\prime}-\frac{1}{r}(v^\prime+\Sigma^\prime)-\nu^\prime\Sigma^\prime-\frac{1}{2}(\nu^\prime)^2+\frac{1}{2}(\Sigma^\prime)^2)=8\pi [T^{\phi }_{\phi }(=T^{\theta}_{\theta})-T^{r}_{r}]=0 \label{new73}
	\end{equation}
		where a dot (a prime) over a variable denotes its derivative with respect to the cosmic time
(radial coordinate). Reading $\Sigma^\prime$ and $\Sigma^{\prime\prime}$ from Eq.~\eqref{new72} and their radial derivatives, and then substituting them into Eq.~\eqref{new73}, shows that this equation is a complete time derivative that can be integrated as \cite{ExactS}:
		\begin{equation} 
		e^{\frac{\nu}{2}}(\nu^{\prime\prime}-\frac{\nu^\prime}{2}-\frac{1}{2}\nu^{\prime 2})=g(r) \label{new74}
	\end{equation}
	where $g(r)$ is an arbitrary function of the radial coordinate. For any choice of $g(r)$, the radial dependence of the corresponding function $\nu(t,r)$ can be obtained.
	To obtain the Glass-Mashhoon geometry, consider $g(r)$ as \cite{glass1976}: 
		\begin{equation} 
		g(r)= \frac{3\lambda_0 r^2 (\beta\gamma - \alpha\delta)^2}{\big((\alpha r^2 + \beta) (\delta + \gamma r^2)\big)^{\frac{5}{2}}} \label{new75}
	\end{equation}
		 With this choice,  the metric coefficients from the Eqs.~\eqref{new72} and \eqref{new74} are obtained as:
				\begin{equation}
\begin{cases}
			e^{{\Sigma(r,t)}/{2}} = \frac{1 - \frac{1}{2}m_0 U(r) a^{-1}(t)}{1 + \frac{1}{2}m_0 U(r) a^{-1}(t)}\\ 
		
			e^{{\nu(r,t)}/{2}} = (1 + \frac{1}{2} m_0 U(r) a^{-1}(t))^2 V^{-1}(r) a(t)\\

			U(r) = \sqrt{\frac{\gamma r^2 + \delta}{\alpha r^2 + \beta}}\\
	
	V(r) = \gamma r^2 + \delta
\end{cases}
\label{new76}
		\end{equation}
		Here, $a(t)$ plays the role of the scale factor of the universe. We assume that $m_0$, $U$, $V$ and $a(t)$ are all positive. The constants $\alpha$, $\beta$, $\gamma$ and $\delta$ are arbitrary. Thus, there is a class of solutions; each solution is identified by a unique set of these parameters.  The spatially curved McVittie metric is given by $\alpha >0$, $\beta = 0$ and $\delta >0$, while the corresponding flat case requires $\gamma=0$ as well. Metric \eqref{new71} can be written in a more convenient form by introducing the new coordinate $\overline{r} \equiv {1}/{U(r)}$ \cite{glass1976}.
		Then, the line element can be written as:
		\begin{equation} 
		ds^2 = -A^2(\overline{r},t) dt^{2}+ B^2(\overline{r},t) d\overline{r}^2 + R^2 d\Omega^2 \label{new78}
	\end{equation}
	where the metric components are given by:
			\begin{equation}
\begin{cases}
			A(\overline{r},t) = \frac{1 - \frac{m_0}{2 a(t) \overline{r}}}{1 + \frac{m_0}{2 a(t) \overline{r}}} \\
		
			B(\overline{r},t) = \overline{r} a(t) (1 + \frac{m_0}{2 a(t) \overline{r}})^2 \frac{1}{\sqrt{(\alpha - \gamma \overline{r}^2) (\delta \overline{r}^2 - \beta)}}\\
			
			R(\overline{r},t) = a(t) \frac{1}{\alpha \delta - \beta \gamma} (1 + \frac{m_0}{2 a(t) \overline{r}})^2 \sqrt{(\alpha - \gamma \overline{r}^2) (\delta \overline{r}^2 - \beta)}
\end{cases}
			\label{new79}
		\end{equation}
			We see that the above coordinate system is still comoving as in \eqref{new71}. 
			\subsection{Density, Pressure, and Ricci Scalar}\label{V.B} 
		By substituting the line element \eqref{new78} into Einstein's equations, we can derive the energy density and pressure of the matter distribution\footnote{To find the density and pressure in the coordinate system \eqref{new71}, it is sufficient to replace $\overline{r} \equiv {1}/{U(r)}$ in Eqs.~\eqref{new80} and \eqref{new81}.}
		\begin{equation} 
		\rho(\overline{r},t) = \frac{3H^2}{8\pi} + \frac{24}{\pi} a^2(t) \frac{[2\overline{r}^5 \delta \gamma a(t) + \alpha \beta m_0]}{(2 \overline{r} a(t) + m_0)^5} \label{new80}
	\end{equation} 
		\begin{equation} 
		P(\overline{r},t) = - \frac{3H^2}{8\pi} - \frac{\dot{H}}{4 \pi} \left(\frac{1 + \frac{m_0}{2 a(t) \overline{r}}}{1 - \frac{m_0}{2 a(t) \overline{r}}}\right) - \frac{8 a^2(t)}{\pi} \frac{(4 a^2(t) \overline{r}^6 \delta \gamma - \alpha \beta m_0^2)}{(2 a(t) \overline{r} - m_0)(2 a(t) \overline{r} + m_0)^5} \label{new81}
	\end{equation}
		where $H=\dot{a}/a$. A special case arises when $\beta = \gamma = 0$, in which the above relations reduce to the density and pressure of the flat McVittie metric, as derived in section \ref{II.A} in Eqs.~\eqref{NM1} and \eqref{NM2}. The second term in Eq.~\eqref{new80} represents the spatial curvature of the Glass-Mashhoon metric\footnote {The spatial part of the line element \eqref{new78} is given by: $^{(3)}\textrm{ds}^{2}= B^{2}(\bar{r},t) d\bar{r}^2 + R^2 d\Omega ^2$, and the corresponding spatial Ricci scalar  can be derived as: $^{(3)}R=  384 a^2(t) {[2\overline{r}^5 \delta \gamma a(t) + \alpha \beta m_0]}/{(2 \overline{r} a(t) + m_0)^5}$. This differs from the second term in Eq.~\eqref{new80} only by a numerical coefficient.}. Setting this term to zero reduces the Glass-Mashhoon metric to the flat McVittie metric. This occurs because the second term in Eq.~\eqref{new80} vanishes identically  when $\beta=\gamma = 0$, which is consistent with the flat McVittie case mentioned earlier.

The pressure diverges at $\bar{r} = {m_0}/(2 a(t))$ \footnote {It coincides exactly with the McVittie spacetime singularity.}, where the second and third terms of the pressure diverge, as does the Ricci scalar $R = 8 \pi (\rho - 3P)$. Therefore, even if the background spacetime is deSitter, i.e. $\dot{H}=0$, the pressure will still diverge due to the third term. This contrasts with the case of a central mass immersed in a de Sitter background, where the pressure is only singular at the origin. The singularity at $\overline{r} = {m_0}/(2 a(t))$ is again a weak gravitational singularity because, the determinant of the spatial part of metric \eqref{new71} shows that an arbitrary differential volume does not tend to zero. Therefore, as Eq.~\eqref{new80} shows, the energy density does not diverge at this point. A simple calculation shows that the energy density and pressure, as in Eqs.~\eqref{new80} and \eqref{new81},  satisfy the following conservation equation: 
	\begin{equation} 
		\dot{\rho} + 3 H \left(\frac{1 - \frac{m_0}{2 a(t) \overline{r}}}{1 + \frac{m_0}{2 a(t) \overline{r}}}\right) (\rho + P) = 0 \label{new82}
	\end{equation} 
	As expected, when $m_0=0$, this reduces to the conservation law for the cosmic fluid of the FLRW background.
	
	Note that the line element \eqref{new78} is singular not only at $\overline{r} = {m_0}/(2 a(t))$ but also has at $r=0$ where the curvature scalar is not singular according to Eqs.~\eqref{new80} and \eqref{new81}. The same is  true for the other curvature scalars.
			To examine the ECs for the Glass–-Mashhoon solution, using Eqs.~\eqref{new80} and \eqref{new81}, one must determine the sign of the following expressions outside the curvature singularity (due to the divergence of the pressure in the vicinity of the curvature singularity):  
			 \begin{align}
			 	\rho+P
			 	&=\;-\frac{1}{4\pi}\,\frac{2\,a(t)\,\overline{r}+m_0}{2\,a(t)\,\overline{r}-m_0}\,\dot{H}
			 	+\frac{16\,a^2(t)}{\pi}\,
			 	\frac{ a(t)\,\overline{r}^{5}\,\delta\gamma\,\big(4\,a(t)\,\overline{r}-3m_0\big)
			 		+ \alpha\beta\,m_0\,\big(3\,a(t)\,\overline{r}-m_0\big)}
			 	{ \big(2\,a(t)\,\overline{r}-m_0\big)\,\big(2\,a(t)\,\overline{r}+m_0\big)^{5} }
			 	\label{G1} \\[6pt]
			 	\rho-P
			 	&=\; \frac{3}{4\pi}\,H^2
			 	+\frac{1}{4\pi}\,\frac{2\,a(t)\,\overline{r}+m_0}{2\,a(t)\,\overline{r}-m_0}\,\dot{H}
			 	+\frac{16\,a^2(t)}{\pi}\,
			 	\frac{ a(t)\,\overline{r}^{5}\,\delta\gamma\,\big(8\,a(t)\,\overline{r}-3m_0\big)
			 		+ \alpha\beta\,m_0\,\big(3\,a(t)\,\overline{r}-2m_0\big)}
			 	{ \big(2\,a(t)\,\overline{r}-m_0\big)\,\big(2\,a(t)\,\overline{r}+m_0\big)^{5} }
			 	\label{G2} \\[6pt]
			 	\rho+3P
			 	&=\; -\frac{3}{4\pi}\,H^2
			 	-\frac{3}{4\pi}\,\frac{2\,a(t)\,\overline{r}+m_0}{2\,a(t)\,\overline{r}-m_0}\,\dot{H}
			 	+\frac{48\,a^3(t)\,m_0}{\pi}\,
			 	\frac{\overline{r}\,\big(\alpha\beta-\overline{r}^{4}\delta\gamma\big)}
			 	{ \big(2\,a(t)\,\overline{r}-m_0\big)\,\big(2\,a(t)\,\overline{r}+m_0\big)^{5} }
			 	\label{G3}
			 \end{align}
			 The NEC in this spacetime is satisfied if Eq.~\eqref{G1} is non--negative.  
			 The WEC is satisfied if both Eqs.~\eqref{new80} and \eqref{G1} are non--negative.  
			 The DEC holds if Eqs.~\eqref{G1},\eqref{G2} are non--negative, and finally, the SEC is satisfied if Eqs.~\eqref{G1},\eqref{G3} are non--negative.  
			 
			 However, note that a curvature singularity exists only if $m_0 > 0$. Therefore, based on the above expressions, the ECs in this spacetime are generally not trivially satisfied. For $\beta = \gamma = 0$ and $\alpha, \delta > 0$, one exactly recovers the ECs of the matter-dominated flat McVittie spacetime studied in Sec.~\ref{II.A}.
			 
			It can be seen that the signs of the two constant parameters $\chi \equiv \gamma \delta$ and $X \equiv \alpha \beta$ are effective in determining the validity of the ECs.
			 			For a matter-dominated background $(a(t) \sim t^{2/3})$, with $\chi, X > 0$, outside the curvature singularity,  the NEC and WEC are always satisfied for $\bar{r} \geq {3m_0}/({4a(t)})$, according to Eqs.~\eqref{new80},\eqref{G1}. Moreover, in this case,  the DEC is also always satisfied  for $\bar{r} \geq {3m_0}/({2a(t)})$ from Eqs.~\eqref{G1},\eqref{G2}.  
			 			 Satisfying the SEC is somewhat more subtle. Assuming $\chi, X > 0$, according to Eqs.~\eqref{G1},\eqref{G3},  the SEC is also satisfied in the interval ${3m_0}/({4a(t)}) \leq \bar{r} \leq \sqrt{{X}/{\chi}}$. 
			 
			 It should be emphasized that the regions mentioned for the matter-dominated background outside the curvature singularity are not the only regions where the ECs hold. Rather, they are special regions where the validity of the ECs can be seen more transparently.  
		\subsection{Misner-Sharp-Hernandez Mass} \label{V.C}		
		The MSH mass defined in Eq.~\eqref{new32} for the Glass-Mashhoon metric \eqref{new78}, can be obtained using the areal radius in Eq.~\eqref{new79}, as
		\begin{equation} 
		M_{MSH} = \frac{1}{2} H^2 R^3 + \frac{32 R^3 a^2(t)}{(m_0 + 2 \overline{r} a(t))^6} [2 \overline{r}^3 a(t) \gamma + \alpha m_0] [2 \overline{r}^3 a(t) \delta + \beta m_0]
		\label{new86}
	\end{equation}	
The first term represents the mass of the cosmic fluid contained within a sphere with a radius of $R$ and the second term illustrates various physical phenomena. For instance, in the flat McVittie metric, the second term is constant, and if $H=\text{cons.}$, then it can be interpreted as the mass of the central object. The MSH mass can also be expressed in terms of the fluid density Eq.~\eqref{new80}, with an additional term: 
		\begin{equation} 
		M_{MSH} = \frac{4}{3} \pi R^3 \rho + \frac{m_0}{(\alpha \delta - \beta \gamma)^3 \overline{r}^5} [(\alpha - \gamma \overline{r}^2) (\delta \overline{r}^2 - \beta)]^{\frac{5}{2}}
		\label{new87}
	\end{equation}
\subsection{Trapping Horizons in isotropic coordinates}\label{V.D}
We start  with the line element \eqref{new78}. Therefore, the radial null rays satisfy the following equation:  
\begin{equation}
\frac{d\bar{r}}{dt} = \pm \frac{A}{B}
\label{new88}
\end{equation} 
where $A$ and $ B $ are given by Eq.~\eqref{new79}. Outside the curvature singularity, i.e., $ \bar{r}={m_0}/{(2 a(t))} $,  we obtain $ {A}/{B}>0 $ from Eq.~\eqref{new79}. Consequently, the outgoing and ingoing null rays are given by:  
	\begin{equation}
				\begin{cases}
					l^p = (\frac{1}{A}, \frac{1}{B}) \\
					n^p = (\frac{1}{A}, -\frac{1}{B})
				\end{cases} 
\label{new89}
\end{equation}
where \( p=0,1 \) and the normalization is set to \( l_p n^p=-2 \).  Finally, using Eq.~\eqref{new6}, the expansion along the null vector fields is obtained as:  
\begin{equation}
\theta_{\pm}=2[H \pm \frac{1}{\bar{r} B} (A + \frac{ \alpha \beta -\bar{r}^4 \gamma \delta}{(\alpha-\gamma \bar{r}^2)(\delta \bar{r}^2 - \beta)})]
\label{new91}
\end{equation}
where \( H={\dot{a}}/{a} \), 
and \( H>0 \) is assumed.

According to Eq.~\eqref{new91},  both the ingoing and outgoing expansions can generally vanish unless the term \( A + { \alpha \beta -\bar{r}^4 \gamma \delta}/({(\alpha-\gamma \bar{r}^2)(\delta \bar{r}^2 - \beta)} \)) remains strictly positive or negative under specific conditions.
Also, Eq.~\eqref{new91} shows that whenever either the ingoing or the outgoing
	expansions vanishes\footnote{On the MTS
		where one expansion vanishes, the other remains strictly positive.}
	, \(\theta_{\pm}|_{\mp}>0\) still holds.
	Hence any THs that may exist in the Glass--Mashhoon
	spacetime are necessarily past--type.
	
	The crucial difference between the two horizons---namely the one satisfying
	\(\theta_{+}|_{\theta_{-}=0}>0\) and the one satisfying
	\(\theta_{-}|_{\theta_{+}=0}>0\) (assuming that the Lie derivative of
	the relevant vanishing expansion has the same sign in both cases)---lies in
	the asymptotic domain to which they are attached.
Since that asymptotic region is, by definition, \emph{untrapped}
	\((\theta_{+}\theta_{-}<0)\), two possibilities can occur:
	\begin{enumerate}
		\item \(\theta_{+}>0,\; \theta_{-}<0\), in which the outgoing null
		congruence diverges while the ingoing congruence converges;
		\item \(\theta_{-}>0,\; \theta_{+}<0\), in which the ingoing null
		congruence diverges while the outgoing congruence converges.
	\end{enumerate}

 As a special case of Glass-Mashhoon solution, in the flat McVittie spacetime, where \( \beta=\gamma=0,\alpha>0,\delta>0 \) \cite{glass1976}, ensuring that \( \theta_+>0 \) and that only \( \theta_- \) can vanish.  
 
 The conclusions reached here regarding the existence, nature, and signature of the THs in the Glass--Mashhoon spacetime are valid only if one can construct smooth congruences of outgoing and ingoing null geodesics simultaneously.
 As previously mentioned in Section \ref{II}, when the Hubble parameter of the flat McVittie asymptotically approaches a positive constant, the late--time evaluation of the expansion parameters associated with these outgoing/ingoing null congruences  breaks down in both isotropic and PG coordinates. A different treatment is therefore required in that regime.

In what follows, we analyse two further special cases of the Glass--Mashhoon solution.

\subsection{Curved-McVittie Spacetime ($\beta=0$, $\alpha>0$, $\delta>0$ and $\gamma \neq 0$)}\label{V.D.1}
A natural starting point is to generalize the flat McVittie solution by setting \(\gamma \neq 0\). 
Choosing \(\alpha, \delta > 0\) ensures that  the solution reduces to the Schwarzschild or FLRW solution in an appropriate limit.

The spatially curved McVittie spacetime is described by the following line element in $\bar{r}$ coordinate defined in section \ref{V.A} : \footnote{In \cite{McVittie1933}, the line element of the spatially curved McVittie metric 
	in isotropic coordinates, after simplification, takes the form
	$ds^2 = -\left(\frac{1 - m/(2 a(t) r)\,(1+kr^2/4)^{1/2}}
	{1 + m/(2 a(t) r)\,(1+kr^2/4)^{1/2}}\right)^2 dt^2 
	+ a(t)^2 (1+m/(2 a(t) r)\,(1+kr^2/4)^{1/2})^4/(1+kr^2/4)^2\,(dr^2 + r^2 d\Omega^2),$
	where $m>0$ is a constant and the sign of $k$ determines the type of spatial curvature. 
	By applying the coordinate transformation 
	$\bar{r} = b r / \sqrt{1+kr^2/4},$
	and introducing the definitions 
	$k/4 := \gamma \delta$, $b^2 := \alpha \delta$, and $m_0 := m b$, 
	one obtains the line element \eqref{86}. Therefore, the sign of $\gamma \delta$ 
	corresponds to the nature of spatial curvature in the spatially curved McVittie spacetime.
}
\begin{equation}
	ds^2= - (\frac{1-\frac{m_0}{2 a(t) \bar{r}}}{1+\frac{m_0}{2 a(t) \bar{r}}})^2 \, dt^2 + a^2(t) \, (1+\frac{m_0}{2 a(t) \bar{r}})^4 \, \frac{1}{\alpha \delta-\gamma \delta \bar{r}^2} \, d\bar{r}^2+ (\frac{a(t)}{\alpha \delta})^2 \, (1+\frac{m_0}{2 a(t) \bar{r}})^4 \, (\alpha \delta-\gamma \delta \bar{r}^2) \, \bar{r}^2 \, d\Omega^2
	\label{86}
\end{equation}
Similar to the flat McVittie solution, this spacetime exhibits a weak gravitational singularity at \( \bar{r}={m_0}/{2 a(t)} \). For the coordinate \( \bar{r} \) and the line element \eqref{86} to be well-defined in the curved McVittie spacetime, the condition $\bar{r}^2 (\alpha \delta - \gamma \delta \bar{r}^2) \geq 0$  must hold. Since $\alpha\delta>0$, if $\gamma\delta>0$, this condition imposes an upper bound on the coordinate $\bar{r}$.

To determine whether this spacetime describes a physically viable solution, we analyze the ECs outside the curvature singularity, i.e., \( \bar{r}={m_0}/{2 a(t)} \). 
According to Eqs.~\eqref{new80} and \eqref{new81}, the energy density and pressure associated with the matter distribution of curved McVittie spacetime, are obtained as follows:\cite{glass1976}
	\begin{align}
		\rho(t,\bar r)
		&= \frac{3\,H^2}{8\pi}
		+ \frac{48\,a(t)^3\,\bar r^{5}\,\gamma \delta}{\pi\,\big(2a(t)\,\bar r + m_0\big)^{5}}
		\label{2003}
		\\[6pt]
		P(t,\bar r)
		&= -\frac{3\,H^2}{8\pi}
		- \frac{\dot H}{4\pi}\,
		(\frac{1+\tfrac{m_0}{2a(t)\,\bar r}}{\,1-\tfrac{m_0}{2a(t)\,\bar r}\,})
		- \frac{32\,a(t)^{4}\,\bar r^{6}\,\gamma \delta}{\pi\,\big(2a(t)\,\bar r - m_0\big)\,\big(2a(t)\,\bar r + m_0\big)^{5}}
		\label{2004}
	\end{align}
	As can be seen, for $\gamma \delta = 0$, the energy density in Eq.~\eqref{NM1} and the pressure in Eq.~\eqref{NM2} of the spatially flat McVittie spacetime are recovered.  
	For simplicity of calculations, we first define the following dimensionless quantities:
	\begin{equation}
		\tilde{\bar{r}} \equiv \frac{\bar{r}}{m_0}, 
		\quad
		m \equiv \frac{m_0}{b},  
		\quad 
		\tilde{t} \equiv \frac{t}{m}, 
		\quad 
		\tilde{\chi} \equiv m^2 \chi
		\label{2002}
	\end{equation}
	where $\chi \equiv \gamma \delta$ and $b^2 = \alpha \delta > 0$.
	Therefore, the expressions defined for evaluating the ECs are also normalized with respect to \(m\). Moreover, we assume that \(\tilde{\chi}>0\) and the background is matter--dominated ($a(t) \sim t^{2/3}$).

	\textbf{NEC:}
	According to Eqs.~\eqref{G1} ,\eqref{2003} and \eqref{2004}, the NEC for this spacetime  holds outside the curvature singularity if:
	\begin{equation}
		\widetilde{\rho+P}=\frac{(y+1)^6+3\,a(\tilde{t})\,\tilde{\chi}(2y-3)y^5}{6\pi\,a(\tilde{t})^3 (y-1)(y+1)^5} \geq 0
		\label{2005}
	\end{equation}
	where $\widetilde{\rho+P} \equiv m^2 (\rho+P)$.\footnote{It can be shown that, according to Eq.~ \eqref{2002}, the pressure and density of the curved matter dominated McVittie are normalized by a factor of $m^2$.}
	
	Let \(y \equiv 2a(\tilde{t})\tilde{\bar{r}}\). Since the spatial curvature is positive, there exists an upper bound on \(\tilde{\bar{r}}\), as well as a lower bound due to the condition of being outside the curvature singularity. This condition, expressed in terms of \(y\), simplifies to  
	$y \in \left(1, {2a(\tilde{t})}/{\sqrt{\tilde{\chi}}}\right)$.
	Considering that the denominator of Eq.~\eqref{2005} is positive, we obtain
	\begin{equation}
		\text{Sgn}[\widetilde{\rho+P}] = \text{Sgn}\left[(y+1)^6+3\,a(\tilde{t})\,\tilde{\chi}(2y-3)y^5\right].
		\label{2006}
	\end{equation}
		Two general cases can occur for the sign of Eq.~\eqref{2006}:  
	
	\noindent
	\textbf{(a) \(0<\tilde{\chi}<\left({128}/{3}\right)^{2/3}\):}  
	
	\textbf{(1-a)} In the interval \({\sqrt{\tilde{\chi}}}/{2}<a(\tilde{t})\leq {64}/{(3\tilde{\chi})}\):  
	for every \(y \in \left(1,{2a(\tilde{t})}/{\sqrt{\tilde{\chi}}}\right)\), we have \(\widetilde{\rho+P}\geq 0\). Hence, the NEC holds.  
	
	\textbf{(2-a)} When \(a(\tilde{t})>{64}/{3\tilde{\chi}}>{\sqrt{\tilde{\chi}}}/{2}\):  
	in this temporal range, one can show that there exists a \(y_{N} \in (1,3/2)\)\footnote{Because for \(y\geq 3/2\), according to Eq.~\eqref{2005}, we always have \(\widetilde{\rho+P}\geq 0\). Similar to Glass-Masshoon case we discussed in section \ref{V.B}.} such that the numerator of Eq.~\eqref{2005} vanishes. Referring to the functional behavior introduced in Eq.~\eqref{2006}:  
	
	 \textbf{(2-a-1)} If \({2a(\tilde{t})}/{\sqrt{\tilde{\chi}}} \leq y_{N}\):  
	then for any \(1<y<{2a(\tilde{t})}/{\sqrt{\tilde{\chi}}}\leq y_{N}<3/2\), we obtain \(\widetilde{\rho+P}<0\), and thus the NEC does not hold.  
	
	\textbf{ (2-a-2)} If \(0<y_{N}<{2a(\tilde{t})}/{\sqrt{\tilde{\chi}}}\):  
	then for \(1<y<y_{N}<{2a(\tilde{t})}/{\sqrt{\tilde{\chi}}}\), we have \(\widetilde{\rho+P}<0\), while at \(y=y_{N}\) it becomes \(\widetilde{\rho+P}=0\). Moreover, in the interval \(y_{N}\leq y<{2a(\tilde{t})}/{\sqrt{\tilde{\chi}}}\), the condition \(\widetilde{\rho+P}\geq 0\) is satisfied.  
	
	\noindent
	\textbf{(b) \(\tilde{\chi}\geq \left({128}/{3}\right)^{2/3}\):}  
	
	In this case, since \(a(\tilde{t})>{\sqrt{\tilde{\chi}}}/{2}\geq{64}/{3\tilde{\chi}}\), only a scenario similar to case (2-a) occurs. In other words, there is no temporal range in which the NEC is satisfied for all values of \(y\).  
	In summary, for positive spatial curvatures, one may expect a violation of the NEC at sufficiently large time scales.

	\textbf{WEC}: Since the spatial curvature for the matter dominated curved McVittie spacetime is assumed to be positive, the energy density in Eq.~\eqref{2003} will be positive. Moreover, since the ECs are considered outside the curvature singularity, the WEC for this spacetime with positive spatial curvature is equivalent to the NEC, which was previously examined.

	\textbf{DEC:} The DEC for this spacetime, according to Eqs.~\eqref{G1} ,\eqref{G2} ,\eqref{2003} and \eqref{2004}, holds outside the curvature singularity if Eqs.~\eqref{2005} and:
		\begin{equation}
		\widetilde{\rho-P} = \frac{(12a(\tilde{t}){\tilde{\chi}}+1)y^6 + (2-9a(\tilde{t}){\tilde{\chi}})y^5 - 5y^4 - 20y^3 - 25y^2 - 14y - 3}{6\pi a(\tilde{t})^3 (y-1)(y+1)^5} \geq 0 
		\label{2007}
	\end{equation}  
	hold.
	The signs of Eqs.~\eqref{2005} and \eqref{2007} can be categorized as follows:  
	
	\textbf{(a) $0 < \tilde{\chi} < (128/3)^{2/3}$}:  
	
	\textbf{(a-1)} ${\sqrt{\tilde{\chi}}}/{2} < a(\tilde{t}) \leq {64}/({3\tilde{\chi}})$:  
	In this interval, it was shown that for every $y \in (1, {2a(\tilde{t})}/{\sqrt{\tilde{\chi}}})$, one has $\widetilde{\rho+P} > 0$. Therefore, the sign of $\widetilde{\rho-P}$ indicates whether the DEC holds or is violated. It can be shown that there exists a $y_D$ such that the numerator of Eq.~\eqref{2007} vanishes. Based on the behavior of the numerator in Eq.~\eqref{2007}:  
	
\textbf{	(a-1-1)} $y_D \geq {2a(\tilde{t})}/{\sqrt{\tilde{\chi}}}$:  
	In this case, for every $1<y<{2a(\tilde{t})}/{\sqrt{\tilde{\chi}}} \leq y_D$, one has $\widetilde{\rho-P}<0$, and thus the DEC will not hold.  
	
\textbf{	(a-1-2)} $0<y_D<{2a(\tilde{t})}/{\sqrt{\tilde{\chi}}}$:  
	For $1<y<y_D<{2a(\tilde{t})}/{\sqrt{\tilde{\chi}}}$, one has $\widetilde{\rho-P}<0$, while at $y=y_D$, $\widetilde{\rho-P}=0$. In the range $y_D \leq y < {2a(\tilde{t})}/{\sqrt{\tilde{\chi}}}$, one finds $\widetilde{\rho-P} \geq 0$, and hence the DEC is satisfied.  
	
\textbf{	(a-2)}  $a(\tilde{t}) > {64}/({3\tilde{\chi}}) > {\sqrt{\tilde{\chi}}}/{2}$:  
	In this interval, it can be shown that for every $y \in (1,{2a(\tilde{t})}/{\sqrt{\tilde{\chi}}})$, according to Eq.~\eqref{2006}, one has $\widetilde{\rho-P} > 0$. Thus, the validity of the DEC depends on the sign of $\widetilde{\rho+P}$, which was analyzed in detail for NEC above. Therefore, in this interval, the DEC is satisfied provided the NEC holds for $y \in (1,{2a(\tilde{t})}/{\sqrt{\tilde{\chi}}})$.  
	
	\textbf{(b) $\tilde{\chi} > (128/3)^{2/3}$:}  
	
	In this case, since $a(\tilde{t}) > {\sqrt{\tilde{\chi}}}/{2} \geq {64}/({3\tilde{\chi}})$, one has $\widetilde{\rho-P} > 0$. Thus, the validity of the DEC depends on the sign of $\widetilde{\rho+P}$. Therefore, in this case as well, the DEC is equivalent to the NEC for every $y \in (1,{2a(\tilde{t})}/{\sqrt{\tilde{\chi}}})$.

	\textbf{SEC:}  According to Eqs.~\eqref{G1} ,\eqref{G3} ,\eqref{2003} and \eqref{2004}, the SEC holds outside the curvature singularity if \eqref{2005} and:
		\begin{equation}
		\widetilde{\rho+3P} = \frac{(y+1)^5 (y+5) \;-\; 9a(\tilde{t})\,\tilde{\chi}\, y^5}{6\pi\, a(\tilde{t})^{3}\,(y-1)\,(y+1)^5} \geq 0 .
		\label{2008}
	\end{equation}
	holds.
	For the sake of computational simplicity, we define the function \(S(y)\) as follows:  
	\begin{equation}
		S(y):=\frac{(y+1)^5(y+5)}{9\,y^5}\, 
		\label{2009}
	\end{equation}  
		Determining the sign of Eq.~\eqref{2008} is equivalent to determining the sign of the expression \(S(y)- a(\tilde{t}) \tilde{\chi}\).  
	The function \(S\) is strictly decreasing in the interval \((1,y_\star)\) and strictly increasing in the interval \((y_\star,\infty)\), such that:  
	\begin{equation}
		y_\star=2+\sqrt{29}, \qquad 
		S_{\min}:=S(y_\star)=\frac{(3+\sqrt{29})^5(7+\sqrt{29})}{9(2+\sqrt{29})^5}.
		\label{2010}
	\end{equation}  
		In the following, we determine the sign of Eq.~\eqref{2008}.  

	\textbf{(a)} $1<{2a(\tilde{t})}/{\sqrt{\tilde{\chi}}}\le y_*$:
	
\textbf{(a-1)} $0<a(\tilde{t})\tilde{\chi}\le S({2a(\tilde{t})}/{\sqrt{\tilde{\chi}}})$: 
		for $y\in(1,{2a(\tilde{t})}/{\sqrt{\tilde{\chi}}})$ we always have $\widetilde{\rho+3P}>0$.
		
		\textbf{(a-2)} ${S(2a(\tilde{t})}/{\sqrt{\tilde{\chi}}})<a(\tilde{t})\tilde{\chi}<{64}/{3}$:
		there exists exactly one simple root $y_1^S$ with $\widetilde{\rho+3P}=0$; 
		on $y\in(1,y_1^S)$ we have $\widetilde{\rho+3P}>0$, and on 
		$y\in(y_1^S,{2a(\tilde{t})}/{\sqrt{\tilde{\chi}}})$ we have $\widetilde{\rho+3P}<0$.
		
	\textbf{(a-3)} $a(\tilde{t})\tilde{\chi}\ge{64}/{3}$:
		for $y\in(1,{2a(\tilde{t})}/{\sqrt{\tilde{\chi}}})$ we always have $\widetilde{\rho+3P}<0$.
	
	\textbf{(b):} ${2a(\tilde{t})}/{\sqrt{\tilde{\chi}}}>y_*$:
	
		\textbf{(b-1)} $0<a(\tilde{t})\tilde{\chi}\le S_{\min}$:
		for $y\in(1,{2a(\tilde{t})}/{\sqrt{\tilde{\chi}}})$ we always have $\widetilde{\rho+3P}>0$; 
		at $a(\tilde{t})\tilde{\chi}=S_{\min}$, a degenerate root occurs at $y=y_*$ with 
		$\widetilde{\rho+3P}|_{y=y_*}=0$.
		
		\textbf{(b-2)} $S_{\min}<a(\tilde{t})\tilde{\chi}\le \min\{{64}/{3},\,S({2a(\tilde{t})}/{\sqrt{\tilde{\chi}}})\}$:
		there are exactly two roots $y_2^S,y_3^S$ with $\widetilde{\rho+3P}=0$; 
		on $y\in(1,y_2^S)$: $\widetilde{\rho+3P}>0$, on $y\in(y_2^S,y_3^S)$: $\widetilde{\rho+3P}<0$, 
		and on $y\in(y_3^S,{2a(\tilde{t})}/{\sqrt{\tilde{\chi}}})$: $\widetilde{\rho+3P}>0$.
		
		\textbf{(b-3)} $\min\{{64}/{3},\,S({2a(\tilde{t})}/{\sqrt{\tilde{\chi}}})\}
		< a(\tilde{t})\tilde{\chi} \le 
		\max\{{64}/{3},\,S({2a(\tilde{t})}/{\sqrt{\tilde{\chi}}})\}$:
		exactly one root $y_4^S$ exists on $(1,{2a(\tilde{t})}/{\sqrt{\tilde{\chi}}})$.\\
		\hspace*{1em}If ${64}/{3}>S({2a(\tilde{t})}/{\sqrt{\tilde{\chi}}})$: 
		$\widetilde{\rho+3P}>0$ on $(1,y_4^S)$ and $\widetilde{\rho+3P}<0$ on 
		$(y_4^S,{2a(\tilde{t})}/{\sqrt{\tilde{\chi}}})$.\\
		\hspace*{1em}If ${64}/{3}=S({2a(\tilde{t})}/{\sqrt{\tilde{\chi}}})$:
		two boundary roots occur and effectively $\widetilde{\rho+3P}<0$ on the interval.\\
		\hspace*{1em}If $64/3<S({2a(\tilde{t})}/{\sqrt{\tilde{\chi}}})$:
		$\widetilde{\rho+3P}<0$ on $(1,y_4^S)$ and $\widetilde{\rho+3P}>0$ on 
		$(y_4^S,{2a(\tilde{t})}/{\sqrt{\tilde{\chi}}})$.
		
\textbf{(b-4)} $a(\tilde{t})\tilde{\chi}>\max\{{64}/{3},\,S({2a(\tilde{t})}/{\sqrt{\tilde{\chi}}})\}$:
		for every $y\in(1,{2a(\tilde{t})}/{\sqrt{\tilde{\chi}}})$ we always have $\widetilde{\rho+3P}<0$.
	
	Not all of the different cases obtained above will hold for a fixed $\tilde{\chi}>0$. 
	For a given $\tilde{\chi}>0$, all the intervals, mentioned above in determining the sign of Eq.~\eqref{2008},  must be examined. 
	It can be shown that:  
	
	For $0<\tilde{\chi}<(2/9)^{2/3}$, only cases (a-1), (b-1), and (b-2) will occur.  
	
	For $(2/9)^{2/3}\leq \tilde{\chi}<\left({2S_{\min}}/{y_*}\right)^{2/3}$, only case (a-1) and all cases (b) cases will occur.  
	
	For $\left({2S_{\min}}/{y_*}\right)^{2/3}\leq \tilde{\chi}<\left({128}/{3y_*}\right)^{2/3}$, 
	cases (a-1), (a-2), and (b-4) will occur, and case (b-3) will occur if $64/3>S\!\left({2a}/{\sqrt{\tilde{\chi}}}\right)$.  
	
	For $\left({128}/{3y_*}\right)^{2/3}\leq \tilde{\chi}<(128/3)^{2/3}$, all cases (a) and only case (b-4) will occur.  
	
	Finally, for $\tilde{\chi}\geq (128/3)^{2/3}$, cases (a-3) and (b-4) will occur.  
	
	Therefore, as $\tilde{\chi}$ increases, the cases in which $\widetilde{\rho+3P}$ is non-negative will become more restricted.  
	If Eqs.~\eqref{2005} and \eqref{2008} are satisfied, the SEC will hold for the curved matter dominated McVittie spacetime outside the curvature singularity. 
	As we observe, with increasing $\tilde{\chi}$ and time, the intervals where the SEC holds become more restricted. For example, for $\tilde{\chi}\geq (128/3)^{2/3}$, the SEC will not hold at any time for $y\in \left(1,{2a}/{\sqrt{\tilde{\chi}}}\right)$.
		
Now, we discuss about the MSH mass of curved McVittie spacetime. Assuming \( \gamma \delta \geq 0 \), the energy density of the matter distribution in the curved McVittie spacetime is positive, as shown by Eqs.~\eqref{new80} and \eqref{2003}. Therefore, according to Eq.~\eqref{new87}, the MSH mass is always positive for this spacetime. This avoids issues related to repulsive gravitational effects arising from a negative MSH mass \cite{faraoni PG}.

We will now compute the expansion of the null vector fields for this solution. Using Eq. \eqref{new91} and assuming \( \beta=0 \), we find that:
\begin{equation}
\theta_{\pm}=2[H \pm \frac{1}{\bar{r} B} (A - \frac{ \chi \bar{r}^2}{b^2-\chi \bar{r}^2})]
\label{new93}
\end{equation}
It appears that, depending on the given constants, \( \theta_{\pm}=0 \) can occur. To determine the location of the THs, we analyze the solutions of \( \theta_{\pm}=0 \) for a matter-dominated background with \( H={2}/{3t} \). For simplicity,using Eq.~\eqref{2002}, we rewrite Eq.~\eqref{new93} as:
\begin{equation}
\tilde{\theta}_{\pm}=2[\frac{2}{3 \tilde{t}} \pm \frac{1}{\tilde{\bar{r}} \hat{B}} (A - \frac{ \tilde{\chi} \tilde{\bar{r}}^2}{1-\tilde{\chi} \tilde{\bar{r}}^2})]
\label{new94}
\end{equation}
where $\hat{B} \equiv b B$.
It can be seen that $\theta_\pm|_{\theta_\mp}>0$. Consequently, in agreement with \ref{V.D} (after imposing the necessary conditions for the well-definedness of the family of future-directed ingoing and outgoing null geodesics), the curved McVittie spacetime admits PTH(s).

 In what follows, we provide a detailed analysis of the number and types of THs in the matter-dominated curved McVittie geometry for $\chi>0$, in the interval, $\tilde{\bar{r}} \in \left({1}/{(2a(\tilde{t}))}, {1}/{\sqrt{\tilde{\chi}}}\right)$.
\subsection*{Case 1: $\theta_+=0 , \; \theta_-|_{\theta_+=0}>0$}
Since the existence of roots of $\theta_+=0$ in the interval $\tilde{\bar{r}} \in \left({1}/({2a(\tilde{t})}), {1}/{\sqrt{\tilde{\chi}}}\right)$ is under consideration, the admissible temporal domain corresponds to
$\tilde{t}>\tilde{t}_{\min}=\left({\sqrt{\chi}}/{2}\right)^{3/2}$.
To investigate the existence of a root of $\theta_+=0$ within the resulting intervals, it is first necessary to determine the sign of $\theta_{+}(\tilde{t},\tilde{\bar{r}})$ at the boundaries $\tilde{\bar{r}}\to {1}/{(2a(\tilde{t}))}^{+}$ and $\tilde{\bar{r}}\to {1}/{\sqrt{\tilde{\chi}}}^{-}$ as a function of time. According to Eq.~\eqref{new94}, it can be shown that, for any time $\tilde{t}$,
\begin{equation}
\tilde{\theta}_{+}\Big|_{\tilde{\bar{r}}\to \tfrac{1}{\sqrt{\tilde{\chi}}}^{-}} \to -\infty, \qquad 
\frac{1}{2}\tilde{\theta}_{+}\Big|_{\tilde{\bar{r}}\to \tfrac{1}{2a(\tilde{t})}^{+}}
= \frac{2}{3\tilde{t}} - \frac{\tilde{\chi}}{4a(\tilde{t})\sqrt{4a(\tilde{t})^{2}-\tilde{\chi}}}.
\label{mw89}
\end{equation}
Therefore, for $\tilde{t}>\tilde{t}_{\min}$, the value of $\theta_+|_{{1}/{(2a(\tilde{t}))}^{+}}$ may be
 positive, negative, or vanish. In the case of positivity, one may expect at least one root
 $\tilde{\bar{r}}_{+}(\tilde{t})$.
To further examine the behavior of $\theta_+$ in the interval $\tilde{\bar{r}} \in \left({1}/{(2a(\tilde{t}))}, {1}/{\sqrt{\tilde{\chi}}}\right)$ with respect to time, we compute
\begin{equation}
\partial_{\tilde{\bar{r}}} \tilde{\theta}_+ 
= \frac{8a(\tilde{t}) (-4\tilde{\bar{r}}^{2} a(\tilde{t})^2 +4a(\tilde{t})\tilde{\bar{r}}\big(3\tilde{\chi}^2\tilde{\bar{r}}^{4}-6\tilde{\chi}\tilde{\bar{r}}^{2}+2\big)-1)}
{\left(1+{2a(\tilde{t})\tilde{\bar{r}}}\right)^{4}\left(1-\tilde{\chi}\tilde{\bar{r}}^{2}\right)^{3/2}}.
\label{mw90}
\end{equation}
By analyzing the sign of the above expression, we find that in the time domain
$4\left(1-{\sqrt{6}}/{3}\right)\leq{\tilde{\chi}}/{a(\tilde{t})^{2}}<4$,
for every $\tilde{\bar{r}} \in \left({1}/{2a(\tilde{t})}, {1}/{\sqrt{\tilde{\chi}}}\right)$, the inequality $\partial_{\tilde{\bar{r}}}\theta_+<0$ always holds.\footnote{When 
${\tilde{\chi}}/{(a(\tilde{t}))^{2}}=4\bigl(1-{\sqrt{6}}/{3}\bigr)$, 
	the condition $\partial_{\tilde{\bar{r}}}\theta_{+}=0$ occurs at $\tilde{\bar{r}}_{(+,D)}=1/2a(\tilde{t})$. 
	Therefore, within the interval $\tilde{\bar{r}}\in(1/2a(\tilde{t}),1/\sqrt{\tilde{\chi}})$, 
	$\partial_{\tilde{\bar{r}}}\theta_{+}$ remains negative.}

On the other hand, for
$0<{\tilde{\chi}}/{a(\tilde{t})^{2}}<4\left(1-{\sqrt{6}}/{3}\right)$,
one finds that for ${1}/{2a(\tilde{t})}<\tilde{\bar{r}}<\tilde{\bar{r}}_{(+,D)}(\tilde{t})$, $\partial_{\tilde{\bar{r}}}\theta_+>0$;
at $\tilde{\bar{r}}=\tilde{\bar{r}}_{(+,D)}(\tilde{t})$, $\partial_{\tilde{\bar{r}}}\theta_+=0$;
and for $\tilde{\bar{r}}_{(+,D)}(\tilde{t})<\tilde{\bar{r}}<{1}/{\sqrt{\tilde{\chi}}}$, $\partial_{\tilde{\bar{r}}}\theta_+<0$. 

Thus, $\partial_{\tilde{\bar{r}}}\theta_+$ admits only a single root at $\tilde{\bar{r}}_{(+,D)}(\tilde{t})$. Consequently, one expects that $\theta_+=0$ within $\tilde{\bar{r}} \in \left({1}/{2a(\tilde{t})} , {1}/{\sqrt{\tilde{\chi}}}\right)$ possesses at most two roots.

Here, we analyze the emergence time of such root, if it exists, as well as its number, based on the sign of $\theta_+$ determined through its radial derivative.

For simplicity, Eq.~\eqref{new94} can be rewritten as follows:
\begin{equation}	
	\frac{1}{2}\,\tilde{\theta}_{+} \;=\; \frac{2}{3 \tilde{t}} \;+\; F(\tilde{\bar{r}},\tilde{t}).
	\label{mw91}
\end{equation}
Based on Eq.~\eqref{mw89}, one has
\begin{equation}
	F\big|_{\tfrac{1}{2a(\tilde{t})}^+} < 0,
	\quad \text{and} \quad
	F\big|_{\tfrac{1}{\sqrt{\tilde{\chi}}}^-} \;\longrightarrow\; -\infty.
	\label{mw92}
\end{equation}
Therefore, it follows that
$F(\tilde{\bar{r}},\tilde{t})$ for $r \in \left({1}/{2a(\tilde{t})}, {1}/{\sqrt{\tilde{\chi}}}\right)
$,
is bounded from above. Consequently, one may uniquely define
\begin{equation}
	M(\tilde{t}) \;:=\; \sup_{\,\tilde{\bar{r}} \in \left(\tfrac{1}{2a(\tilde{t})}, \tfrac{1}{\sqrt{\tilde{\chi}}}\right)}
	F(\tilde{\bar{r}},\tilde{t}).
	\label{mw93}
\end{equation}
Taking into account the behavior of $\theta_{+}$ at the boundary points
$\tilde{\bar{r}} = 1/(2a(\tilde{t}))$ and $\tilde{\bar{r}} = 1/\sqrt{\tilde{\chi}}$,
together with the properties of $\partial_{\tilde{\bar{r}}}\theta_{+}$ in the interval
$\tilde{\bar{r}} \in (1/2a(\tilde{t}), 1/\sqrt{\tilde{\chi}})$, it follows that the existence of at least one root
$\tilde{\bar{r}}_{+}(\tilde{t})$ of the equation $\theta_{+}=0$ is possible if, for some
$\tilde{t}>\tilde{t}_{\min}$, the maximum value of $\theta_{+}$ within the interval is non-negative:
\begin{equation}
	\frac{2}{3\tilde{t}} + M(\tilde{t}) \;\geq\; 0.
	\label{mw94}
\end{equation}
Equivalently, the set
\begin{equation}
	S_{+} := \Big\{\, \tilde{t}>\tilde{t}_{\min} \;\big|\; M(\tilde{t}) \geq -\tfrac{2}{3\tilde{t}} \,\Big\}
	\label{mw95}
\end{equation}
guarantees the existence of at least one root of $\theta_{+}=0$.
It can be shown that $S_{+}$ is bounded below, and hence
$\tilde{t}_{(+)} := \inf S_{+}, \, \tilde{t}_{(+)} \geq \tilde{t}_{\min}$,
exists uniquely. This time $\tilde{t}_{(+)}$ corresponds to the earliest occurrence of a root of
$\theta_{+}=0$. Explicitly, at $\tilde{t}=\tilde{t}_{(+)}$ from Eq.~\eqref{mw94}, we expect
\begin{equation}
	M(\tilde{t}_{(+)}) \;=\; -\frac{2}{3 \tilde{t}_{(+)}}.
	\label{mw96}
\end{equation}
For $\tilde{t}<\tilde{t}_{(+)}$, no admissible root of $\theta_{+}=0$ exists in the interval
$\tilde{\bar{r}} \in (1/2a(\tilde{t}),\,1/\sqrt{\tilde{\chi}})$, whereas for $\tilde{t}>\tilde{t}_{(+)}$ one may obtain either one or two roots.
Since for any fixed $\tilde{\chi}>0$, the value of $\tilde{t}_{(+)}$ is unique, if two distinct roots of
$\theta_{+}=0$ exist for some $\tilde{t}>\tilde{t}_{(+)}$, they must both have originated at the same
$\tilde{t}_{(+)}$ corresponding to that particular $\tilde{\chi}$.
Moreover, since
$	\tilde{t}_{(+)} \;\geq\; \tilde{t}_{\min} \;=\; \left({\sqrt{\tilde{\chi}}}/{2}\right)^{3/2}$,
it follows that $\tilde{t}_{(+)}$ is an increasing function of $\tilde{\chi}$.
Thus, the number of roots for $\tilde{t} \geq \tilde{t}_{(+)}$ is also a function of $\tilde{\chi}$.
In order to determine the number of such roots, it suffices to analyze the sign of
$\theta_{+}\big|_{{1}/{2a(\tilde{t})}^+}$ as a function of $\tilde{t}$.
It can be shown that from Eq.~\eqref{mw89}
\begin{equation}
	\theta_{+}\Big|_{\tfrac{1}{2a(\tilde{t})}^+}
	\begin{cases}
		> 0, & \tilde{t}>\tilde{t}_{\mathrm{le}}, \\[6pt]
		= 0, & \tilde{t}=\tilde{t}_{\mathrm{le}}, \\[6pt]
		< 0, & \tilde{t}<\tilde{t}_{\mathrm{le}},
		\label{Y}
	\end{cases}
\end{equation}
where $\tilde{t}_{\mathrm{le}}$ is determined from the relation
\begin{equation}
	a_{\mathrm{le}} = \tilde{t}_{\mathrm{le}}^{2/3}
	= \frac{9 \tilde{\chi}^{2} + \sqrt{81\tilde{\chi}^{4} + 65536 \tilde{\chi}}}{512}.
	\label{MW98}
\end{equation}
At $\tilde{t} = \tilde{t}_{le} > \tilde{t}_{min}$, one obtains 
$\theta_+\big|_{{1}/{2a(\tilde{t})}^+} = 0$.
This implies that $\tilde{\bar{r}} = \left({1}/{2a(\tilde{t})}\right)^+$ corresponds to one of the roots at $\tilde{t}_{le}$. Therefore, a more detailed investigation of this time $\tilde{t}_{le}$ is required.  
Using Eq.~\eqref{MW98}, it can be shown that
\begin{equation}
\begin{cases}
	0 < \tilde{\chi} < \tilde{\chi}_{\text{critical}} 
	\iff  
	\dfrac{\tilde{\chi}}{a_{le}^2} \in \left( 4\!\left(1-\tfrac{\sqrt{6}}{3}\right),\, 4 \right), \\[1em]
	\tilde{\chi} \geq \tilde{\chi}_{\text{critical}} 
	\iff  
	\dfrac{\tilde{\chi}}{a_{le}^2} \in \left( 0,\, 4\!\left(1-\tfrac{\sqrt{6}}{3}\right)\right],
\end{cases}
\end{equation}
where
\begin{equation}
\dfrac{\tilde{\chi}_{\text{critical}}}{a_{le}^2} = 4\!\left(1 - \tfrac{\sqrt{6}}{3}\right) 
\quad \Longleftrightarrow \quad \tilde{\chi}_{\text{critical}} \approx 9.02.
\end{equation}
Next, we examine the behavior of $\theta_+$ in the interval  
$\tilde{\bar{r}} \in \left( {1}/{2a(\tilde{t})}, {1}/{\sqrt{\tilde{\chi}}} \right), 
\, \tilde{t} > \tilde{t}_{min}$,
based on the partitioning with respect to $\tilde{\chi} > 0$.

Consider the case $0<\tilde{\chi}<\tilde{\chi}_\mathrm{critical}$.
In the regime $4\left(1-{\sqrt{6}}/{3}\right)<{\tilde{\chi}}/{a_{le}^{2}}<{\tilde{\chi}}/{a(\tilde{t})^{2}}<4$, from Eq.~\eqref{mw89}, 
we find that $\theta_{+}|_{\tilde{\bar{r}}=(1/2a(\tilde{t}))^{+}}<0$ while $\theta_{+}|_{\tilde{\bar{r}}=1/\sqrt{\tilde{\chi}}}\to -\infty$. 
Since ${\tilde{\chi}}/{a(\tilde{t})^{2}}>4\left(1-{\sqrt{6}}/{3}\right)$, from Eq.~\eqref{mw90}, we have 
$\partial_{\tilde{\bar{r}}}\theta_{+}<0$. Consequently, for every 
$\tilde{\bar{r}}\in(1/2a(\tilde{t}),1/\sqrt{\tilde{\chi}})$, the function $\theta_{+}$ remains negative. Equivalently 
$M(\tilde{t})<-{2}/{3\tilde{t}}$, and therefore no solution of the form $\theta_{+}=0$ exists. 

At $\tilde{t}=\tilde{t}_{le}$, however, from Eqs.~\eqref{mw89} and \eqref{Y}, one has $\theta_{+}|_{(1/2a(\tilde{t}))^{+}}=0$ and 
$\theta_{+}|_{1/\sqrt{\tilde{\chi}}}\to -\infty$. Since $4\left(1-{\sqrt{6}}/{3}\right)<{\tilde{\chi}}/{a_{le}^{2}}<4$, 
it again follows that from Eq.~\eqref{mw90}, $\partial_{\tilde{\bar{r}}}\theta_{+}<0$, so that $\theta_{+}(\tilde{\bar{r}},\tilde{t}_{le})\leq 0$ for 
$\tilde{\bar{r}}\in(1/2a(\tilde{t}),1/\sqrt{\tilde{\chi}})$. Thus, at $\tilde{t}=\tilde{t}_{le}$, the maximum of 
$\theta_{+}(\tilde{\bar{r}},\tilde{t}_{le})$ is zero, and the relation ${2}/{3\tilde{t}_{le}}+M(\tilde{t}_{le})=0$ 
holds at $\tilde{\bar{r}}_{+}\approx(1/2a(\tilde{t}))^{+}$, which corresponds to the appearance of the first simple root of $\theta_{+}=0$ 
(simple because $\partial_{\tilde{\bar{r}}}\theta_{+}|_{\tilde{\bar{r}}=(1/2a(\tilde{t}))^{+},\,\tilde{t}=\tilde{t}_{le}}<0$). 

In the range $4\left(1-{\sqrt{6}}/{3}\right)<{\tilde{\chi}}/{a(\tilde{t})^{2}}<{\tilde{\chi}}/{a_{le}^{2}}<4$, using Eq.~\eqref{mw89}, 
we obtain $\theta_{+}|_{(1/2a(\tilde{t}))^{+}}>0$ while $\theta_{+}|_{1/\sqrt{\tilde{\chi}}}\to -\infty$. This shows that 
the root $\tilde{\bar{r}}_{+}(\tilde{t})$ drifts away from $\tilde{\bar{r}}=1/(2a(\tilde{t}))$ as time evolves. Likewise, in the domain 
$0<{\tilde{\chi}}/{a(\tilde{t})^{2}}\leq 4\left(1-{\sqrt{6}}/{3}\right)<{\tilde{\chi}}/{a_{le}^{2}}<4$, one still finds 
$\theta_{+}|_{(1/2a(\tilde{t}))^{+}}>0$ and $\theta_{+}|_{1/\sqrt{\tilde{\chi}}}\to -\infty$. Here, besides the simple root 
$\tilde{\bar{r}}_{+}(\tilde{t})$, based on Eq.~\eqref{mw90}, a local maximum of $\theta_{+}$ appears  at $\tilde{\bar{r}}_{(+,D)}(\tilde{t})$ with 
$1/2a(\tilde{t})<\tilde{\bar{r}}_{(+,D)}(\tilde{t})<\tilde{\bar{r}}_{+}(\tilde{t})$. 

In summary, if $0<\tilde{\chi}<\tilde{\chi}_{\mathrm{critical}}$, then within 
$\tilde{\bar{r}}\in(1/2a(\tilde{t}),1/\sqrt{\tilde{\chi}})$, $\theta_{+}=0$ admits a simple root. This root corresponds to a PTH associated with the matter-dominated curved McVittie spacetime. This spacetime emerges emerges at $\tilde{t}_{(+)}\approx\tilde{t}_{le}$ with 
$\tilde{\bar{r}}_{+}(\tilde{t}_{(+)})\approx 1/(2a_{le}))$. 

Now, consider the case $\tilde{\chi}\geq \tilde{\chi}_{\mathrm{critical}}$. 
For these values of $\tilde{\chi}$, the inequality 
$0<{\tilde{\chi}}/{a_{le}^{2}} \leq 4\bigl(1-{\sqrt{6}}/{3}\bigr)$ holds.

In the regime 
$0<{\tilde{\chi}}/{a_{le}^{2}} \leq 4\bigl(1-{\sqrt{6}}/{3}\bigr)<{\tilde{\chi}}/{a(\tilde{t})^{2}}<4$, from Eqs.~\eqref{mw89} and \eqref{mw90}, 
one finds that $\theta_{+}|_{(1/2a(\tilde{t}))^{+}}<0$, 
$\theta_{+}|_{(1/\sqrt{\tilde{\chi}})^{-}}\to -\infty$, and 
$\partial_{\tilde{\bar{r}}}\theta_{+}<0$. 
Consequently, for all $\tilde{\bar{r}}\in(1/2a(\tilde{t}),1/\sqrt{\tilde{\chi}})$, the function $\theta_{+}$ remains strictly negative, 
and has no roots within this interval.  

In contrast, in the interval 
$0<{\tilde{\chi}}/{a_{le}^{2}}<{\tilde{\chi}}/{a(\tilde{t})^{2}}<4\bigl(1-{\sqrt{6}}/{3}\bigr)<4$,using Eqs.~\eqref{mw89} and \eqref{Y},
we still have $\theta_{+}|_{(1/2a(\tilde{t}))^{+}}<0$ and 
$\theta_{+}|_{(1/\sqrt{\tilde{\chi}})^{-}}\to -\infty$. 
However, based on Eq.~\eqref{mw90}, in this regime the sign of the derivative changes: 
for $1/2a(\tilde{t})<\tilde{\bar{r}}\leq \tilde{\bar{r}}_{(+,D)}$ one finds $\partial_{\tilde{\bar{r}}}\theta_{+}\geq 0$, 
while for $\tilde{\bar{r}}_{(+,D)}<\tilde{\bar{r}}<1/\sqrt{\tilde{\chi}}$ one has $\partial_{\tilde{\bar{r}}}\theta_{+}<0$. 
The following theorem establishes the existence of a degenerate root of $\theta_{+}=0$ in this situation. 
 
\textbf{Theorem 1:}
	For $\tilde{\chi}\geq \tilde{\chi}_{\mathrm{critical}}$, the condition $\theta_{+}=0$ admits a degenerate root at $\tilde{t}_{(+)}$, when
	$0<{\tilde{\chi}}/{a_{le}^{2}} \leq {\tilde{\chi}}/{a_{(+)}^{2}} < 4\left(1-{\sqrt{6}}/{3}\right)<4$,
	and this root lies within $\tilde{\bar{r}}\in(1/2a(\tilde{t}),1/\sqrt{\tilde{\chi}})$.
	\footnote{At $\tilde{\chi}=\tilde{\chi}_{\mathrm{critical}}$, using Eqs.~\eqref{mw89}, \eqref{mw90} and \eqref{Y}, a degenerate root occurs at 
		$\tilde{t}_{(+)}=\tilde{t}_{le}\approx 6.5656$. Consequently, 
		$\tilde{\bar{r}}_{+}(\tilde{t}_{(+)})=(1/2a_{le})^{+}$. 
		Therefore, for $\tilde{t}>\tilde{t}_{le}=\tilde{t}_{(+)}$ 
		(i.e. ${\tilde{\chi}}/{a_{le}^{2}}={\tilde{\chi}}/{a_{(+)}^{2}}
		> {\tilde{\chi}}/{a(\tilde{t})^{2}}$, we have 
		$\theta_{+}|_{(1/2a(\tilde{t}))^{+}}>0$, and thus only a simple root 
		corresponding to $\theta_{+}=0$ remains.}
		
\textbf{Proof 1:} 
In the range 
$0<{\tilde{\chi}}/{a_{le}^{2}}<{\tilde{\chi}}/{a(\tilde{t})^{2}}<4\bigl(1-{\sqrt{6}}/{3}\bigr)<4$, 
we have already shown that from Eqs.~\eqref{mw89} and \eqref{mw90}, $\theta_{+}|_{(1/2a(\tilde{t}))^{+}}$ and $\theta_{+}|_{(1/\sqrt{\tilde{\chi}})^{-}}$ are both negative, 
whereas $\partial_{\tilde{\bar{r}}}\theta_{+}$ can be positive, negative, or zero.  
Since
\begin{equation}
\frac{1}{2}\tilde{\theta}_{+}|_{\tilde{\bar{r}}=(1/2a(\tilde{t}))^{+}, \, \tilde{t}=\tilde{t}_{(+)}} 
= \frac{2}{3\tilde{t}_{(+)}} + F(1/2a(\tilde{t}),\tilde{t}_{(+)}) ,
\end{equation}
and at $\tilde{t}=\tilde{t}_{(+)}$ we also have 
$M(\tilde{t}_{(+)})=-{2}/{3\tilde{t}_{(+)}}$, 
and,
\begin{equation}
M(\tilde{t}_{(+)}) \;\geq\; F(\tilde{\bar{r}},\tilde{t}_{(+)})
\quad\text{for }\quad \tilde{\bar{r}}\in(1/2a(\tilde{t}),1/\sqrt{\tilde{\chi}}).
\end{equation}
Therefore,
\begin{equation}
\frac{1}{2}\tilde{\theta}_{+}|_{\tilde{\bar{r}}=(1/2a(\tilde{t}))^{+},\, \tilde{t}=\tilde{t}_{(+)}} 
= -M(\tilde{t}_{(+)}) + F(1/2a(\tilde{t}),\tilde{t}_{(+)}) \leq 0 ,
\end{equation}
and consequently $\tilde{t}_{(+)}\leq \tilde{t}_{le}$. 
Equality holds when $\tilde{\chi}=\tilde{\chi}_{\mathrm{critical}}$, 
while for $\tilde{\chi}>\tilde{\chi}_{\mathrm{critical}}$ the inequality is strict.  
Since at $\tilde{t}_{(+)}\leq \tilde{t}_{le}$ the function $\theta_{+}$ attains a local maximum at 
$\tilde{\bar{r}}_{(+,D)}(\tilde{t}_{(+)})$, with
\begin{equation}
\partial_{\tilde{\bar{r}}}\theta_{+}|_{\tilde{\bar{r}}=\tilde{\bar{r}}_{(+,D)}(\tilde{t}_{(+)})}=0 ,
\qquad
\max_{\tilde{\bar{r}}=\tilde{\bar{r}}_{(+,D)}(\tilde{t}_+)}\theta_{+}=0 ,
\end{equation}
it follows that $\tilde{\bar{r}}_{(+,D)}(\tilde{t}_{(+)})$ constitutes a degenerate root of $\theta_{+}=0$ which forms at $\tilde{t}_{(+)}$, when
${\tilde{\chi}}/{a_{(+)}^{2}}\in\big(0,\,4(1-{\sqrt{6}}/{3})\big]$.

In the interval , 
$0 < {\tilde{\chi}}/{a_{le}^2} < {\tilde{\chi}}/{a(\tilde{t})^2} < {\tilde{\chi}}/{a_{(+)}^2} < 4\left(1-{\sqrt{6}}/{3}\right) < 4$,
we have
$M(\tilde{t}) > -{2}/{3\tilde{t}} \, \text{or equivalently} \, \max \theta_{+} > 0$,
while simultaneously $\theta_{+}|_{(1/2a(\tilde{t}))^{+}} < 0, 
\, \theta_{+}|_{1/\sqrt{\tilde{\chi}}^{-}} \to -\infty$.
Consequently, in the range 
$\tilde{\bar{r}} \in \left({1}/{2a(\tilde{t})}, {1}/{\sqrt{\tilde{\chi}}}\right)$,
the equation $\theta_{+}=0$ admits two distinct roots 
$\tilde{\bar{r}}^{u}_{+}(\tilde{t})$ and $\tilde{\bar{r}}^{l}_{+}(\tilde{t})$.  
At $\tilde{t}=\tilde{t}_{(+)}$, these coincide with the degenerate root 
$\tilde{\bar{r}}_{(+,D)} = \tilde{\bar{r}}^{u}_{+} = \tilde{\bar{r}}^{l}_{+}$,
and satisfy 
$\tilde{\bar{r}}^{l}_{+}(\tilde{t}) < \tilde{\bar{r}}_{(+,D)} (\tilde{t}) < \tilde{\bar{r}}^{u}_{+}(\tilde{t})$
throughout the above time interval.

As $\tilde{t} \to \tilde{t}_{le}$, the separation between the two roots,
$\big(\tilde{\bar{r}}^{u}_{+} - \tilde{\bar{r}}^{l}_{+}\big)(\tilde{t})$,
increases, while $\tilde{\bar{r}}^{l}_{+} \to (1/2a(\tilde{t}))^{+}$.  
At $\tilde{t}=\tilde{t}_{le}$ we then have 
$\theta_{+}|_{(1/2a(\tilde{t}))^{+}} = 0$.
For $\tilde{t}>\tilde{t}_{le}$ (equivalently, 
${\tilde{\chi}}/{a_{le}^2} > {\tilde{\chi}}/{a(\tilde{t})^2}$), it follows that
$\theta_{+}\big|_{(1/2a(\tilde{t}))^{+}} > 0$,
and thus no lower root $\tilde{\bar{r}}^{l}_{+}$ exists in the interval 
$\tilde{\bar{r}} \in (1/2a(\tilde{t}) , 1/\sqrt{\tilde{\chi}})$.  
However, the upper root $\tilde{\bar{r}}^{u}_{+}$ remains as a simple root of $\theta_{+}=0$.

In summary: for $\tilde{\chi} \geq \tilde{\chi}_{\text{critical}}$, there is a degenerate root at $\tilde{\bar{r}}_{(+,D)}(\tilde{t}_{(+)})$ where $\theta_{+}=0$, occurring in the interval
${\tilde{\chi}}/{a(\tilde{t})^2} \in \Big(0,\; 4\left(1-{\sqrt{6}}/{3}\right)\Big]$.
At this time, a PITH associated with the matter-dominated curved McVittie spacetime begins to form.\footnote{Before the formation time of the degenerate root $\tilde{t}_{(+)}$, we have
	$M(\tilde{t})<-{2}/{3\tilde{t}}$, whereas for $\tilde{t}>\tilde{t}_{(+)}$,
	$M(\tilde{t})>-{2}/{3\tilde{t}}$. As a result, 
	$\partial_{\tilde{t}}\theta_{+}\big|_{\tilde{\bar{r}}=\tilde{\bar{r}}_{(+,D)}(\tilde{t}_{(+)})}>0$. 
	Therefore, at the point of formation of the degenerate root, we obtain
	$\mathcal{L}_{-}\theta_{+}\big|_{\theta_{+}(\tilde{\bar{r}}_{(+,D)}(\tilde{t}_{(+)}))=0} 
	= \left({1}/{A}\,\partial_{\tilde{t}}\theta_{+}\right)\big|_{\tilde{\bar{r}}=\tilde{\bar{r}}_{(+,D)}(\tilde{t}_{(+)})}>0 $,
	which corresponds to a PITH.} 

- If $\tilde{\chi} = \tilde{\chi}_{\text{critical}}$, then for $\tilde{t}>\tilde{t}_{(+)}=\tilde{t}_{le}$ (with 
$\theta_{+}|_{(1/2a_{le})^{+}}=0$), there exists a single simple root of $\theta_{+}=0$, corresponding to a PITH.  

- If $\tilde{\chi} > \tilde{\chi}_{\text{critical}}$, then in the interval 
$\tilde{t}_{(+)}<\tilde{t}<\tilde{t}_{le}$ there are two distinct roots 
$\tilde{\bar{r}}^{u}_{+}$ and $\tilde{\bar{r}}^{l}_{+}$, corresponding to two PTHs.  
As time progresses, the separation between these horizons increases until, at 
$\tilde{t}=\tilde{t}_{le}$, the lower root $\tilde{\bar{r}}^{l}_{+}$ (the smaller PTH) leaves the range 
$\tilde{\bar{r}}\in (1/2a(\tilde{t}), 1/\sqrt{\tilde{\chi}})$.  
For $\tilde{t}>\tilde{t}_{le}$, only a single simple root remains, corresponding to one PTH.
\subsection*{Case 2: $\theta_-=0 , \; \theta_+|_{\theta_-=0}>0$}
In this section, we investigate the root(s) of $\theta_{-}=0$ in the interval 
$r \in (1/2a(\tilde{t}),\,1/\sqrt{\tilde{\chi}})$, which corresponds to the time domain 
$\tilde{t}>\tilde{t}_{\min}=({\sqrt{\tilde{\chi}}}/{2})^{3/2}$. 
We begin by determining the sign of $\theta_{-}(\tilde{\bar{r}},\tilde{t})$ at the boundaries 
$\tilde{\bar{r}}\to (1/2a(\tilde{t}))^{+}$ and $\tilde{\bar{r}}\to (1/\sqrt{\tilde{\chi}})^{-}$.  
According to Eq.~\eqref{new94}, we obtain
\begin{equation}
\tilde{\theta}_{-}\Big|_{\tilde{\bar{r}}\to \frac{1}{\sqrt{\tilde{\chi}}}^{-}} \;\longrightarrow\; +\infty, 
\qquad 
\tilde{\theta}_{-}\Big|_{\tilde{\bar{r}}\to \tfrac{1}{2a(\tilde{t})}^{+}} = \frac{2}{3\tilde{t}}+\frac{\tilde{\chi}}{4a(\tilde{t})\sqrt{4a(\tilde{t})^{2}-\tilde{\chi}}} > 0 .
	\label{nw105}
\end{equation}
Therefore, for every $\tilde{t}>\tilde{t}_{\min}$, the value of $\theta_{-}$ at the boundaries 
$(1/2a(\tilde{t}))^{+}$ and $(1/\sqrt{\tilde{\chi}})^{-}$ (for $\tilde{\chi}>0$) will be positive. 
From Eq.~\eqref{new94}, the derivative $\partial_{\tilde{\bar{r}}}\tilde{\theta}_{-}$ is obtained as  
\begin{equation}
	\partial_{\tilde{\bar{r}}} \tilde{\theta}_{-}
	= \frac{8a(\tilde{t}) (4\tilde{\bar{r}}^{2} a(\tilde{t})^2 - 4a(\tilde{t})\tilde{\bar{r}}\big(3\tilde{\chi}^{2}\tilde{\bar{r}}^{4}-6\tilde{\chi}\tilde{\bar{r}}^{2}+2\big)+1)}
	{\left(1+2a(\tilde{t})\tilde{\bar{r}}\right)^{4}\left(1-\tilde{\chi}\tilde{\bar{r}}^{2}\right)^{3/2}} .
	\label{nw106}
\end{equation}
By determining the sign of Eq.~\eqref{nw106}, one can see that in the time interval 
$4\left(1-{\sqrt{6}}/{3}\right)\leq{\tilde{\chi}}/{a(\tilde{t})^{2}}<4$, 
for every $\tilde{\bar{r}}\in\bigl(1/2a(\tilde{t}),\,1/\sqrt{\tilde{\chi}}\bigr)$, 
we always have $\partial_{\tilde{\bar{r}}}\theta_{-}>0$. 
(At ${\tilde{\chi}}/{a(\tilde{t})^{2}}=4(1-{\sqrt{6}}/{3})$, 
$\partial_{\tilde{\bar{r}}}\theta_{-}=0$ occurs at $\tilde{\bar{r}}_{(-,D)}=1/2a(\tilde{t})$, 
and therefore within $\tilde{\bar{r}}\in(1/2a(\tilde{t}),1/\sqrt{\tilde{\chi}})$ the derivative remains positive.)  

On the other hand, for $0<{\tilde{\chi}}/{a(\tilde{t})^{2}}<4(1-{\sqrt{6}}/{3})$, 
one finds that for ${1}/{2a(\tilde{t})}<\tilde{\bar{r}}<\tilde{\bar{r}}_{(-,D)}(\tilde{t})$, 
$\partial_{\tilde{\bar{r}}}\theta_{-}<0$; at 
$\tilde{\bar{r}}=\tilde{\bar{r}}_{(-,D)}(\tilde{t})$, $\partial_{\tilde{\bar{r}}}\theta_{-}=0$; 
and for $\tilde{\bar{r}}_{(-,D)}(\tilde{t})<\tilde{\bar{r}}<1/\sqrt{\tilde{\chi}}$, 
$\partial_{\tilde{\bar{r}}}\theta_{-}>0$.  

Thus, $\partial_{\tilde{\bar{r}}}\theta_{-}$ admits only a single root at 
$\tilde{\bar{r}}_{(-,D)}(\tilde{t})$. Consequently, one expects that $\theta_{-}=0$ within 
$\tilde{\bar{r}}\in\bigl({1}/{2a(\tilde{t})},\,1/\sqrt{\tilde{\chi}}\bigr)$ possesses at most two roots.

By rewriting Eq.~\eqref{new94} in the form 
\begin{equation}
	\frac{1}{2}\,\tilde{\theta}_{-} = \frac{2}{3 \tilde{t}} - F(\tilde{\bar{r}}, \tilde{t}) ,
	\label{nw107}
\end{equation}
and as was shown in Eq.~\eqref{mw92}, one has $F|_{({1}/{2a(\tilde{t})})^{+}}<0$, and 
$F|_{{1}/{\sqrt{\tilde{\chi}}}^{-}} \to -\infty$. Therefore, $F(\tilde{\bar{r}},\tilde{t})$ 
is bounded above and admits a supremum in the interval 
$\tilde{\bar{r}}\in(1/2a(\tilde{t}),\,1/\sqrt{\tilde{\chi}})$ (see Eq.~\eqref{mw93}). 
Accordingly, $-F$ admits an infimum in the same interval :
\begin{equation}
	-M(\tilde{t}) = \inf_{\tilde{\bar{r}}\in(1/2a(\tilde{t}),\,1/\sqrt{\tilde{\chi}})} 
	\big(-F(\tilde{\bar{r}},\tilde{t})\big) .
	\label{nw108}
\end{equation}
Taking into account the positivity of $\theta_{-}$ at both ends of the interval 
$(1/2a(\tilde{t}),\,1/\sqrt{\tilde{\chi}})$, from Eq.~\eqref{nw105}, and the sign of $\partial_{\tilde{\bar{r}}}\theta_{-}$ in Eq.~\eqref{nw106}, it is possible that there is at least one root
 $\tilde{\bar{r}}_{-}(\tilde{t})$ of $\theta_{-}=0$ in 
$\tilde{\bar{r}}\in(1/2a(\tilde{t}),\,1/\sqrt{\tilde{\chi}})$, provided that the minimum of $\theta_{-}$ is nonpositive for 
$\tilde{t}>\tilde{t}_{\min}$ . This condition can be expressed as
\begin{equation}
	\frac{2}{3\tilde{t}} - M(\tilde{t}) \leq 0 .
	\label{nw109}
\end{equation}
Equivalently, the set
\begin{equation}
	S_{-} := \Big\{\, \tilde{t}>\tilde{t}_{\min} \;\big|\; M(\tilde{t}) \geq \tfrac{2}{3\tilde{t}} \,\Big\}
	\label{nw110}
\end{equation}
guarantees the existence of at least one root of $\theta_{-}=0$. 
It can be shown that $S_{-}$ is bounded below, and hence
$\tilde{t}_{(-)} := \inf S_{-}$ exists uniquely and $\tilde{t}_{(-)} \geq \tilde{t}_{\min}$.
 This time $\tilde{t}_{(-)}$ corresponds to the earliest occurrence of a root of 
$\theta_{-}=0$. Explicitly, at $\tilde{t}=\tilde{t}_{(-)}$, we expect
\begin{equation}
	M(\tilde{t}_{(-)}) = \frac{2}{3\tilde{t}_{(-)}} .
	\label{nw111}
\end{equation}
For $\tilde{t}<\tilde{t}_{(-)}$, no admissible root of $\theta_{-}=0$ exists in the interval 
$\tilde{\bar{r}}\in(1/2a(\tilde{t}),\,1/\sqrt{\tilde{\chi}})$, whereas for $\tilde{t}>\tilde{t}_{(-)}$ 
one may obtain either one or two roots.

As in  the case of $\theta_{+}=0$, since $\tilde{t}_{(-)}\geq \tilde{t}_{\min}$, 
the function $\tilde{t}_{(-)}$ is increasing with respect to $\tilde{\chi}$. 
Therefore, for every $\tilde{\chi}>0$, we expect  
$\tilde{t}_{(-)}(\tilde{\chi}>0)>\tilde{t}_{(-)}(\tilde{\chi}=0)=2\sqrt{3}$, 
as seen previously in  Sec.~\ref{II.A} for the matter-dominated flat McVittie spacetime.

In the time interval $4\bigl(1-{\sqrt{6}}/{3}\bigr)\leq{\tilde{\chi}}/{a(\tilde{t})^{2}}<4$, using Eq.~\eqref{nw105},
we have shown that $\theta_{-}|_{(1/2a(\tilde{t}))^{+}}>0$ and 
$\theta_{-}|_{(1/\sqrt{\tilde{\chi}})^{-}}\to +\infty$. 
In this regime, since $\partial_{\tilde{\bar{r}}}\theta_{-}>0$ for 
$\tilde{\bar{r}}\in(1/2a(\tilde{t}),\,1/\sqrt{\tilde{\chi}})$ as seen in Eq.~\eqref{nw106}, it follows that 
$\theta_{-}(\tilde{\bar{r}},\tilde{t})$ is strictly increasing. Thus, $\theta_{-}$ does not satisfy 
$\theta_{-}=0$ for any $\tilde{\bar{r}}\in(1/2a(\tilde{t}),\,1/\sqrt{\tilde{\chi}})$.

In the interval $0<{\tilde{\chi}}/{a(\tilde{t})^{2}}<4\bigl(1-{\sqrt{6}}/{3}\bigr)$, using Eq.~\eqref{nw105},
both $\theta_{-}|_{(1/2a(\tilde{t}))^{+}}$ and $\theta_{-}|_{(1/\sqrt{\tilde{\chi}})^{-}}$ are positive.~In this case, however, using Eq.~\eqref{nw106},  $\partial_{\tilde{\bar{r}}}\theta_{-}$ changes sign: 
for $1/2a(\tilde{t})<\tilde{\bar{r}}\leq\tilde{\bar{r}}_{(-,D)}$ we have $\partial_{\tilde{\bar{r}}}\theta_{-}\leq 0$, 
whereas for $\tilde{\bar{r}}_{(-,D)}<\tilde{\bar{r}}<1/\sqrt{\tilde{\chi}}$ 
we have $\partial_{\tilde{\bar{r}}}\theta_{-}>0$.

For $\tilde{t}<\tilde{t}_{(-)}$ (equivalently, 
$0<{\tilde{\chi}}/{a_{(-)}^{2}}<{\tilde{\chi}}/{a(\tilde{t})^{2}}<4\bigl(1-{\sqrt{6}}/{3}\bigr)$), 
one finds $M(\tilde{t})<{2}/{3\tilde{t}}$, or equivalently
$\min_{\tilde{\bar{r}}=\tilde{\bar{r}}_{(-,D)}(\tilde{t})}\theta_{-}>0 $.

Finally, at $\tilde{t}=\tilde{t}_{(-)}$, where 
$0<{\tilde{\chi}}/{a_{(-)}^{2}}<4\bigl(1-{\sqrt{6}}/{3}\bigr)$, 
we obtain $M(\tilde{t}_{(-)})={2}/{3\tilde{t}_{(-)}}$. 
At this moment, one has
\begin{equation}
	\partial_{\tilde{\bar{r}}}\theta_{-}\Big|_{\tilde{\bar{r}}=\tilde{\bar{r}}_{(-,D)}(\tilde{t}_{(-)})}=0, 
	\qquad 
	\min_{\tilde{\bar{r}}=\tilde{\bar{r}}_{(-,D)}(\tilde{t}_{(-)})}\theta_{-}=0 ,
	\label{nw112}
\end{equation}
so that $\tilde{\bar{r}}_{(-,D)}(\tilde{t}_{(-)})$ is a degenerate root of $\theta_{-}=0$.

For times $\tilde{t}>\tilde{t}_{(-)}$, one has 
$M(\tilde{t})>{2}/{3\tilde{t}}$, and consequently
$\min_{\tilde{\bar{r}}=\tilde{\bar{r}}_{(-,D)}(\tilde{t})}\theta_{-}<0 $.
Since $\theta_{-}$ remains positive at the boundaries of the interval 
$(1/2a(\tilde{t}),\,1/\sqrt{\tilde{\chi}})$, there exist two distinct roots 
$\tilde{\bar{r}}_{-}^{\,l}$ and $\tilde{\bar{r}}_{-}^{\,u}$ of $\theta_{-}=0$. 
Both roots emerge from $\tilde{\bar{r}}=\tilde{\bar{r}}_{(-,D)}(\tilde{t}_{(-)})$ 
at $\tilde{t}=\tilde{t}_{(-)}$, and then separate from each other as $\tilde{t}$ increases. 
In fact, the separation $(\tilde{\bar{r}}_{-}^{\,u}-\tilde{\bar{r}}_{-}^{\,l})(\tilde{t})$ 
grows monotonically for $\tilde{t}>\tilde{t}_{(-)}$.  
Because $\theta_{-}$ remains positive at the boundaries of the interval 
$(1/2a(\tilde{t}),\,1/\sqrt{\tilde{\chi}})$, the two roots $\tilde{\bar{r}}_{-}^{\,u}$ and 
$\tilde{\bar{r}}_{-}^{\,l}$ continue to exist in this domain, as time evolves,.  

To investigate the behavior of the two roots of $\theta_{-}=0$, namely $\tilde{\bar{r}}_{-}^{\,u}$ and $\tilde{\bar{r}}_{-}^{\,l}$, at sufficiently large times, we expand $\theta_{-}=0$ in powers of $1/\sqrt{a(\tilde{t})}=t^{-1/3}$\footnote{ This is true aslong as, the two null vector fields 
defined in Eq.~\eqref{new89} remain well-defined for the corresponding families of outgoing/ingoing 
null geodesics in the matter-dominated curved McVittie spacetime.}. 
This yields
\begin{equation}
\tilde{\bar{r}}_{-}^{\,l} \;\approx\; (\tfrac{1}{2a(\tilde{t})})^{+}, 
\qquad 
\tilde{\bar{r}}_{-}^{\,u} \;\approx\; \frac{1}{\sqrt{2\tilde{\chi}}} 
- \frac{1}{6\sqrt{2}\,\tilde{\chi}}\,\frac{1}{\sqrt{a(\tilde{t})}} .
\label{nw113}
\end{equation}
Hence,  we expect the larger root (i.e. $\tilde{\bar{r}}_{-}^{\,u}$) 
to asymptotically approach the constant value $(1/\sqrt{2\tilde{\chi}})^{-}$ at sufficiently late times\footnote{
	In fact, according to Eq.~\eqref{new94}, in order for $\theta_{-}=0$ to occur 
	in the region $\tilde{\bar{r}}>1/2a(\tilde{t})$, one must have
	$A>{\tilde{\chi}\,\tilde{\bar{r}}_{-}^{2}}/{1-\tilde{\chi}\,\tilde{\bar{r}}_{-}^{2}} $.
	Moreover, from Eq.~\eqref{new79}, for $\tilde{\bar{r}}>1/2a(\tilde{t})$ we obtain $A<1$. 
	Hence, we expect that the roots of $\theta_{-}=0$ satisfy the inequality
	$0<\tilde{\bar{r}}_{-}<{1}/{\sqrt{2\tilde{\chi}}}$ .}, 
while the smaller root (i.e. $\tilde{\bar{r}}_{-}^{\,l}$) tends toward $1/2a(\tilde{t})$. 
Clearly, the separation between these two roots increases with time.  

In summary: For $\tilde{\chi}>0$, there exists a degenerate root of $\theta_{-}=0$ 
at $\tilde{\bar{r}}_{(-,D)}(\tilde{t}_{(-)})$, which forms in the time interval 
${\tilde{\chi}}/{a_{(-)}^{2}} \in \bigl(0,\,4(1-{\sqrt{6}}/{3})\bigr]$. 
At this instant, a POTH begins to form in the matter-dominated 
curved McVittie spacetime\footnote{Before the formation of the degenerate root at $\tilde{t}_{(-)}$, one has 
	$M(\tilde{t})<{2}/{3\tilde{t}}$, whereas for $\tilde{t}>\tilde{t}_{(-)}$ one finds 
	$M(\tilde{t})>{2}/{3\tilde{t}}$. Consequently, 
	$\partial_{\tilde{t}}\theta_{-}|_{\tilde{\bar{r}}=\tilde{\bar{r}}_{(-,D)}(\tilde{t}_{(-)})}<0$. 
	Therefore, at the moment of formation of the degenerate root when $\theta_{-}=0$, we obtain
	$\mathcal{L}_{+}\theta_{-}|_{\theta_{-}(\tilde{\bar{r}}_{(-,D)}(\tilde{t}_{(-)}))=0} 
	= \left({1}/{A}\,\partial_{\tilde{t}}\theta_{-}\right)_{\tilde{\bar{r}}=\tilde{\bar{r}}_{(-,D)}(\tilde{t}_{(-)})}<0 $,
	which corresponds to a POTH.}.

For times $\tilde{t}>\tilde{t}_{(-)}$, two distinct roots 
$\tilde{\bar{r}}_{-}^u$ and $\tilde{\bar{r}}_{-}^l$, analogous to two PTHs, 
appear in the interval $\tilde{\bar{r}}\in(1/2a(\tilde{t}),\,1/\sqrt{\tilde{\chi}})$ 
for the matter-dominated McVittie spacetime.  

In the special case $\tilde{\chi}=0$, for $\tilde{t}<+\infty$, the results obtained in Section~\ref{II.A} for the matter-dominated flat McVittie model, are recovered.
In this case, a degenerate root of $\theta_{-}=0$, corresponding to a POTH, 
is formed at $\tilde{t}_{(-)}(\tilde{\chi}=0)=\tilde{t}_{*}=2\sqrt{3}$. 
As time progresses, two roots $\tilde{\bar{r}}_{-}^u>\tilde{\bar{r}}_{-}^l$, 
analogous to two PTHs, emerge. 
For large times (still with $\tilde{t}<+\infty$), one finds that 
$\tilde{\bar{r}}_{-}^l \to (1/2a(\tilde{t}))^{+}$, while 
$\tilde{\bar{r}}_{-}^u \sim \tilde{t}^{1/3}$.  

Finally, one may state a useful theorem concerning the comparison 
between the formation times of horizons associated with $\theta_{\pm}=0$.  

\textbf{Theorem 2:} 
	The TH(s) corresponding to $\theta_{+}=0$ always forms earlier than 
	the TH(s) corresponding to $\theta_{-}=0$, for every $\tilde{\chi}>0$. 
	Equivalently, one has $\tilde{t}_{(+)}<\tilde{t}_{(-)}$.

\textbf{Proof 2:} 
From Eqs.~\eqref{mw95} and \eqref{nw110} we have $S_{-}\subseteq S_{+}$. 
Since both sets admit an infimum, it follows that $\inf S_{-}\geq \inf S_{+}$. 
Consequently, $\tilde{t}_{(-)}\geq \tilde{t}_{(+)}$. 
However, equality can never occur. Indeed, by Eq.~\eqref{mw96}, $M(\tilde{t}_{(+)})=-{2}/{3\tilde{t}_{(+)}}$ and by Eq.~\eqref{nw111}, $M(\tilde{t}_{(-)})={2}/{3\tilde{t}_{(-)}}$. 
Since $M<0$ at $\tilde{t}_{(+)}$ and $M>0$ at $\tilde{t}_{(-)} $, 
the equality $\tilde{t}_{(-)}=\tilde{t}_{(+)}$ is impossible for $\tilde{t}<+\infty$. 
Therefore, for every $\tilde{\chi}>0$ one must have $\tilde{t}_{(-)}>\tilde{t}_{(+)}$. 
\subsection*{A look at the case $\tilde{\chi}<0$}
The curved McVittie spacetime with $\tilde{\chi}<0$ exhibits properties very different from the case $\tilde{\chi}\geq 0$. Equivalently, this corresponds to negative spatial curvature, i.e., $\gamma\delta<0$.
As noted in Eq.~\eqref{86}, a weak gravitational singularity still exists at 
$\tilde{\bar{r}}=1/2a(t)$. However,  unlike the case $\gamma\delta>0$, 
there is no upper bound for $\tilde{\bar{r}}$ since $\gamma\delta<0$.

To study the physical domain of the curved McVittie solution with negative spatial curvature, 
and to analyze the possible existence of TH(s), we restrict our attention to the region 
outside the gravitational singularity, i.e.\ $\tilde{\bar{r}}>1/2a(t)$.

According to Eq.~\eqref{new80}, the energy density associated with the curved McVittie spacetime 
with $\tilde{\chi}<0$ is not always non-negative. In fact, at sufficiently large times (which, 
 scale as ${1}/{|\tilde{\chi}|^{3/2}}$ for a matter-dominated background), 
the density becomes negative for every $\tilde{\bar{r}}$. 
Furthermore, the appearance of regions with negative density, based on Eqs.~\eqref{new86} and \eqref{new87}, implies that the MSH mass of the curved McVittie spacetime with negative spatial curvature is not always non-negative.  
Indeed, regions with negative MSH mass appear. This suggests the presence of repulsive gravitational 
effects for this solution~\cite{faraoni PG}.

Concerning the THs of the matter-dominated curved McVittie spacetime with 
negative spatial curvature, Eq.~\eqref{new94} shows that $\theta_{+}=0$ never occurs. 
One only finds a PTH of the type $\theta_{-}=0$ with 
$\theta_{+}|_{\theta_{-}=0}>0$.
According to Eqs.~\eqref{new94} and \eqref{nw106}, the behavior of the roots of 
$\theta_{-}=0$ for $\tilde{\chi}<0$ in the interval 
$\tilde{\bar{r}}\in(1/2a(\tilde{t}),\,+\infty)$ with $a(\tilde{t})=\tilde{t}^{2/3}$ can be analyzed 
analogously to the case $\tilde{\chi}>0$. 
In summary, for $\tilde{\chi}<0$, 
the roots of $\theta_{-}=0$ in $\tilde{\bar{r}}\in(1/2a(\tilde{t}),\,+\infty)$ with 
$a(\tilde{t})=\tilde{t}^{2/3}$ (corresponding to PTH(s) in the matter–dominated curved 
McVittie spacetime with negative spatial curvature) behave as follows:

\textbf{(a)}  $\chi<0$ \& $0<|\tilde{\chi}|\leq|\tilde{\chi}_{m}|$: 
For $0<|\tilde{\chi}|<|\tilde{\chi}_{m}|$, where numerically 
$\tilde{\chi}_{m}\in(-0.08,-0.07)$, there exists a time $\tilde{t}_{app}$ 
(which increases with $|\tilde{\chi}|$) at which a degenerate root
of $\theta_{-}=0$ appears. For $0<\tilde{t}<\tilde{t}_{app}$, no root of 
$\theta_{-}=0$ exists in $\tilde{\bar{r}}\in(1/2a(\tilde{t}),+\infty)$. In the interval 
$\tilde{t}_{app}<\tilde{t}<\tilde{t}_{crit1}={1}/{27|\tilde{\chi}|^{3/2}}$, 
there exist two simple roots. Moreover, there is a time $\tilde{t}_{crit2}$ 
such that for $\tilde{t}_{crit1}<\tilde{t}<\tilde{t}_{crit2}$, there are three 
simple roots, while for $\tilde{t}>\tilde{t}_{crit2}$ only one simple root 
remains in $\tilde{\bar{r}}\in(1/2a(\tilde{t}),+\infty)$.

At $\tilde{\chi}=\tilde{\chi}_{m}$, for $0<\tilde{t}<\tilde{t}_{crit1}\approx 1.75$, 
there are no roots, while for $\tilde{t}>\tilde{t}_{crit1}$ there is exactly one 
simple root in $\tilde{\bar{r}}\in(1/2a(\tilde{t}),+\infty)$. However, at two special times, 
namely $\tilde{t}_{\chi=\chi_{c},1}\approx 2.96$ and $\tilde{t}_{\chi=\chi_{c},2}\approx 27$, 
the root becomes \emph{degenerate}. At all other times it remains simple.

\textbf{(b)}  $\chi<0$ \& $|\tilde{\chi}_{m}|<|\tilde{\chi}|\leq|\tilde{\chi}_{c}|=({256}/{9})^{2/3}$: 
For this range, in $0<\tilde{t}\leq \tilde{t}_{crit1}={1}/{27|\tilde{\chi}|^{3/2}}$, 
there are no roots of $\theta_{-}=0$ in $\tilde{\bar{r}}\in(1/2a(\tilde{t}),+\infty)$. 
For $\tilde{t}>\tilde{t}_{crit1}$, exactly one simple root exists.

At the critical value $\tilde{\chi}=\tilde{\chi}_{c}$, the behavior is similar, 
except that there exists a special time $\tilde{t}_{edge}=(|\tilde{\chi}|/4)^{3/4}>\tilde{t}_{crit1}$ 
at which no root of $\theta_{-}=0$ exists. Hence, for $0<\tilde{t}<\tilde{t}_{crit1}$, 
no root appears; for $\tilde{t}\in(\tilde{t}_{crit1},\tilde{t}_{edge})$ and 
$\tilde{t}\in(\tilde{t}_{edge},+\infty)$, exactly one simple root is present.

\textbf{(c)}   $\chi<0$ \& $|\tilde{\chi}|>|\tilde{\chi}_{c}|$:
In this regime, based on the sign of $\theta_{-}|_{(1/2a(\tilde{t}))^{+}}$, one finds that for 
$0<\tilde{t}\leq \tilde{t}_{crit1}={1}/{27|\tilde{\chi}|^{3/2}}$, no root of 
$\theta_{-}=0$ exists in $\tilde{\bar{r}}\in(1/2a(\tilde{t}),+\infty)$. 
There exist $\tilde{t}_{1\pm}>\tilde{t}_{crit1}$, determined by
\begin{equation}
\tilde{t}_{1\pm}^{2/3}=\frac{9|\tilde{\chi}|^{2}\pm 
	\sqrt{81|\tilde{\chi}|^{4}-65536|\tilde{\chi}|}}{512}.
	\label{mw114}
\end{equation}
For the positive sign in Eq.~\eqref{mw114} and negative $\tilde{\chi}$, the above expression is equal to Eq.~\eqref{MW98}.
For $\tilde{t}_{crit1}<\tilde{t}<\tilde{t}_{1-}$, there is exactly one simple root. 
For $\tilde{t}_{1-}\leq \tilde{t}\leq \tilde{t}_{1+}$, no roots exist. 
Finally, for $\tilde{t}>\tilde{t}_{1+}$, one simple root reappears.

In \cite{nolan2017}, the global properties of the curved McVittie spacetime with negative spatial curvature\footnote{$\tilde{\chi}<0$, or, in the notation of \cite{nolan2017}, $K<0$.} in a dust-filled expanding universe with $H>0$, $\lim_{t\to\infty}H(t)=0$, and $\Lambda=0$\footnote{At early times, this coincides with our model of curved McVittie with negative spatial curvature in a matter-dominated background.} are addressed in detail. First, as mentioned earlier in Section~\ref{II.A}, and in agreement with our quasi-local results, for an expanding background there are no trapped regions. This is confirmed again in \cite{nolan2017} for the curved McVittie spacetime with $K<0$, $H>0$, $\lim_{t\to\infty}H(t)=0$, and $\Lambda=0$. In addition, \cite{nolan2017} shows that for $K<0$ and $\Lambda=0$ there exists a BH event horizon, whose generators approach the boundary $(t\to+\infty,\ R\to 2m_0)$. It is also shown that a finite-radius cosmological event horizon does not exist for curved McVittie with $K<0$ and $\Lambda=0$; instead, future null infinity is present as $\mathscr{I}^+$, rather than a spacelike future boundary that would indicate a de Sitter--type cosmological event horizon.
	
As a result, within our Hayward-inspired quasi-local framework, the curved McVittie spacetime in an expanding universe and more specifically in early matter-dominated universe,  does not describe a cosmological BH.

\begin{figure}[H]
 \centering
 \includegraphics[width=\textwidth]{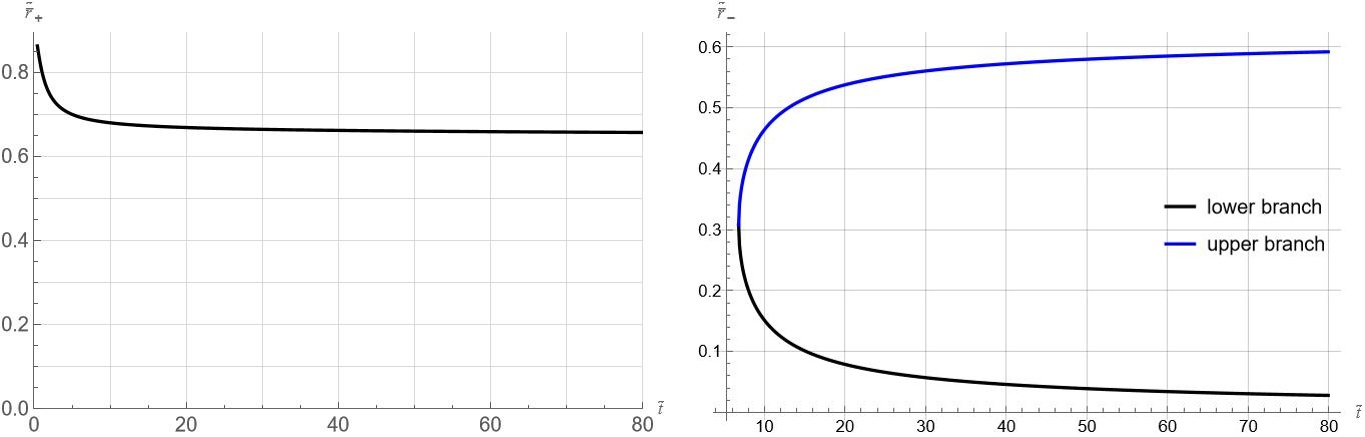}
 \caption{The time evolution of the THs in the matter-dominated curved McVittie solution for $\tilde{\chi}=1.25,$  
 $ a(\tilde{t})=\tilde{t}^{{2}/{3}} $ and ${1}/({2 a(\tilde{t}))} < \tilde{\bar{r}} < \sqrt{{1}/{\bar{\chi}}}$. The $\tilde{\bar{r}}_{+}$ denotes the solution corresponding to $\theta_+ = 0$ , while $\tilde{\bar{r}}_{-}$ denotes the solutions corresponding to $\theta_- = 0$.\protect }
\label{fig5}
\end{figure}

In the following,  we will examine the existence, nature, and signature of the THs and the ECs of the matter-dominated curved McVittie spacetime for a specific value $\tilde{\chi}=1.25$.  
According to the calculations carried out in this section, we expect that for this value of $\tilde{\chi}$, after a certain time, one PTH attached to the second asymptotic region (with $\theta_{+}=0,\;\theta_{-}>0$) and two PTHs attached to the first asymptotic region (with $\theta_{-}=0,\;\theta_{+}>0$) will appear.
Based on Eq.~\eqref{new94},the time evolution of the expansions \( \theta_{\pm}=0 \) are plotted  in Fig. \ref{fig5}.
The inequality ${1}/{2a(\tilde{t})} < \tilde{\bar{r}} < {1}/{\sqrt{\tilde{\chi}}}$ 
ensures that the THs lie outside the curvature singularity and within the region where Eq.~\eqref{new94} is well-defined.  
Fig. \ref{fig5} shows that, as expected, for $\tilde{\chi}=1.25$, there exists a simple root corresponding to 
$\theta_{+}=0,\,\theta_{-}>0$, which is a PTH. According to Eq.~\eqref{mw114}, this horizon forms at 
$\tilde{t}_{(+)} \approx \tilde{t}_{le}(\tilde{\chi}=1.25) \approx 0.45$. At large times (as long as the expansion parameters 
remain well-defined), it asymptotes to $(1/\sqrt{2\tilde{\chi}})^{+}$.  

In addition, a degenerate root corresponding to $\theta_{-}=0,\,\theta_{+}>0$ forms at $\tilde{t}_{(-)} \approx 6.81$. 
As expected, $\tilde{t}_{(-)}>\tilde{t}_{(+)}$, meaning that the PTH associated with $\theta_{-}=0$ forms 
later than the one associated with $\theta_{+}=0$. Clearly, the PTH associated with $\theta_{-}=0$ in this spacetime 
forms later than the PTH of the matter-dominated spatially flat McVittie spacetime, i.e. 
$\tilde{t}_{(-)} (\tilde{\chi}=1.25)>\tilde{t}_{(-)} (\tilde{\chi}=0)=\tilde{t}_{*}=2\sqrt{3}$. For $\tilde{t}>\tilde{t}_{(-)}$, 
there will be two distinct roots, corresponding to two PTHs in this spacetime.  

According to the calculations in the previous sections regarding the asymptotic behavior of THs associated with 
$\theta_{-}=0$ in the matter-dominated curved McVittie spacetime with $\tilde{\chi}>0$, one can see that the smaller 
horizon, i.e. $\tilde{\bar{r}}_{-}^{l}$, at large times (as long as the expansion parameters remain well-defined) 
approaches $(1/2a(\tilde{t}))^{+}$, while the larger horizon, i.e. $\tilde{\bar{r}}_{-}^{u}$, approaches 
$(1/\sqrt{2\tilde{\chi}})^{-}$.  
Moreover, as shown in Fig. \ref{fig5}, it is evident that the PTH corresponding to $\theta_{+}=0$ always lies 
outside the two PTHs corresponding to $\theta_{-}=0$.  

In the previous section, we discussed the ECs of the matter-dominated curved McVittie spacetime with $\tilde{\chi}>0$ in detail. Here, we examine the validity of the ECs on the THs of this spacetime for $\tilde{\chi}=1.25$:

\textbf{NEC:} First, we examine the NEC on the PTH which is characterized by 
$\theta_{+}=0, \theta_{-}>0$. 
Using Eq.~\eqref{2005} and setting $\tilde{\chi}=1.25$ and $\theta_{+}=0$, 
we obtain the following diagram:

\begin{figure}[H]
	\centering
	\includegraphics[width=0.6\textwidth]{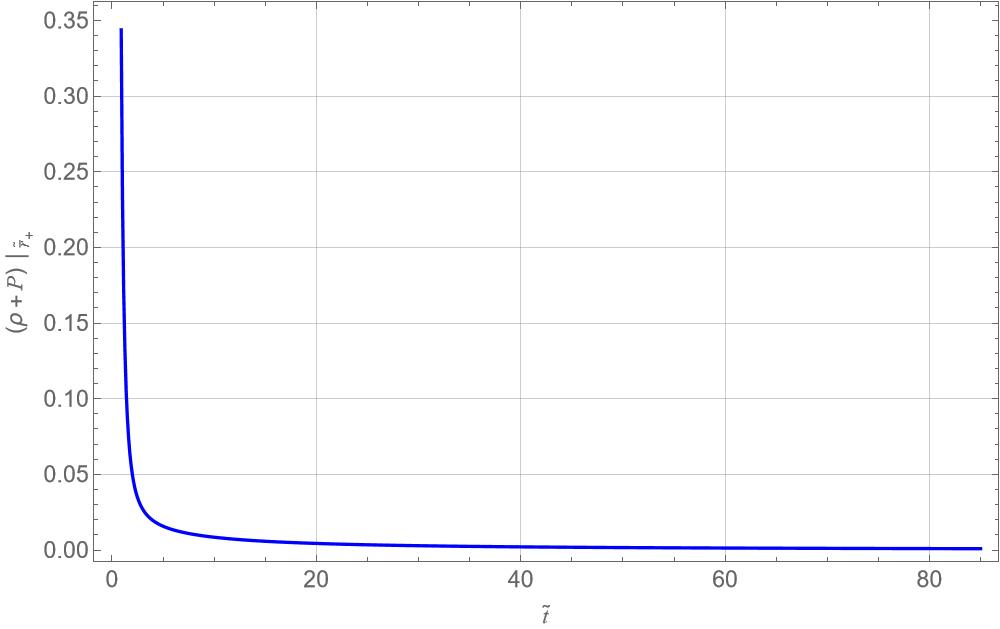}
	\caption{
		The time evolution of 
		$\widetilde{\rho+P}$ at $\theta_+=0$ (for $\tilde{t} \geq \tilde{t}_{(+)}$)
		in the matter-dominated curved McVittie solution for 
		$\tilde{\chi} = 1.25$ in the range  
		${1}/{2 a(\tilde{t})} < \tilde{\bar{r}} < \sqrt{1 / \tilde{\chi}} $.
	}
	\label{fig6}
\end{figure}
According to Fig.~\ref{fig6}, the NEC on the PTH (associated with $\theta_{+}=0$) is always satisfied. Based on the previous analysis of the NEC in curved, matter-dominated McVittie spacetime with $\tilde{\chi}>0$, it is evident that this condition is valid for 
$\tilde{\chi}=1.25$.  
Since in this case $\tilde{\chi}=1.25<(128/3)^{2/3}$, it follows that for times
$\left({\sqrt{\tilde{\chi}}}/{2}\right)^{3/2} < \tilde{t}_{(+)} < \tilde{t} < \left({64}/{3\tilde{\chi}}\right)^{3/2} \approx 70.51$,
the NEC will always hold for any $\tilde{\bar{r}} \in \left({1}/{2a(\tilde{t})}, {1}/{\sqrt{\tilde{\chi}}}\right)$. According to Fig.~\ref{fig5}, after $\tilde{t}\geq \left({64}/{3\tilde{\chi}}\right)^{3/2}$,  one always has $\tilde{\bar{r}}_{+}>{3}/{4a(\tilde{t})}$. Therefore, noting Eq.~\eqref{2005}, the NEC is also satisfied in this time interval. Consequently, on the PTH corresponding to $\theta_{+}=0$, the NEC is always satisfied for $\tilde{\chi}=1.25$.

Now we examine the NEC on the PTHs corresponding to $\theta_{-}=0$. The NEC is always satisfied on the larger PTH, i.e., $\tilde{\bar{r}}_{-}^u$,  
since according to Fig.~\ref{fig6}, it can be shown that $\tilde{\bar{r}}_{-}^u > {3}/{4a(\tilde{t})}$. 
Consequently, according to Eq.~\eqref{2005}, the NEC holds on this TH for all $\tilde{t} \geq \tilde{t}_{(-)}$.
\begin{figure}[H]
	\centering
	\includegraphics[width=1\textwidth]{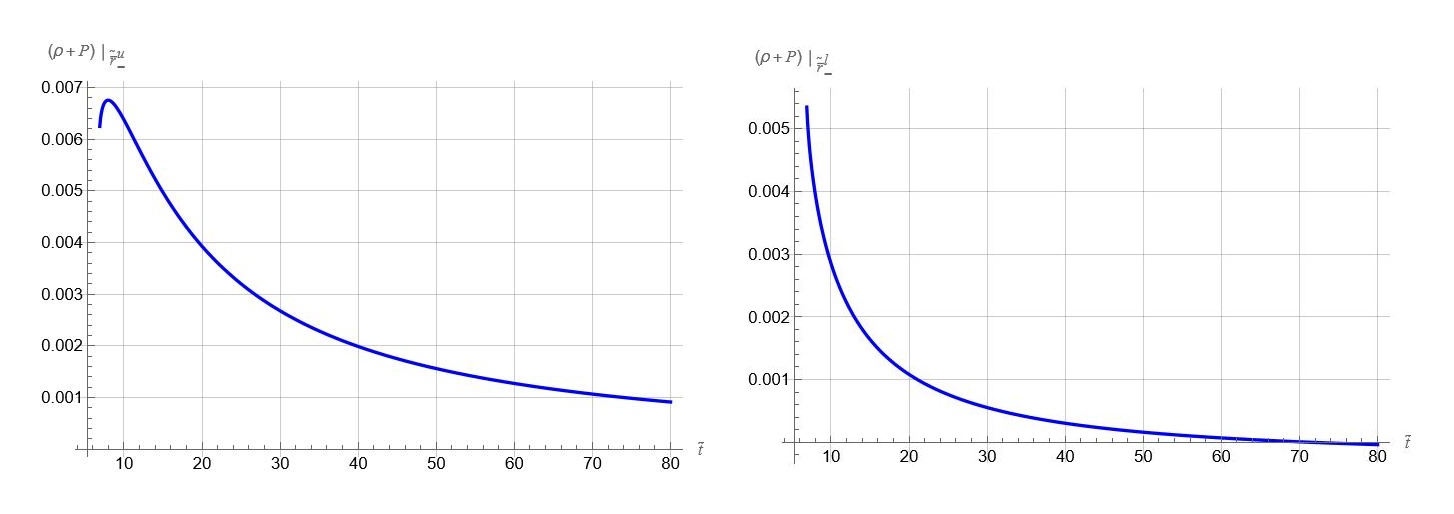}
	\caption{
		The time evolution of 
		$\widetilde{\rho+P}$ at $\tilde{\bar{r}}_{-}^{u}$ and $\tilde{\bar{r}}_{-}^{l}$ (for $\tilde{t}>\tilde{t}_{(-)}$)
		in the matter-dominated curved McVittie solution for 
		$\tilde{\chi} = 1.25$ in the range 
		${1}/{2 a(\tilde{t})} < \tilde{\bar{r}} < \sqrt{1 / \tilde{\chi}} $.
	}
	\label{fig7}
\end{figure}
For the smaller PTH, $\tilde{\bar{r}}_{-}^l$, it can be shown that the NEC is satisfied for $\tilde{t}_{(-)} < \tilde{t} \lesssim 71.17$  (slightly after $(64/3\tilde{\chi})^{3/2} \approx 70.51$), while it is violated for $\tilde{t} \gtrsim 71.17$.

\textbf{WEC:} As previously shown for the matter-dominated curved McVittie spacetime with $\tilde{\chi}>0$, the validity of the WEC is equivalent to the validity of the NEC.  According to Eq.~\eqref{2003}, the energy density of the matter filed associated with this spacetime is positive for every $\tilde{\chi}>0$. Therefore, the WEC on the PTHs associated with both $\theta_{+}=0$ and $\theta_{-}=0$ is also equivalent to the NEC, which was examined above.

\begin{figure}[H]
	\centering
	\includegraphics[width=0.6\textwidth]{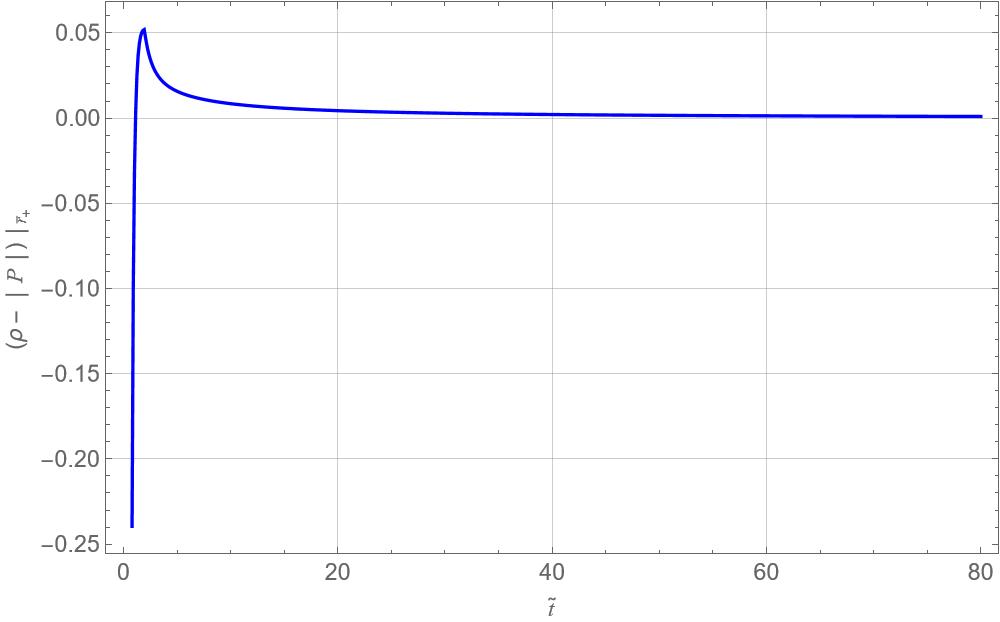}
	\caption{
		The time evolution of 
		$\widetilde{\rho- |P|}$ at $\theta_+=0$ (for $\tilde{t} \geq \tilde{t}_{(+)}$)
		in the matter-dominated curved McVittie solution for 
		$\tilde{\chi} = 1.25$ in the range  
		${1}/{2 a(\tilde{t})} < \tilde{\bar{r}} < \sqrt{1 / \tilde{\chi}} $.
	}
	\label{fig8}
\end{figure}

\textbf{DEC:} As we have shown, the DEC for the curved McVittie spacetime is equivalent to the validity of Eqs.~\eqref{2005} and \eqref{2007}. 
 According to Eq.~\eqref{2003}, however, the validity of the DEC is equivalent to $\rho - |P| \geq 0$, since the energy density of the matter field associated with this spacetime is positive.  
The time evolution of $\rho - |P|$ on the PTH corresponding to $\theta_{+}=0$ for $\tilde{\chi}=1.25$ is plotted in Fig.~\ref{fig8}. 
According to it, the DEC on the PTH corresponding to 
$\theta_{+}=0$ will always hold for $\tilde{t} \gtrsim 1.12 > \tilde{t}_{(+)}$. 
Furthermore, when $\tilde{t} \gtrsim 3.32$ it can be shown that 
$\tilde{\bar{r}}_{+} \geq {3}/{2a(\tilde{t})}$, indicating the validity of the DEC in this time interval, as previously discussed.
In the intermediate range $1.12 \lesssim \tilde{t} \lesssim 3.32$, the DEC is also satisfied based on calculations carried out in matter-dominated curved McVittie spacetime.
In the following, we examine the DEC on the PTHs corresponding to $\theta_{-}=0$.
\begin{figure}[H]
	\centering
	\includegraphics[width=1\textwidth]{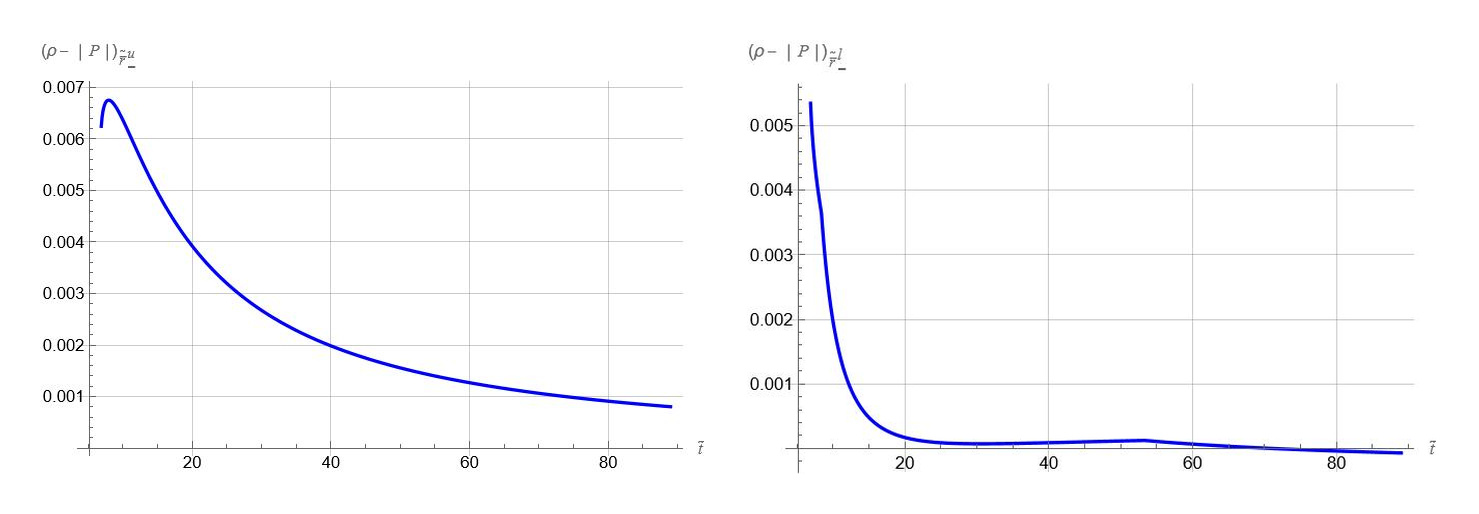}
	\caption{
		The time evolution of 
		$\widetilde{\rho- |P|}$ at $\tilde{\bar{r}}_{-}^{u}$ and $\tilde{\bar{r}}_{-}^{l}$ (for $\tilde{t}>\tilde{t}_{(-)}$)
		in the matter-dominated curved McVittie solution for 
		$\tilde{\chi} = 1.25$in the range 
		${1}/{2 a(\tilde{t})} < \tilde{\bar{r}} < \sqrt{1 / \tilde{\chi}} $.
	}
	\label{fig9}
\end{figure}
For the larger PTH, i.e., $\tilde{\bar{r}}_{-}^u$, the DEC always holds. 
However, on the smaller PTH, i.e., $\tilde{\bar{r}}_{-}^l$, for $\tilde{t} \gtrsim 71.17$, the DEC is violated, which was expected according to the calculations performed to determine the intervals in which the matter-dominated curved McVittie spacetime with $\tilde{\chi}>0$ satisfies the DEC. 
This is because for $\tilde{t} > \left({64}/{3\tilde{\chi}}\right)^{3/2} \approx 70.57$, we showed that the validity of the DEC is equivalent to the validity of the NEC or Eq.~\eqref{2005}. 
Since we demonstrated that the NEC is violated during this time interval, the DEC is also violated on the smaller PTH.

\textbf{SEC:} As we have shown, the SEC for the curved McVittie spacetime is equivalent to the validity of Eqs.~\eqref{2005} and \eqref{2008}. 
We examined the NEC on the PTHs of the curved, matter-dominated McVittie spacetime with $\tilde{\chi}=1.25$ in detail above.
Therefore, this section focuses on the validity of Eq.~\eqref{2008}.  
The time evolution of $\widetilde{\rho+3P}$ on the PTH corresponding to $\theta_{+}=0$ for $\tilde{\chi}=1.25$ is plotted below.
\begin{figure}[H]
	\centering
	\includegraphics[width=0.6\textwidth]{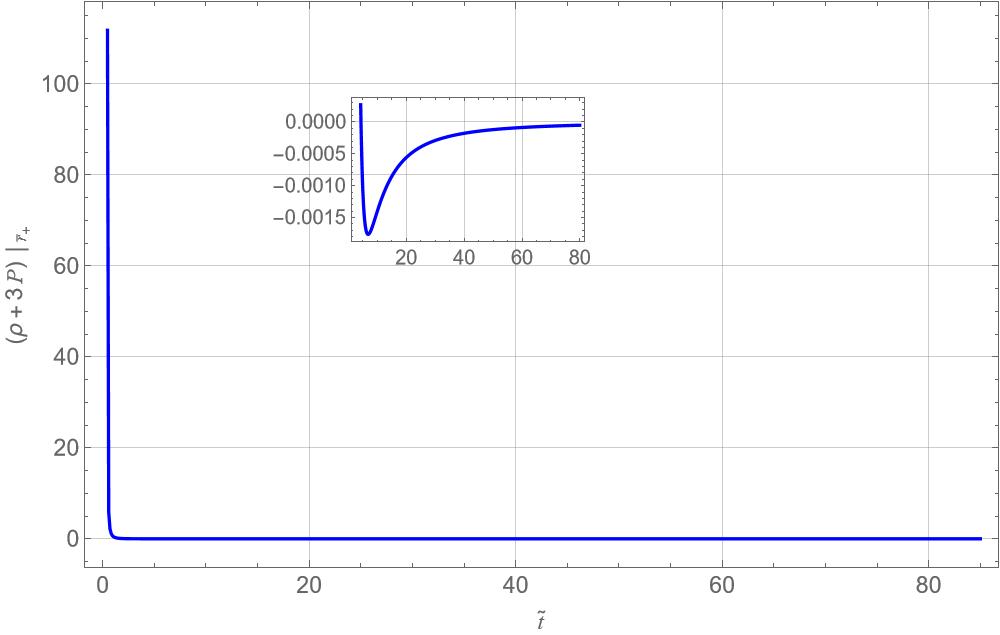}
	\caption{
		The time evolution of 
		$\widetilde{\rho+3P}$ at $\theta_+=0$ (for $\tilde{t} \geq \tilde{t}_{(+)}$)
		in the matter-dominated curved McVittie solution for 
		$\tilde{\chi} = 1.25$ in the range 
		${1}/{2 a(\tilde{t})} < \tilde{\bar{r}} < \sqrt{1 / \tilde{\chi}} $.
	}
	\label{fig10}
\end{figure}
Since $\tilde{\chi}=1.25$, it falls within the classification  
$\tilde{\chi} \in \left(\left({2S_{\min}}/{y_*}\right)^{2/3}, \; \left({128}/{3y_*}\right)^{2/3}\right)$, 
and therefore includes specific cases of the sign determination carried out for Eq.~\eqref{2009}. 
As shown in Fig.~\ref{fig10}, one can demonstrate that $\widetilde{\rho+3P}$ on the PTH corresponding to $\theta_{+}=0$ is always negative for $\tilde{t}\gtrsim 4.31$, and always non-negative for $t_{(+)} \leq \tilde{t} \lesssim 4.31$. 
Therefore, in the time interval $t_{(+)} \leq \tilde{t} \lesssim 4.31$, Eq.~\eqref{2005} holds and the SEC is satisfied. After this interval, however, it is violated.
In fact, considering $\tilde{\chi}=1.25$, the admissible cases for the sign of Eq.~\eqref{2009} allow the validity of Eqs.~\eqref{2005} and \eqref{2008}, and consequently the SEC, only on the PTH corresponding to $\theta_{+}=0$ within small time intervals. The following figure examines the SEC on the PTHs corresponding to $\theta_{-}=0$. 
\begin{figure}[H]
	\centering
	\includegraphics[width=1\textwidth]{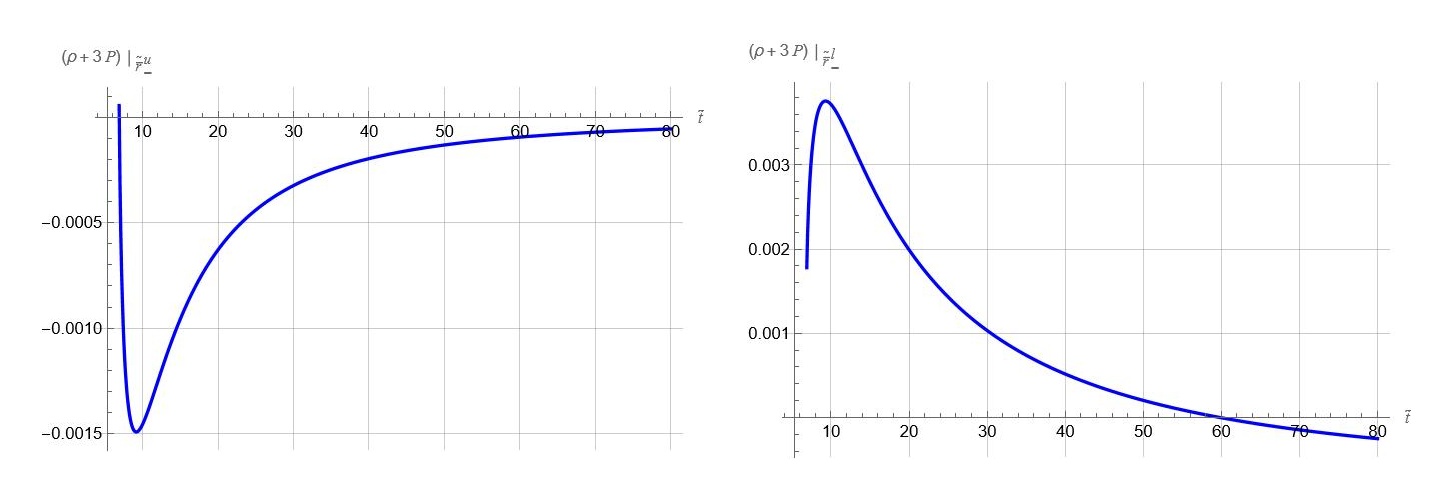}
	\caption{
		The time evolution of 
		$\widetilde{\rho+3P}$ at $\tilde{\bar{r}}_{-}^{u}$ and $\tilde{\bar{r}}_{-}^{l}$ (for $\tilde{t}>\tilde{t}_{(-)}$)
		in the matter-dominated curved McVittie solution for 
		$\tilde{\chi} = 1.25$ in the range  
		${1}/{2 a(\tilde{t})} < \tilde{\bar{r}} < \sqrt{1 / \tilde{\chi}} $.
	}
	\label{fig11}
\end{figure} 
Based on Fig.\ref{fig11}, on the larger PTH, i.e., $\tilde{\bar{r}}_{-}^u$, the SEC only holds in the very short time interval $\tilde{t}_{(-)} \approx 6.81 \leq \tilde{t} \lesssim 6.91$. Thereafter,  the SEC no longer holds due to the violation of Eq.~\eqref{2008} first, and then due to the violation of Eqs.~\eqref{2005} and \eqref{2008}. 
On the smaller PTH, i.e., $\tilde{\bar{r}}_{-}^l$, in the interval $\tilde{t}_{(-)} \leq \tilde{t} \lesssim 59.83$, since both Eqs.~\eqref{2005} and \eqref{2008} are satisfied, the SEC holds; beyond this range, it is violated.  
Indeed, as observed in the sign analysis of Eq.~\eqref{2009}, for smaller values of $y$, one can expect that the SEC remains valid over a larger time interval.

\begin{figure}[H]
 \centering
 \includegraphics[width=0.6\textwidth]{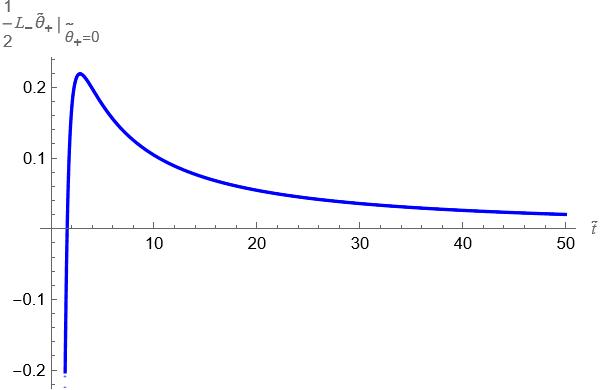}
 \caption{
     The time evolution of 
     $\mathcal{L}_- \tilde{\theta}_+ |_{\tilde{\theta}_+ = 0}$ i.e., at $\tilde{\bar{r}}_{+}$ (for $\tilde{t} \geq \tilde{t}_{(+)}$)
     in the matter-dominated curved McVittie solution for 
     $\bar{\chi} = 1.25$ in the range 
     ${1}/{2a(\tilde{t})} < \tilde{\bar{r}} < \sqrt{1 / \bar{\chi}} $.
      }
\label{fig12}
\end{figure}

In the following we proceed to determine the nature and signature of the matter-dominated curved McVittie for $\tilde{\chi}=1.25$.
To determine the nature of the TH corresponding to $\theta_+=0$, we examine the signs of $\mathcal{L}_{-} \theta_{+} |_{\theta_{+}=0}$ 
 Fig. \ref{fig12} plots the time evolution of this quantity.
As shown in Fig. \ref{fig12}, for the PTH corresponding to $\theta_+=0$ , we have $\mathcal{L}_{-} \theta_{+}|_{\theta_+=0} < 0$ during the time interval $\tilde{t}_{(+)} < \tilde{t} \lesssim 1.57$,  indicating that the PTH is an OTH. This implies that during this period, the matter-dominated curved McVittie solution contains a WH-type TH (POTH) attached to the second asymptotic region.\footnote{On this PTH, according to Eq.~\eqref{new94}, we have $\theta_+=0$ and $\theta_->0$.}
At $\tilde{t} \approx 1.57$, a past degenerate TH (PDTH) emerges that satisfies $\mathcal{L}_{-} \theta_{+}|_{\theta_+=0} = 0$ . For times $\tilde{t} \gtrsim 1.57$, the nature of the PTH changes, becoming a cosmic-type TH (PITH), attached to the second asymptotic region, where $\mathcal{L}_{-} \theta_{+}|_{\theta_+=0} > 0$.
The following figure determines the nature of PTHs corresponding to $\theta_-=0$ for $\tilde{\chi}=1.25$.
\begin{figure}[H]
	\centering
	\includegraphics[width=1\textwidth]{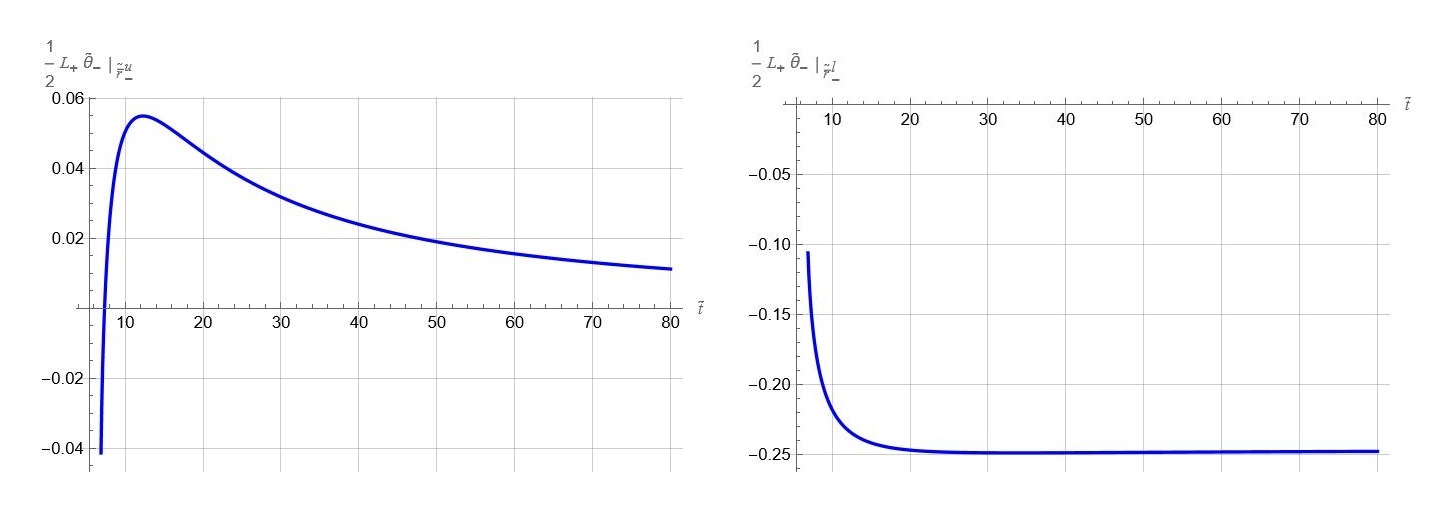}
	\caption{
		The time evolution of 
		$\mathcal{L}_+ \tilde{\theta}_- |_{\tilde{\theta}_- = 0}$ i.e., at $\tilde{\bar{r}}^{u}_{-}$ and $\tilde{\bar{r}}^{l}_{-}$(for $\tilde{t} \geq \tilde{t}_{(+)}$)
		in the matter-dominated curved McVittie solution for 
		$\bar{\chi} = 1.25$ in the range
		${1}/{2a(\tilde{t})} < \tilde{\bar{r}} < \sqrt{1 / \bar{\chi}} $.
	}
	\label{fig13}
\end{figure}
As shown in Fig.~\ref{fig13}, on the larger PTH, i.e. $\tilde{\bar{r}}^{u}_{-}$, 
for $t_{(-)} \leq \tilde{t} \lesssim 7.35$, we have 
$\mathcal{L}_{+}\theta_{-}|_{\theta_{-}=0} < 0$, which corresponds to a POTH 
attached to the first asymptotic region \footnote{On this PTH, according to Eq.~\eqref{new94}, 
we also have $\theta_{-}=0$ and $\theta_{+}>0$.}.  
At $\tilde{t} \approx 7.35$, $\mathcal{L}_{+}\theta_{-}|_{\theta_{-}=0}=0$, 
which corresponds to a PDTH. For $\tilde{t} \gtrsim 7.35$, 
$\mathcal{L}_{+}\theta_{-}|_{\theta_{-}=0} > 0$ on the larger PTH,  which is equivalent to a PITH 
attached to the first asymptotic region.  
On the other hand, for the smaller PTH, i.e. $\tilde{\bar{r}}^{l}_{-}$, for all 
$\tilde{t} \geq t_{(-)}$, we have $\mathcal{L}_{+}\theta_{-}|_{\theta_{-}=0} < 0$. 
Therefore, in this spacetime, the smaller PTH is always a POTH attached to the first 
asymptotic region.
As we just saw, the two PTHs corresponding to $\theta_{-}=0$, coincide at $\tilde{t} = t_{(-)}$. Based on Fig.\ref{fig13},  a POTH is obtained at this point. 

Moreover, as was shown earlier, for every $\tilde{\chi}>0$, the PTH corresponding to $\theta_{-}=0$ 
emerges as a degenerate root at $\tilde{t}=\tilde{t}_{(-)}$, which corresponds to a POTH. 
Beyond this point, two distinct PTHs corresponding to $\theta_{-}=0$, appear in the 
matter-dominated curved McVittie spacetime with $\tilde{\chi}>0$. 
Thus, the results obtained here for $\tilde{\chi}=1.25$ regarding the nature of the PTH(s) 
corresponding to $\theta_{-}=0$ are consistent with the general form derived earlier.

The following figure determines the signature of the THs by analyzing the sign of \( ds^2 \) computed in Eq.~\eqref{86}, at the location of the THs in the matter-dominated curved McVittie spacetime for $\chi=1.25$.
\begin{figure}[H]
	\centering
	\includegraphics[width=0.6\textwidth]{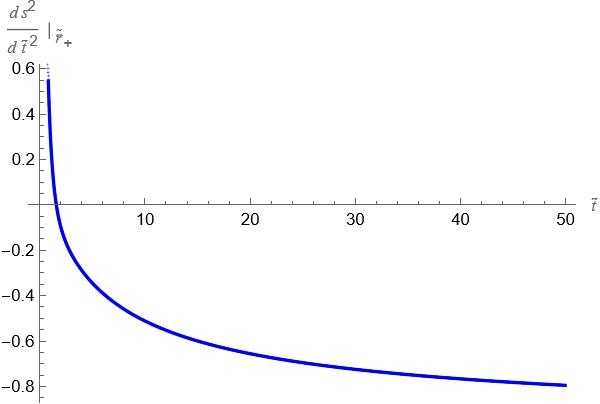}
	\caption{
		The time evolution of 
		${ds^2}/{d\tilde{t}^2}|_{\tilde{\bar{r}}=\tilde{\bar{r}}_{+}}$ (for $\tilde{t} \geq \tilde{t}_{(+)}$)
		in the matter-dominated curved McVittie solution for 
		$\bar{\chi} = 1.25$ in the range
		${1}/{2 a(\tilde{t})} < \tilde{\bar{r}} < \sqrt{1 / \bar{\chi}}$. 
	}
	\label{fig14}
\end{figure}

As expected, the NEC based on Fig.\ref{fig6}, always holds on the PTH, \( \tilde{\bar{r}}_{+} (\tilde{t}) \). Thus,  this TH is a spacelike POTH (\( ds^2|_{\tilde{\bar{r}}_{+}} > 0 \)) for $\tilde{t}_{(+)} \leq \tilde{t} \lesssim 1.57$. At \( \tilde{t} \approx 1.57 \), it becomes null (\( ds^2|_{\tilde{\bar{r}}_{+}} = 0 \)), and subsequently, a timelike PITH (\( ds^2|_{\tilde{\bar{r}}_{+}} < 0 \)) appears.\cite{Hayward}
The following figure examines the signatures of the PTHs corresponding to $\theta_-=0$.
\begin{figure}[H]
	\centering
	\includegraphics[width=1\textwidth]{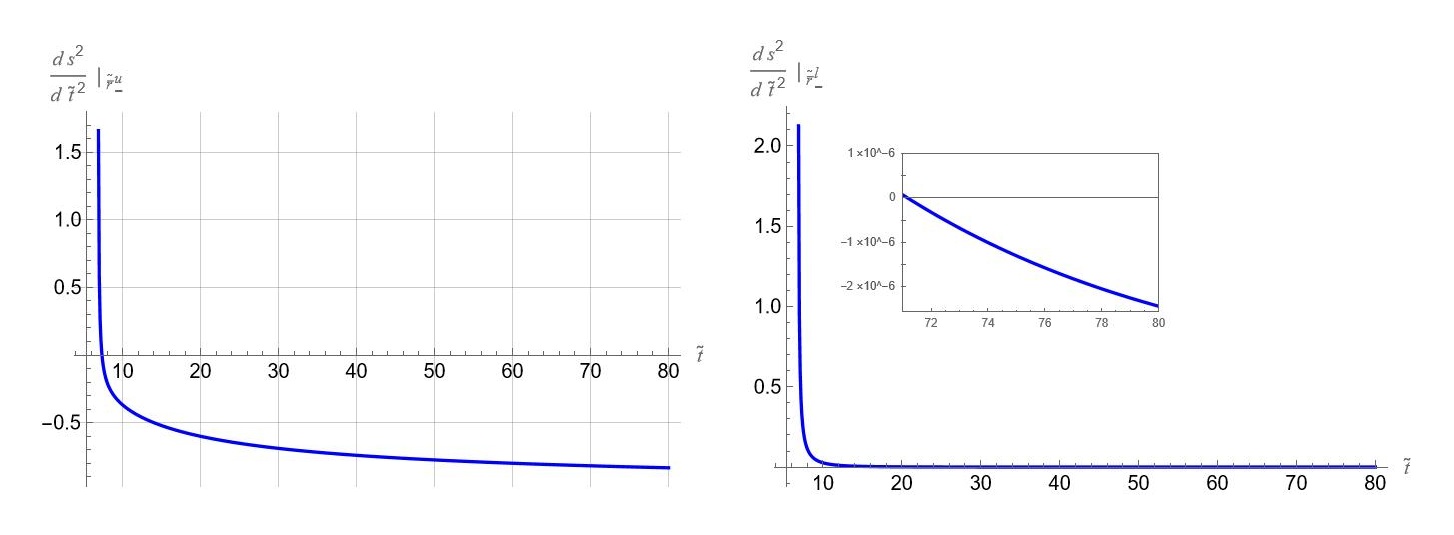}
	\caption{
		The time evolution of 
		${ds^2}/{d\tilde{t}^2}|_{\tilde{\bar{r}}=\tilde{\bar{r}}^{u}_{-}}$ and ${ds^2}/{d\tilde{t}^2}|_{\tilde{\bar{r}}=\tilde{\bar{r}}^{l}_{-}}$(for $\tilde{t} \geq \tilde{t}_{(-)}$)
		in the matter-dominated curved McVittie solution for 
		$\bar{\chi} = 1.25$ in the range 
		${1}/{2 a(\tilde{t})} < \tilde{\bar{r}} < \sqrt{1 / \bar{\chi}}$. 
	}
	\label{fig15}
\end{figure}
As shown in Fig.~\ref{fig15}, for $\tilde{t}_{(-)} \leq \tilde{t} \lesssim 7.35$, the larger PTH, i.e., $\tilde{\bar{r}}^{u}_{-}$, 
behaves as a spacelike POTH 
\((ds^{2}|_{\tilde{\bar{r}}^{u}_{-}}>0)\). 
At $\tilde{t}\approx 7.35$, it becomes null \((ds^{2}|_{\tilde{\bar{r}}^{u}_{-}}=0)\), 
and eventually a timelike PITH emerges. 
According to Fig.~\ref{fig7}, the NEC always holds on the larger PTH, thus, 
the obtained results regarding the signature of this PTH were expected.  
However, on the smaller PTH, i.e., $\tilde{\bar{r}}^{l}_{-}$, 
for $\tilde{t}_{(-)} \leq \tilde{t} \lesssim 71.17$, it is a spacelike POTH \((ds^{2}|_{\tilde{\bar{r}}^{l}_{-}}>0)\). 
At $\tilde{t}\approx 71.17$, it becomes null \((ds^{2}|_{\tilde{\bar{r}}^{l}_{-}}=0)\), and for $\tilde{t} \gtrsim 71.17$, it is a timelike POTH. 
The presence of a timelike POTH \((ds^{2}|_{\tilde{\bar{r}}^{l}_{-}}<0)\) at later times is in fact the consequence of violating the NEC 
for $\tilde{t} \gtrsim 71.17$, as shown in Fig.~\ref{fig7}.  

The table below summarizes the signatures and natures of the THs in the matter-dominated curved McVittie spacetime with $\tilde{\chi}=1.25$:
\begin{table}[h!]
	\begin{scriptsize}
	\begin{tabular}{|c|c|c|c|c|c|c|c|c|c|}
		\hline
		& $0<\smash{\tilde{t}} \lesssim 0.45$ & $0.45 \lesssim \smash{\tilde{t}}< 1.57$ & $\smash{\tilde{t}} \approx 1.57$ & $1.57 \lesssim \smash{\tilde{t}} < 6.81$ & $6.81 \lesssim \smash{\tilde{t}} < 7.35$ & $\smash{\tilde{t}} \approx 7.35$ & $7.35 \lesssim \smash{\tilde{t}} < 71.17$ & $\smash{\tilde{t}} \approx 71.17$ & $\smash{\tilde{t}} \gtrsim 71.17$  \\
		\hline
		$\tilde{\bar{r}}_{+}(\tilde{t})$ & N/A & POTH-SL & PDTH-Null & PITH-TL & PITH-TL & PITH-TL & PITH-TL & PITH-TL & PITH-TL \\
		\hline
		$\tilde{\bar{r}}^{u}_{-}(\tilde{t})$ & N/A & N/A & N/A & N/A & POTH-SL & PDTH-Null & PITH-TL & PITH-TL & PITH-TL \\
		\hline
		$\tilde{\bar{r}}^{l}_{-}(\tilde{t})$ & N/A & N/A & N/A & N/A & POTH-SL & POTH-SL & POTH-SL & POTH-Null & POTH-TL \\
		\hline
	\end{tabular}
		\end{scriptsize}
	\caption{The sign and nature of the THs of the matter-dominated curved McVittie spacetime for different time intervals $\tilde{t}$.
		SL and TL refer to spacelike and timelike, respectively. $\tilde{t}_{(+)} \approx 0.45$ and $\tilde{t}_{(-)} \approx 6.81$ are the creation times of the PTHs corresponding to $\theta_{+}=0$ and $\theta_{-}=0$, respectively.
		}
\end{table}

\subsection{Trapping Horizons in Painlev\'e--Gullstrand coordinates} \label{V.E}
So far, we have analyzed the generalized spherically symmetric solution of Einstein's equations under the McVittie assumption. We now turn to writing the Glass-Mashhoon metric in PG coordinates by applying two coordinate transformations starting from Eq.~\eqref{new78} \footnote{In this section, we assume the non-negativity of the Glass-Mashhoon MSH mass as defined in section ~\ref{V.C}}.

First, we perform a coordinate transformation from $(t, \bar{r})$ to $(t, R)$  in the line element \eqref{new78}, where $R= a(t)  (1 + {m_0}/{(2 a(t) \overline{r})})^2 \sqrt{(\alpha - \gamma \overline{r}^2) (\delta \overline{r}^2 - \beta)}/{(\alpha \delta - \beta \gamma)}$ is the areal coordinate. It gives: 
\begin{equation}
\begin{cases}
ds^2=-L^2 dt^2-2M dRdt+N^2 dR^2+R^2 d\Omega^2\\
L^2\equiv A^2-B^2 (\frac{\partial _{t}R}{\partial_{\bar{r}}R})^2 \\
M\equiv (\frac{B}{\partial_{\bar{r}}R})^2 \partial_{t}R \\
N \equiv \frac{B}{\partial_{\bar{r}}R} 
\end{cases}
\label{new98}
\end{equation}
Next, we perform a second coordinate transformation from $(t, R)$ to $(\tau, R)$ where $dt=f d\tau - \sigma dR$  and the condition $g_{11}=1$ is satisfied in the final form of the metric. Applying this transformation, $\sigma$ is fixed, while the integration factor $f$ remains unfixed. Substituting $dt $ in Eq.~\eqref{new98}, we obtain:  
\begin{equation}
ds^2=-L^2f^2 d\tau^2+2f(L^2\sigma-M)d\tau dR+dR^2+R^2 d\Omega^2 
\label{new99}
\end{equation}  
where $ L, M $ are obtained from Eq.~\eqref{new98}, and  by applying $g_{11}=1$, $ \sigma$ is given by the following equation:  
\begin{equation}
-L^2 \sigma^{2}+2M\sigma+N^2=1
\label{new100}
\end{equation}  
Also, comparing Eq.~\eqref{new14} with Eq.~\eqref{new99}, we get:
\begin{equation}
\begin{cases}
c(R,\tau)=\sqrt{L^2 f^2 +v^2}\\
v(R,\tau)=f(L^2 \sigma-M)
\end{cases}
\label{new101}
\end{equation}
where without loss of generality, we choose the positive sign for $ c(R,\tau)$. Furthermore,  using Eq.~\eqref{new100}, we can write:  
\begin{equation}
\frac{v}{f}=L^2 \sigma-M=\pm \sqrt{M^2+L^2(N^2-1)}
\label{new102}
\end{equation}  
The location of the THs for the Glass-Mashhoon metric is given by
\begin{equation}
1-\frac{2M_{MSH}}{R}=\nabla_{c}R \nabla^{c}R=(\frac{\partial_{\bar{r}} R}{B})^2-(\frac{\partial_{t} R}{A})^2
\label{new103}
\end{equation}
where $c = 0, 1 $. Thus, using Eqs.~\eqref{new78}, \eqref{new98}, and \eqref{new102}, and assuming $H\neq 0$, we find that on the TH one has $L^2=0$. If, in addition, $N^2-1$ and $\sigma$ are finite at $L^2=0$, then \footnote{Similar to the flat McVittie case, Eq.~\eqref{new104} is only valid for $H>0$. In the Schwarzschild limit, where $a(t)\to 1$ (and hence $H\to 0$), the assumptions that $\sigma\vert_{L^2=0}$ and $(N^2-1)\vert_{L^2=0}$ are finite break down due to Eqs.~\eqref{new78}, \eqref{new98}, and \eqref{new102}. Therefore, Eq.~\eqref{new104} is not valid in the $H=0$ limit. In this limit, to determine which sign in Eq.~\eqref{new102} is admissible, one should follow the same procedure as in the flat McVittie discussion in Section \ref{II.B}.
}
\begin{equation}
\frac{v}{f}|_{L^2=0}=-M|_{L^2=0}=\pm M|_{L^2=0}.
\label{new104}
\end{equation}
 Based on Eq.~\eqref{new98} and the definition of the areal radius $R$, we have $\partial_{t}R=RHA$, so with the assumption $ H > 0 $, we have $ M > 0 $, and the negative sign in Eq.~\eqref{new104}  is acceptable. We can see that for TH, $M|_{L^2=0}={A^2}/{(\partial_{t} R)}={A}/{(RH)}>0$\footnote{By determining the free coefficients of the Glass-Mashhoon metric such that it reduces to the flat McVittie metric and according to ${(\frac{1-{m_0}/{(2\ \bar{r}\ a\left(t\right)})}{1+{m_0}/{(2\ \bar{r}\ a\left(t\right)})})}^2=1-{2m_0}/{R}$, we have ${v}/{f}|_{L^2=0}=-M|_{L^2=0}=-1$ which is consistent with Eq.~\eqref{new22}.}. Therefore, after simplification, $c(R,\tau)$ and $ v(R,\tau)$  are written in their final forms as:
\begin{equation}
\begin{cases}
c(R,\tau)=f \frac{L}{L'} \\
v(R,\tau)=-c \sqrt{\frac{2M_{MSH}}{R}} \\
L'=\sqrt{\nabla_{c}R \nabla^{c}R} 
\end{cases}
\label{new105}
\end{equation}  
Finally, the null expansion parameters for the Glass-Mashhoon spacetime in PG coordinates are obtained as:  
\begin{equation}
\begin{cases}
 \theta_{+}=\frac{2}{R}(1+\sqrt{\frac{2 M_{MSH}}{R}})\\
 \theta_{-}=- \frac{2}{R}(1-\sqrt{\frac{2 M_{MSH}}{R}})
\end{cases}
\label{new106}
\end{equation}  
Thus, the results of the PG coordinate calculations show that the generalized solution for the Glass-Mashhoon spacetime in an expanding universe, does not describe a cosmological BH within our Hayward-Inspired quasi-local framework, due to the absence of an FTH, which is consistent with the result of  the special case result of the flat McVittie solution obtained in Section~\ref{II.B}. In that section, we showed that the solution does not describe a cosmological BH.

Based on Eq.~\eqref{new91}, both $\theta_{\pm}$ can vanish in the Glass--Masshoon spacetime. 
However, the analysis in Glass--Masshoon PG coordinates (for non-negative MSH mass cases), 
based on Eq.~\eqref{new106}, shows that only $\theta_{-}$ can vanish. 
Thus, both PG and isotropic-like coordinates ($\bar{r}$) admit PTH(s) corresponding to 
$\theta_{-}=0$ and $\theta_{+}>0$. Nevertheless, PG coordinates do not cover the PTH(s) attached to the 
second asymptotic region associated with $\theta_{+}=0$ and $\theta_{-}>0$. The general 
calculation for a dynamical spherically symmetric solution with non-negative MSH mass in PG coordinates (presented in Section \ref{II.B} and based on Eq.~\eqref{new17}) demonstrates that according to the analysis of null expansions in PG coordinates, the only TH(s) present are those corresponding to FTH ($\theta_{+}=0$ and $\theta_{-}<0$) and PTH ($\theta_{-}=0,\,\theta_{+}>0$ ).

Consequently, the THs attached to the second asymptotic region (i.e., FTH corresponding to 
$\theta_{-}=0,\,\theta_{+}<0$ and PTH corresponding to $\theta_{+}=0,\,\theta_{-}>0$) are not 
captured by the null expansion analysis in PG coordinates\footnote{This holds for both ingoing and outgoing PG coordinates.}. Therefore, there is no inconsistency 
between our results in Sections \ref{V.D} and \ref{V.E}, as both PG and $\bar{r}$ coordinates 
(the latter of which differ from isotropic coordinates only by a smooth coordinate transformation) 
admit PTH(s) corresponding to $\theta_{-}=0$ and $\theta_{+}>0$.  The expansion analysis in PG coordinates does not cover the case of $\theta_{+}=0,\,\theta_{-}>0$ because TH(s) attached to the second asymptotic region are not covered by these coordinates. \footnote{From the definition of the MSH mass, we know that $1-{2M_{MSH}}/{R}=\nabla_{c}R \nabla^{c}R \equiv L'^2$. Thus, a line element corresponding to a general spherically symmetric metric, assuming the non-negativity of the MSH mass, takes the form $ds^2=-F'^2 L'^2 dT^2+ dR^2/{(L'^2)}+R^2 d\Omega^2$ in the Schwarzschild-like coordinates. Here, $F'$ represents the integration factor.
	By applying a coordinate transformation to the line element \eqref{new98} in Schwarzschild-like coordinates, we obtain ${1}/{L'^2}=N^2+{M^2}/{L^2}$. Furthermore, the combination of Eqs.~\eqref{new101} and \eqref{new102} leads directly to Eq.~\eqref{new105}.}
\section{Concluding Remarks}\label{VI}

In this paper we analyze the THs and the physical viability of several spacetimes in order to assess their potential as candidates for cosmological BH models within our Hayward-inspired quasi-local framework for expanding backgrounds. In this framework, a candidate model is evaluated in terms of whether it admits at least an FOTH as a BH-type TH and a P(I)TH as a cosmological TH.

First, we examined the flat McVittie spacetime in isotropic coordinates. 
We demonstrated that in the flat McVittie spacetime, the WEC and, consequently, the NEC, outside the McVittie curvature singularity, are always satisfied, for a decelerating universe by considering a perfect fluid with vanishing radial energy flux (the McVittie condition) and computing its energy density and pressure. 
We then interpreted the solution for a matter-dominated, flat McVittie spacetime, in which the SEC is also always satisfied. As long as the calculations of outgoing and ingoing null ray congruences in isotropic coordinates remain valid, we showed that the flat McVittie spacetime with a positive Hubble parameter outside the curvature singularity contains an outer, inner, or degenerate PTH. 

In the matter-dominated flat McVittie spacetime, at a given time, the ingoing null expansion parameter has a degenerate root, causing two horizons to coincide, both are POTH. 
Over time, the larger horizon becomes the cosmic-type PITH, while the smaller horizon becomes the WH-type POTH. 
Since the NEC is always satisfied in this case, the smaller horizon is always spacelike, while the larger horizon is initially spacelike (becoming null at a specific instant) and then timelike. 

Subsequently, we introduced the line element of a spherically symmetric dynamical spacetime with a non-negative MSH mass in PG coordinates. We obtained the ingoing and outgoing null expansions in the general case. As long as the relevant calculations remain valid, we showed that PG coordinates cannot describe a family of THs attached to the second asymptotic region. 
For the flat McVittie spacetime, whose MSH mass is always non-negative, both THs are attached to the first asymptotic region. As expected, the result in PG coordinates is similar to the isotropic description: the flat McVittie spacetime with an asymptotically vanishing Hubble parameter has no FTH. In this case, the flat McVittie spacetime does not correspond to a cosmological BH.

Next, we investigated the Culetu spacetime in an expanding background. We rewrote the Culetu line element in terms of the cosmic time and derived explicitly the future-directed outgoing and ingoing radial null vector fields. From these, we computed the corresponding null expansions and used them to locate the THs in an expanding background ($H>0$). We found that the Culetu spacetime admits two THs: a PTH and an FTH, with the areal radius of the FTH always smaller than that of the PTH. Moreover, the FTH is always located inside the event horizon of this spacetime. For a matter-dominated background, we analyzed the time evolution of the PTH and FTH and showed that, because the PTH can lie outside the event horizon over after a finite time interval, the NEC must be violated in some region(s) of the matter-dominated Culetu spacetime. We then classified the horizons for $H>0$ and showed that there is an FOTH as a BH-type TH and a cosmological TH as a POTH or a PITH, and can become a PDTH at isolated instants depending on the behavior of $H(t)$.
	
We also evaluated the ECs on both THs for the matter-dominated Culetu spacetime. On the PTH, the NEC, WEC, and SEC are satisfied, while the DEC holds only over a restricted range of radii. In contrast, on the FOTH the WEC and DEC are always violated, whereas the NEC and SEC can hold only for specific radii. Finally, we reformulated the Culetu spacetime in PG coordinates and showed that the TH structure obtained in PG coordinates agrees with the results in the PG-Schild coordinates. We also found that at early times, when the WEC holds, the outgoing PG patch is compatible with culetu spacetime and the relevant TH is a PTH. At later times, and the WEC is violated, the ingoing PG patch becomes compatible and the relevant TH is FTH. This provides a concrete demonstration of how physical conditions on the THs constrain the admissible PG patch and help determine the TH type in these coordinates.
Finally, within our Hayward-inspired framework, we showed that the Culetu spacetime in an expanding universe describes a cosmological BH, in agreement with \cite{sato}.


We then turned to the Sultana--Dyer spacetime in Kerr--Schild coordinates. 
We began by rewriting the Sultana--Dyer line element in terms of the cosmic time and derived explicitly the future-directed outgoing and ingoing radial null vector fields. From these, we computed the corresponding ingoing and outgoing null expansions and used them to locate the THs in an expanding background ($H>0$). We found that for $H>0$ the Sultana--Dyer spacetime admits two THs: a PTH and an FTH. We obtained closed-form expressions for both THs and showed that the areal radius of the FTH is always smaller than that of the PTH. Moreover, the FTH is always located inside the event horizon. For the matter-dominated Sultana--Dyer case, we also examined the time evolution of the TH radii and showed that, because the PTH can lie outside the event horizon after a finite time interval, the NEC must be violated in some region(s) of the matter-dominated Sultana--Dyer spacetime.
We then analyzed the nature and signature of the THs for $H>0$. We found that the FTH is an FOTH and spacelike, whereas the PTH is a past-type TH whose signature can be spacelike, timelike, or (instantaneously) null, depending on the time interval and on the behavior of $H(t)$. In the matter-dominated background, this identifies the PTH as the cosmological TH and the FTH as the BH-type TH.

We also evaluated the ECs on the THs of the matter-dominated Sultana--Dyer spacetime. On the PTH, the NEC and SEC are always satisfied, while the WEC and DEC are satisfied when the PTH lies outside the event horizon. In contrast, on the FTH (the BH-type TH) we found that all ECs are violated in the matter-dominated Sultana--Dyer spacetime.
Finally, we studied the Sultana--Dyer spacetime in PG coordinates and investigated the THs for $H>0$. We determined the signs of the ingoing and outgoing null expansions in PG coordinates as functions of the areal radius $R$ and showed that the PG description captures the existence of both the smaller FTH and the larger PTH. As in the expanding flat McVittie case, to select the appropriate PG patch we evaluated $g_{\tau R}$ for the Sultana--Dyer spacetime in PG coordinates on the THs under the condition $H>0$, and used its sign to determine the admissible PG representation.
As a result, within our Hayward-inspired framework, we showed that the Sultana--Dyer spacetime in an expanding universe describes a cosmological BH, in agreement with \cite{sato}.


We then examined the general features of the Glass-Mashhoon class of spacetimes, a generalized cosmological metric. We demonstrate that, provided the congruence of null rays remains well-defined, this spacetime possesses only PTHs in an expanding universe, even though it allows both ingoing and outgoing expansion parameters to vanish. Considering a perfect fluid with vanishing radial energy flux, 
We then calculated the energy density and pressure associated with the Glass--Mashhoon spacetime. We discussed the ECs and identified certain regions in a decelerating universe where these ECs are satisfied. Next, we generalized the matter-dominated flat McVittie spacetime as a special case of the Glass--–Mashhoon solution and studied the curved McVittie spacetime, which asymptotically approaches the FLRW spacetime with nonzero spatial curvature in the appropriate limit. We then computed the ECs for this solution. 
Unlike in the flat McVittie case, the energy density depends on the spatial coordinate. For negative spatial curvature, it is not always positive.
The pressure function also takes a more complicated form.  
We examined the NEC, WEC, DEC, and SEC in the context of a matter-dominated, curved McVittie spacetime with positive spatial curvature. Unlike the flat McVittie case, the regions where these ECs apply are much more complex. However, these ECs generally hold at early times.

Using comoving coordinates, we demonstrated that the ingoing and outgoing null expansion parameters of the curved McVittie spacetime outside the curvature singularity can both vanish. However, only PTHs (outer, inner, or degenerate) can exist. 
We demonstrated that, in curved matter-dominated McVittie spacetime, the roots of the outgoing null expansion parameter indicate the presence of a PTH after a specific time if the normalized spatial curvature is smaller than a certain threshold. If the spatial curvature exceeds that threshold, a degenerate root first appears at a specific time corresponding to a PITH. Subsequently, two THs arise. As time progresses, the smaller TH moves out of the admissible interval, leaving only one TH for larger, but still finite, times.
The analysis of the roots of the ingoing null expansion parameter shows that initially there is a degenerate root\footnote{This root appears at a time that is always greater than the formation time of the PTH corresponding to the root of the outgoing null expansion parameter.}, corresponding to a POTH. Then, two PTHs appear ,similar to the flat McVittie case.  
The curved matter-dominated McVittie spacetime with negative spatial curvature exhibits entirely different characteristics. In fact, depending on the absolute value of the normalized spatial curvature parameter, one can observe a variety of PTHs in certain time intervals.

Next, we confirmed that the results are consistent with the general analysis by considering the computations in the general case and analyzing the existence and properties of the THs and the ECs on these THs. We focused on the specific case of the curved matter-dominated McVittie spacetime with positive spatial curvature and a fixed value. The conclusion of the calculations is that the McVittie spacetime, based on Hayward's formalism, can not describe a cosmological BH.

Finally, we derived the PG\footnote{Applicable in cases with non-negative MSH mass.} 
and Schwarzschild line elements for the general Glass--Mashhoon solution. 
The results for the ingoing and outgoing expansion parameters in Glass--Mashhoon PG coordinates allow for the existence of PTHs attached to the first asymptotic region. 
This is consistent with what we obtained in comoving coordinates for the Glass--Mashhoon spacetime in an expanding universe. 
However, due to the limitations of PG coordinates, they do not display PTHs attached to the second asymptotic region.

There are additional special cases of the Glass--Mashhoon solution that are  worth discussing. 
One such case is when $\alpha,\delta>0$, $\beta\neq 0$, and $\gamma=0$. 
This solution contains a weak curvature singularity and establishes a minimum value for the comoving radial coordinate when $X\equiv\alpha\beta$ is positive. 
In this case, the NEC, WEC, and SEC are always satisfied outside the curvature singularity in a matter--dominated universe. The DEC is also satisfied in certain regions. 
For this case and outside the curvature singularity, it can be shown that only the ingoing expansion parameter can vanish. This corresponds to a PTH attached to the first asymptotic region. 
For $X<0$, the ECs become more intricate and both the ingoing and outgoing expansion parameters can vanish. Thus, there exist PTHs that are attached to both asymptotic regions.
As expected, this class of solutions also does not describe a cosmological BH.
\section*{Acknowledgement}
The authors would like to express their sincere gratitude to Professor Mashhoon for his interest in the content of this paper. M.E and F.S would like to thank the Iran National Science Foundation (INSF) for supporting this research under grant
number 4043302. F.S. is grateful to the University of Tehran for supporting this work under
a grant provided by the University Research Council.

\end{document}